\documentclass{article}

\usepackage[main,final]{neurips2026/neurips_2026}

\usepackage[utf8]{inputenc} 
\usepackage[T1]{fontenc}    
\usepackage[hidelinks]{hyperref} 
\usepackage{url}            
\usepackage{booktabs}       
\usepackage{amsfonts}       
\usepackage{nicefrac}       
\usepackage{xcolor}         

\usepackage{longtable}
\usepackage{amsmath}
\usepackage{amssymb}
\usepackage{mathtools}
\usepackage{amsthm}
\usepackage{microtype}
\usepackage{graphicx}
\usepackage{subcaption}
\usepackage{booktabs} 
\usepackage{lipsum}
\usepackage{enumitem}
\usepackage{multirow}
\usepackage{colortbl}
\usepackage{caption}
\usepackage{float}
\usepackage{algorithm}
\usepackage{xcolor}
\usepackage{placeins}
\usepackage{tikz}
\usetikzlibrary{arrows.meta, positioning, fit, calc, backgrounds, decorations.pathreplacing, matrix}
\tikzset{
  archfont/.style={font=\scriptsize},
  blk/.style={draw=black!60, rounded corners=2pt, align=center, inner sep=4pt, minimum height=9mm},
  arr/.style={-Latex, thick},
  note/.style={font=\scriptsize, align=center},
  grp/.style={draw=black!40, rounded corners=2pt, inner sep=5pt},
  brace/.style={decorate, decoration={brace, amplitude=4pt}},
}
\usepackage{pgfplots}
\usepgfplotslibrary{groupplots}
\pgfplotsset{compat=1.18}

\usepackage{xspace}
\newcommand{\cider}{\textsc{CIDER}\xspace}
\definecolor{xblue}{HTML}{4169E1}
\definecolor{xgreen}{HTML}{036C3A}
\definecolor{xpurple}{HTML}{9838B1}
\definecolor{xslategray}{HTML}{70818F}
\definecolor{xorange}{HTML}{FF8C00}
\definecolor{xcyan}{HTML}{06AEEF}
\definecolor{xred}{HTML}{FF0000}
\definecolor{xgray}{HTML}{808080}
\definecolor{xxgreen}{HTML}{009F86}
\definecolor{xsienna}{HTML}{8B4512}
\definecolor{xxgreen}{HTML}{009F86}
\definecolor{xxpurple}{HTML}{623E99}
\definecolor{xolive}{HTML}{556B2F}

\definecolor{CIDERGreenFill}{RGB}{210,245,220}

\usepackage[capitalize,noabbrev]{cleveref}
\usepackage{wrapfig}
\usepackage{tabularx}
\newcolumntype{Y}{>{\raggedright\arraybackslash}X}

\theoremstyle{plain}

\theoremstyle{definition}

\theoremstyle{remark}

\newcommand{\mask}{[\textsf{Mask}]}
\usetikzlibrary{positioning,arrows.meta,calc,fit,decorations.pathreplacing}
\tikzset{
  archfont/.style={font=\scriptsize},
  blk/.style={draw=black!60, rounded corners=2pt, align=center, inner sep=4pt, minimum height=9mm},
  arr/.style={-Latex, thick},
  note/.style={font=\scriptsize, align=center},
  grp/.style={draw=black!40, rounded corners=2pt, inner sep=5pt},
  brace/.style={decorate, decoration={brace, amplitude=4pt}},
}

\title{Structured Masked Diffusion for Shared-Codebook Unsourced Random Access Decoding}

\author{%
  Taekyun Lee\textsuperscript{1,*} \quad
  Jiyoung Yun\textsuperscript{2,*} \quad
  Jeffrey G. Andrews\textsuperscript{1} \quad
  Hyeji Kim\textsuperscript{1} \\
  \textsuperscript{1}Dept. of ECE, The University of Texas at Austin \quad
  \textsuperscript{2}Dept. of ECE, Seoul National University \\
  \texttt{taekyun@utexas.edu} \quad \texttt{jyyun423@snu.ac.kr} \\
  \texttt{jandrews@ece.utexas.edu} \quad \texttt{hyeji.kim@austin.utexas.edu} \\
  \textsuperscript{*}Equal contribution.
}

\begin{document}

\maketitle

\begin{abstract}
In shared-codebook unsourced random access (URA), a receiver must recover an unordered set of messages from simultaneous transmissions. Its outer decoder assembles ambiguous slot-wise evidence into distinct codewords while enforcing parity constraints. We propose \cider, a learned outer decoder that combines masked-diffusion refinement with cross-row demixing and parity-aware propagation on the code's Tanner graph. At higher per-bin loads, quality-guided remasking selectively re-decodes low-confidence rows. On the evaluated GF(64) LDPC codes at $K=2$, \cider has lower measured symbol and codeword error rates than SIC-BP and FFT-BP at all four code lengths. In same-GPU decoder-only measurements, the deployed mixed-precision implementation is $8.2\times$ to $167\times$ faster than our optimized FFT-BP implementation across those lengths. We also evaluate higher loads, estimated user counts, and detector mismatch. Code is available at \url{https://github.com/jiyunyoung/CIDER/tree/neurips-2026}.
\end{abstract}

\section{Introduction}

Modern large-scale networked systems---from massive IoT and dense sensing to distributed learning and autonomous systems \citep{schwarting2018planning,bonawitz2019towards,kaloer2024wireless6g}---increasingly involve many sporadically active devices with short payloads, making per-device coordination costly. When several devices transmit simultaneously, the receiver must recover multiple coded messages from one noisy aggregate observation \citep{wang1999iterative,boutros2002iterative,polyanskiy2017perspective}.

Unsourced random access (URA) makes this task particularly challenging: multiple devices share a common codebook and transmit without prior coordination, possibly into the same bin \citep{polyanskiy2017perspective}. The receiver must recover the unordered set of transmitted messages from a noisy mixture without knowing which devices were active or which device sent each codeword. Because the shared-codebook observation carries no per-user labels, decoding requires both assigning plausible slot-wise symbols to distinct message hypotheses and enforcing global code constraints. Practical receivers commonly partition users into bins and perform joint decoding within each bin \citep{liva2024unsourced,marshakov2019polar,pradhan2022sparse}. We focus on this within-bin outer-decoding stage, conditioned on an estimated per-bin load and a known code construction.

The receiver we study has two stages. A symbol-level soft detector maps the raw channel observation to a soft evidence matrix $S \in \mathbb{R}^{L \times Q}$, in which each of the $L$ positions (``slots'' in URA) carries a length-$Q$ score vector over candidate symbols. A channel decoder (``decoder'' in this paper) then consumes $S$ to recover the unordered set of $K$ transmitted codewords under the code constraints. Because all users share a single codebook and the symbols transmitted in each slot are superposed at the receiver, $S$ aggregates contributions across users and cannot be decoded one user at a time: a \emph{multiuser decoder} is required, which must resolve ownership ambiguity---which (slot, symbol) candidates belong to the same user---while enforcing global codeword consistency under collisions, missed symbols, and false candidates.

Classical multiuser decoders tackle the association problem through three mechanisms. \emph{Joint factor-graph message passing} couples user-separation and code constraints into a single graph and runs BP over all users simultaneously \citep{pradhan2022sparse,amalladinne2022integrating,ebert2022coded}. \emph{Stitching-based list-recovery} recovers per-slot top candidates and combinatorially assembles valid codewords across slots \citep{amalladinne2020coded,andreev2022list}. \emph{Successive interference cancellation} (SIC) decodes users in successive stages and subtracts each stage's estimated contribution from the residual evidence \citep{andreev2020polar,vem2019sic,yun2024erasure}. These methods face distinct limitations. SIC is order-dependent and can propagate early errors. Joint BP may require many iterations with costly non-binary check updates, while the candidate space for stitching-based list recovery grows combinatorially with per-slot ambiguity. In short-packet random access, where decoding latency is a practical constraint, these computational bottlenecks directly limit the practicality of the receiver.

Learned decoders offer another route, but a generic one-shot model does not explicitly encode the permutation invariance or global code constraints of shared-codebook decoding \citep{choukroun2024foundation}. Without mechanisms for assigning evidence to distinct hypotheses, it can produce duplicate rows or invalid codewords. A multiuser neural decoder must therefore separate competing user hypotheses while checking that each recovered row is consistent with the code.

\begin{figure*}[h]
  \centering
  \includegraphics[width=\textwidth]{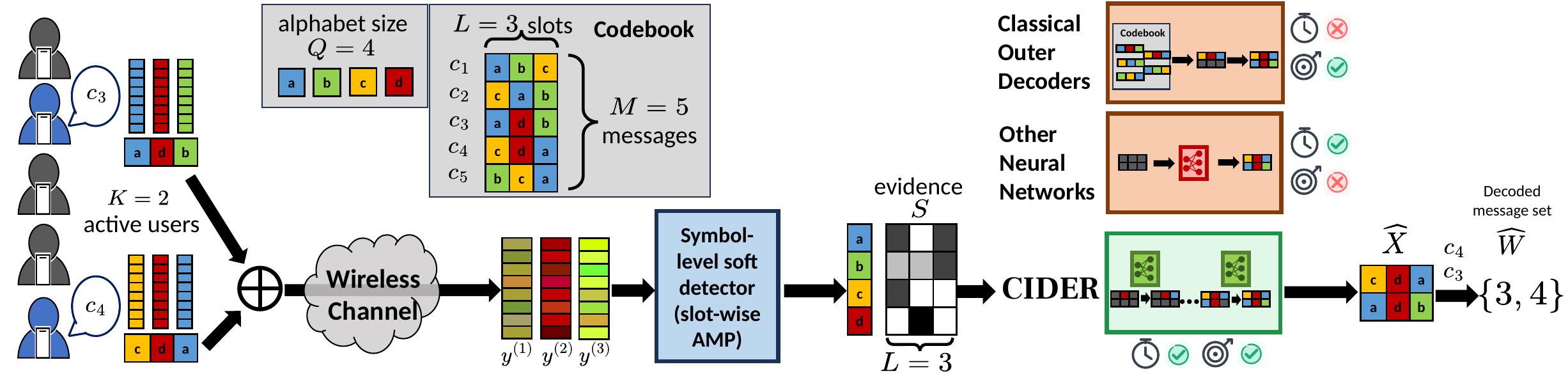}
  \caption{Two-stage receiver pipeline. Active users transmit codewords from a shared codebook over a shared channel. A fixed AMP-based symbol detector
converts slot observations into the evidence matrix $S$, and the multiuser decoder
maps $S$ to an unordered set of decoded codewords.
}
  \label{fig:ura_pipeline}
\end{figure*}

We introduce \cider (\textbf{C}onstraint-aware \textbf{I}terative Masked \textbf{D}iffusion Decoding for \textbf{E}rror-correcting \textbf{R}efinement), a learned shared-codebook multiuser decoder. \cider replaces rule-based decoding with fixed-step masked-diffusion refinement. It turns the ambiguous evidence matrix $S$ into $K$ codewords using two structured operations: demixing, which makes hypothesis rows compete for high-evidence symbols, and parity-aware propagation, which injects sparse code constraints during refinement. This yields a parallel neural analogue of iterative multiuser decoding. We evaluate its reliability and latency against specified neural and classical baselines under a shared evidence interface. Our contributions are as follows.

\paragraph{Contributions.}
\begin{itemize}[leftmargin=1.2em,itemsep=0.5pt,topsep=0.5pt]
  \item We develop a masked-diffusion outer decoder for within-bin, shared-codebook URA. Row competition and parity-aware propagation address duplicate hypotheses and invalid codewords under this set-valued decoding objective.
  \item We compare with task-aligned classical decoders, tuned neural baselines, and non-diffusion controls on the same evidence interface. The reported error and latency advantages are specific to these implementations and operating points; same-GPU measurements are included in Appendix~\Cref{app:response_results}.
  \item We provide extensive scaling experiments and modulewise ablations that isolate why the design is necessary
(demixing, parity-aware propagation and remasking) and
how performance changes with problem size and load.
\end{itemize}

\section{Related Work}

\paragraph{Shared-codebook access.}
Unsourced random access studies recovering an unordered set of messages from a noisy superposition, and is widely analyzed under the Gaussian many-access channel and related models \citep{polyanskiy2017perspective,kowshik2021fundamental}.
Many practical constructions adopt a modular two-stage pipeline,
where a front end produces per-slot soft evidence and a later stage assembles code-consistent messages \citep{amalladinne2020coded,fengler2021sparcs}.

\paragraph{Rule-based multiuser decoding.}
Rule-based multiuser decoders address the set-assembly problem using stitching/list recovery across slots or cancellation/peeling-style procedures combined with iterative decoding \citep{amalladinne2020coded,andreev2022list,yun2024erasure,vem2019sic,ordentlich2017low}.
These methods can be effective but may become brittle or expensive under high ambiguity, motivating alternatives that are iterative yet parallelizable.

\paragraph{Learning-based decoding and discrete diffusion.}
Learned decoders have been explored for channel decoding by augmenting message passing (e.g., neural belief propagation) \citep{nachmani2016learning,nachmani2017improved} and with attention-based architectures \citep{choukroun2022ecct}, as well as end-to-end learned coding \citep{jiang2019turboae}.
In parallel, discrete diffusion and masked denoising models provide iterative refinement mechanisms for categorical variables \citep{austin2021d3pm,chang2022maskgit,lou2024sedd,sahoo2024mdlm}.
Our work connects these threads by using masked denoising for shared-codebook URA outer decoding. \citet{tarizzo2025learned} jointly learn a URA codebook and AMP-style receiver for digital over-the-air aggregation in federated learning. Their receiver targets fragment-level aggregate recovery, whereas our outer decoder recovers complete, distinct codewords from a fixed slot-wise evidence interface. The different receiver outputs and code constructions preclude a like-for-like error-rate comparison.
Additional background on wireless random access and related work details are deferred to Appendix~\Cref{app:wireless_context,app:rw_appendix}.

\section{Preliminaries}
\label{sec:prelim}
This section introduces
the receiver interface used throughout the paper. 
We first define the two-stage joint decoding pipeline and the evidence matrix $S$ produced by the fixed symbol-wise soft detector (\Cref{sec:prelim_ura_framework}).
We then review masked discrete diffusion for discrete infilling (\Cref{sec:prelim_maskdiff}), which we later use as the basis for our learned multiuser decoder.

\paragraph{Notation.}

Uppercase symbols denote random variables (or random objects), and the corresponding lowercase symbols denote realizations; we also use calligraphic letters (e.g., $\mathcal{C},\mathcal{W}$) for sets. Subscripts index components (e.g., $v_a$, $M_{i,j}$), and parenthesized superscripts denote slot/time indices (e.g., $Y^{(\ell)}$). When a quantity is naturally an entry of a matrix or tensor with a slot dimension, the slot index is written as a subscript alongside the other component indices (e.g., codeword slot symbol $c_{m,\ell}$, evidence entry $S_{\ell,a}$).

\subsection{Shared-codebook joint decoding and the two-stage receiver}
\label{sec:prelim_ura_framework}

\Cref{fig:ura_pipeline} illustrates the complete two-stage receiver pipeline through a toy example. Suppose two active users transmit the shared-codebook codewords \(c_3=\texttt{adb}\) and \(c_4=\texttt{cda}\), where the alphabet is \(\{\texttt{a},\texttt{b},\texttt{c},\texttt{d}\}\) and the codeword length is \(L=3\). Because the users transmit simultaneously, the receiver does not observe two labeled codewords. Instead, it observes a noisy superposition over the three slots. The \textbf{symbol-level soft detector} processes this superposed observation and outputs an evidence matrix \(S\in\mathbb{R}^{L\times Q}\), where \(S_{\ell,a}\) scores how plausible it is that symbol \(a\) appeared in slot \(\ell\).

The \textbf{multiuser decoder} then solves the global assembly problem. Starting from an all-\mask{} grid \(X^{(0)}\in([Q]\cup\{\mask\})^{K\times L}\), \cider iteratively refines \(X^{(t)}\) over \(T\) masked-diffusion steps. During this unmasking process, \cider must decide both which symbols to place and how to assign them to the \(K\) user-hypothesis rows. For the example in \Cref{fig:ura_pipeline}, one valid final decoded grid is $
\hat X =
\begin{bmatrix}
\text{\texttt{c}} & \text{\texttt{d}} & \text{\texttt{a}}\\
\text{\texttt{a}} & \text{\texttt{d}} & \text{\texttt{b}}
\end{bmatrix},$ up to row permutation. Mapping these rows back to the shared codebook gives the decoded message set \(\hat{\mathcal W}=\{3,4\}\). Thus, the detector provides local slot evidence, while the multiuser decoder assembles it into complete, code-consistent messages.

\paragraph{Frame, shared codebook, and unordered target.} We study a wireless uplink where many devices send short reports to a common receiver. A frame is divided into \(L\) slots, and each active device transmits one symbol per slot, so each message corresponds to a length-\(L\) codeword. All devices use the same public codebook $
\mathcal{C}=\{c_m\}_{m\in[M]}, c_m\in[Q]^L,$ where \(m\) is a payload index and \([Q]\) is the \(Q\)-ary alphabet. In a given frame, an unknown set of active payloads
\(\mathcal{W}^\star\subseteq[M]\) with \(|\mathcal{W}^\star|=K\) is transmitted simultaneously over the shared channel. The receiver observes a noisy superposition \(Y\) of these transmissions and must recover the unordered set of transmitted messages, not their device identities.

\paragraph{Symbol-level soft detector: local slot evidence.} The first stage handles the physical-layer mixture. It maps the raw channel observation \(Y\) to a slot-wise evidence table \(S\in\mathbb{R}^{L\times Q}\), where \(S_{\ell,a}\) is the log-posterior score that symbol \(a\in[Q]\) is active in slot \(\ell\). Multiple symbols can be active in one slot. This evidence is local: it tells us which symbols look plausible at each slot, but not which symbols should be grouped together into the same user's codeword. In our experiments, this detector is fixed and implemented by AMP; details are provided in Appendix~\Cref{app:amp_mmse}.

\paragraph{Multiuser decoder: global assembly.} The later
stage takes only \(S\) and produces an unordered decoded grid \(\hat X\in[Q]^{K\times L}\), whose rows are candidate codewords. Since the target is an unordered message set, the row order of \(\hat X\) is immaterial. The decoded message set is obtained by mapping valid rows of \(\hat X\) back to the shared codebook, yielding \(\hat{\mathcal W}\subseteq[M]\).
Thus, the multiuser decoder must solve two coupled problems: it must resolve ownership ambiguity across rows and enforce global code constraints across slots. Our goal is to learn this 
decoder while keeping the detector and evidence interface \(S\) fixed. A full formal model is given in Appendix~\Cref{app:ura_formal}.

\subsection{Masked discrete diffusion for discrete infilling}
\label{sec:prelim_maskdiff}

We briefly review masked (absorbing-state) discrete diffusion, which we use as a conditional discrete infilling mechanism \citep{chang2022maskgit,lou2024sedd,sahoo2024mdlm}.
Starting from a fully masked sequence, the model repeatedly predicts missing tokens and reveals a subset of them over a fixed number of refinement steps.
In our joint-decoding setting, the side information is the detector evidence \(S\), and the sequence is reshaped as a \(K\times L\) grid.

\paragraph{Masked forward process.}
Let \([Q]=\{0,1,\ldots,Q-1\}\) be the token alphabet and let \(\mask\) denote an absorbing mask token.
For a clean sequence \(X_{1:N_{\mathrm{seq}}}\in[Q]^{N_{\mathrm{seq}}}\), the step-\(t\) corrupted sequence
\(\tilde X^{(t)}\in([Q]\cup\{\mask\})^{N_{\mathrm{seq}}}\) is obtained by independently masking each position with probability \(\gamma_t\), where \(\gamma_0=1\) and \(\gamma_T=0\):
\begin{equation}
\label{eq:masked_forward}
q\!\left(\tilde X^{(t)}_{j} \,\middle|\, X_j=x_j\right)
=
\begin{cases}
1-\gamma_t, & \tilde X^{(t)}_{j}=x_j,\\
\gamma_t,   & \tilde X^{(t)}_{j}=\mask,\\
0,          & \text{otherwise}.
\end{cases}
\end{equation}

\paragraph{Conditional denoising objective.}
At step \(t\), a denoiser receives the masked sequence \(\tilde x^{(t)}\), side information \(\xi\), and the step index \(t\), and predicts a categorical distribution
\(\hat x_{\theta,j}(\tilde x^{(t)},\xi,t)\) over \([Q]\) for each position \(j\).
Following the standard masked-diffusion objective, we train only on masked positions:
\begin{equation}
\label{eq:masked_ce_objective}
\mathcal{L}(\theta)
=
\mathbb{E}_{t, \tilde X^{(t)}\sim q(\cdot\mid X=x)}
\sum_{j\in\Omega_t}
\mathrm{CE}\Big(x_j,\hat x_{\theta,j}(\tilde x^{(t)},\xi,t)\Big),
\end{equation}
where \(\Omega_t=\{j:\tilde X^{(t)}_j=\mask\}\) and \(\mathrm{CE}(\cdot,\cdot)\) is categorical cross-entropy \citep{sahoo2024mdlm}.

\paragraph{Inference.}
At inference, we initialize \(\tilde x^{(0)}=(\mask,\ldots,\mask)\) and iteratively reveal high-confidence predictions according to a cosine reveal schedule \citep{chang2022maskgit}. In \cider, we set \(\xi\equiv S\) and use this refinement process to fill a \(K\times L\) decoded grid; implementation details are given in Appendix~\Cref{app:maskdiff_infer}.

\section{\cider framework}
\label{sec:method_cider_denoiser}

\begin{figure}[t]
\centering
\captionsetup{font=footnotesize}
\includegraphics[width=\linewidth]{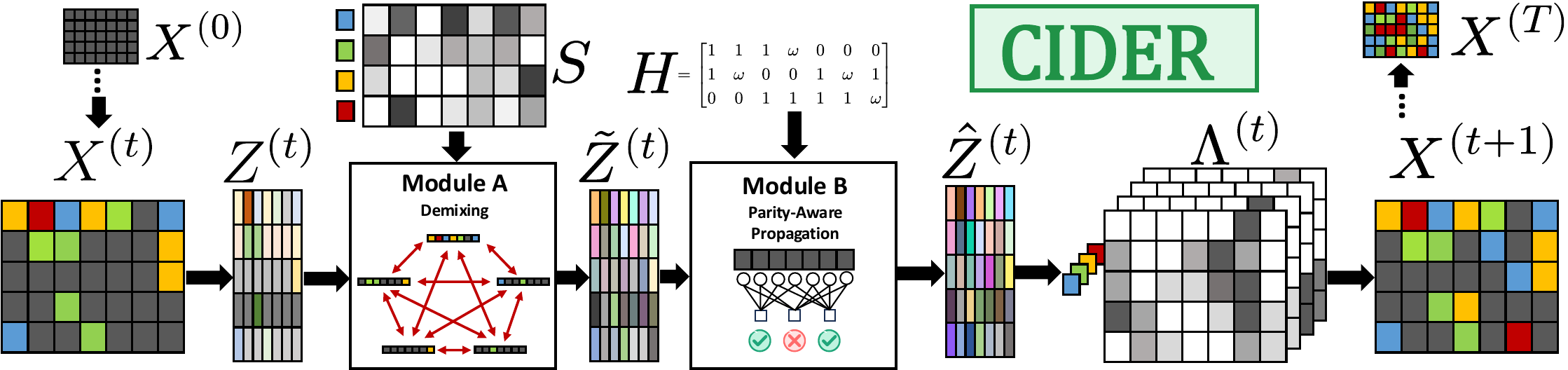}

\caption[CIDER structured denoiser]{
\textbf{CIDER structured denoiser.}
At refinement step $t$, the partially revealed grid $X^{(t)}$ is embedded into token latents $Z^{(t)}$.
Module A demixes the row hypotheses by making them compete for the shared slot-wise evidence $S$, producing demixed latents $\tilde Z^{(t)}$.
Module B then propagates parity information using the parity-check matrix $H$, producing refined latents $\hat Z^{(t)}$ and logits $\Lambda^{(t)}$ over the $Q$ symbols at each grid position.
The reveal rule uses $\Lambda^{(t)}$ to update the discrete grid from $X^{(t)}$ to $X^{(t+1)}$, and repeated refinement produces the final decoded grid $X^{(T)}=\hat X$.
}
\label{fig:arch_cider_main}

\end{figure}

\paragraph{Design principle.} The decoder receives only a slot-wise evidence matrix \(S\): it tells us which symbols are locally plausible in each slot, but not how those symbols should be grouped into \(K\) complete codewords. A generic masked-diffusion decoder can refine missing tokens, but it does not know the two structures that make shared-codebook joint decoding difficult: \textbf{(A) user hypotheses must be demixed}, so that different rows explain different transmitted codewords rather than collapsing to the same high-evidence symbols, and \textbf{(B) each row must be code-consistent}, so that the assembled sequence satisfies the code constraints. \cider keeps the standard masked-diffusion refinement loop, but replaces the generic per-step denoiser with a structured denoiser designed around these two requirements: \textbf{Module~A} separates competing rows through evidence-level competition, while \textbf{Module~B} propagates parity information to guide each row toward a valid codeword.

\paragraph{Overall refinement loop.} Figure~\ref{fig:arch_cider_main} summarizes the decoder. Starting from an all-\mask{} grid \(X^{(0)}\in([Q]\cup\{\mask\})^{K\times L}\), \cider repeats a fixed number of refinement steps. At step \(t\), the current grid is embedded into token features, the structured denoiser uses the evidence \(S\) and the parity-check matrix \(H\) to produce token logits, and the most confident masked entries are revealed:
\[
\Lambda^{(t)} = f_\theta(X^{(t)},S,H,t),
\hat a^{(t)}_{k,\ell}=\arg\max_{a\in[Q]}\Lambda^{(t)}_{k,\ell,a},
X^{(t+1)}_{k,\ell}=
\begin{cases}
\hat a^{(t)}_{k,\ell}, & (k,\ell)\in\mathcal U_t,\\
X^{(t)}_{k,\ell}, & \text{otherwise},
\end{cases}
\]
where \(\mathcal U_t\) is the set of currently masked sites selected for reveal at step \(t\) by the confidence-ranked cosine schedule. After \(T\) steps, the fully revealed grid \(\hat X=X^{(T)}\) is mapped back to an unordered decoded message set through the shared codebook. The exact embedding, logit projection, reveal schedule, and first-reveal stabilization rule are given in Appendix~\Cref{app:arch_cider,app:maskdiff_infer}.

\paragraph{Module A: demixing by row competition.}
The first failure mode is \emph{duplicate-row collapse}: because the same evidence matrix \(S\) is visible to every row, a generic denoiser can assign the same high-evidence slot symbols to multiple user hypotheses. Module~A prevents this by turning local evidence into a row-competitive assignment problem. 
We distinguish between intermediate logits used for demixing and final logits used for token reveal. Before Module~A, we compute intermediate logits from the current embedded state,
\[
\bar\Lambda^{(t)}_{k,\ell,a}=w_a^\top Z^{(t)}_{k,\ell}.
\]
These intermediate logits are used only to compute demixing responsibilities. Each slot-symbol candidate \((\ell,a)\) is softly allocated across the \(K\) rows:
\begin{equation}
\label{eq:method_resp_short}
r^{(t)}_{k,\ell,a}
=
\frac{\exp(\bar\Lambda^{(t)}_{k,\ell,a}/\tau_{\mathrm{demix}})}
{\sum_{k'=0}^{K-1}\exp(\bar\Lambda^{(t)}_{k',\ell,a}/\tau_{\mathrm{demix}})},
\qquad
\sum_{k=0}^{K-1} r^{(t)}_{k,\ell,a}=1 .
\end{equation}
The responsibility \(r^{(t)}_{k,\ell,a}\) measures how much row \(k\) claims symbol \(a\) in slot \(\ell\). We then form a row-specific evidence embedding
\begin{equation}
\label{eq:method_evidence_embed_short}
e^{(t)}_{k,\ell}
=
\sum_{a\in[Q]} r^{(t)}_{k,\ell,a}\, S_{\ell,a}\, v_a ,
\end{equation}
where \(v_a\in\mathbb{R}^D\) is a learnable symbol embedding. Finally, this evidence is fused into the token latent,
\begin{equation}
\label{eq:method_A_update_short}
\tilde Z^{(t)}_{k,\ell}
=
\Phi_A\!\left(Z^{(t)}_{k,\ell}, e^{(t)}_{k,\ell}\right).
\end{equation}
Thus, high-evidence symbols are not independently copied into every row; they are softly divided among competing rows. This creates a repulsive effect between user hypotheses and encourages different rows to explain different parts of the shared evidence. Implementation details are provided in Appendix~\Cref{app:arch_cider}, with overlap diagnostics in Appendix~\Cref{app:failure_modes}.

\paragraph{Module B: parity-aware propagation for code consistency.}

The second failure mode is \emph{invalid assembly}: even after rows are separated, each row is still assembled from local slot-wise choices and may violate the code constraints. Module~B injects global code structure through the sparse parity-check matrix \(H\). Let \(\mathcal{N}_{\mathrm{var}}(j)\) be the slots participating in parity check \(j\), and let \(\mathcal{N}_{\mathrm{chk}}(\ell)\) be the checks involving slot \(\ell\). For each row \(k\), Module~B computes an extrinsic check-to-slot signal
\begin{equation}
\label{eq:method_check_msg_short}
n^{(t)}_{k,j\to \ell}
=
\Psi_j\!\left(
\left\{
T_{H_{j,\ell'}}\!\left(\tilde Z^{(t)}_{k,\ell'}\right)
:\ell'\in\mathcal{N}_{\mathrm{var}}(j)\setminus\{\ell\}
\right\}
\right),
\end{equation}
where \(T_{H_{j,\ell'}}\) applies the finite-field coefficient action associated with the nonzero parity coefficient \(H_{j,\ell'}\), and \(\Psi_j\) aggregates the neighboring slot features for check \(j\). 
In the implementation, $\Psi_j$ uses neighbor-restricted multi-head attention that excludes the target edge, followed by a residual GELU MLP, dropout, and LayerNorm. The Tanner topology and finite-field coefficient permutations are fixed; the message transformations and fusion are learned. The incoming check messages are inverse-permuted, scatter-added, degree-normalized, and fused with the token latent by a learned gate and residual feed-forward network:
\begin{equation}
\label{eq:method_B_update_short}
\hat Z^{(t)}_{k,\ell}
=
\Phi_B\!\left(
\tilde Z^{(t)}_{k,\ell},
\sum_{j\in\mathcal{N}_{\mathrm{chk}}(\ell)}
n^{(t)}_{k,j\to\ell}
\right).
\end{equation}
After Module~B, the final prediction logits are computed from the refined latent state:
\[
\Lambda^{(t)}_{k,\ell,a}=w_a^\top \hat Z^{(t)}_{k,\ell}.
\]
Only these final logits \(\Lambda^{(t)}\) are used by the reveal rule during masked-diffusion inference. Each of the $N_{\mathrm{layer}}$ structured layers applies Module~A and two Tanner message-passing passes within one refinement step; the masked-refinement step count $T$ controls a separate depth.
This update is soft: Module~B does not hard-project a row onto the codebook. Instead, it biases each refinement step toward completions that are compatible with the parity checks. The finite-field coefficient actions, Tanner-graph aggregation block, and LDPC-code conventions are detailed in Appendix~\Cref{app:arch_cider,app:ldpc_details}; complexity implications are discussed in Appendix~\Cref{app:complexity_derivation}.

\paragraph{Why both modules are needed.} The two modules solve different parts of the ambiguity. Module~A makes the \(K\) rows explain different users; without it, rows can duplicate each other even if parity information is present. Module~B makes each row obey the code constraints; without it, rows may be distinct but still not valid codewords. The full denoiser therefore combines demixing and code propagation at every refinement step, allowing the masked-diffusion loop to gradually reveal a set of distinct and globally consistent codewords. LayerNorm rescales features but does not assign row ownership; Module~A's cross-row softmax couples the hypotheses directly.

\paragraph{Quality-guided remasking at higher loads.} For larger \(K\), some decoded rows may remain low-confidence after the first pass. A lightweight quality head is trained post hoc to predict token correctness from the frozen backbone. Its token scores are averaged within each row; low-confidence rows are then remasked and re-decoded while high-confidence rows remain fixed. This PRISM-style \citep{prism2025} step is used only as an inference-time reliability boost in higher-load experiments; details are in Appendix~\Cref{app:prism_remasking}.
 
\section{Experiments}
\label{sec:experiments}

We evaluate \cider as a shared-codebook multiuser decoder under a common slot-wise evidence interface. The main experiments compare decoding reliability and latency across LDPC settings, with additional tests of higher per-bin load and a stochastic-binning wrapper. Full setup and implementation details are in Appendix~\Cref{app:exp_details}.

\subsection{Experimental setup}
\label{sec:exp_setup}

Unless stated otherwise, experiments use an AWGN channel without per-user fading; a fixed AMP-based detector produces evidence \(S\in\mathbb{R}^{L\times Q}\), and all methods decode \(S\mapsto\hat X\in[Q]^{K\times L}\). The main benchmark uses non-binary LDPC codes over GF(64) with \(K=2\), rate \(R=1/3\), and \(L\in\{12,18,24,48\}\). This setting enables fair comparison with strong classical decoders (SIC-BP, FFT-BP, and Top-\(J\) search), whose cost already becomes large as \(K\) or \(L\) grows. We complement it with single-bin scaling up to \(K=8\), stochastic-binning scaling up to \(K_{\mathrm{tot}}=100\), and additional PEG-LDPC/tree-code results. We focus on sparse-graph codes because Module~B performs Tanner-graph propagation. Full simulation details are in Appendix~\Cref{app:exp_details,app:exact_configs,app:other_codes,app:protocol_scaling}.

\subsection{Baselines, training, and metrics}
\label{sec:exp_models_metrics}

We compare \cider against three groups of decoders under the same evidence interface \(S\): 
(i) classical decoders (Top-\(J\) exhaustive search, SIC-BP, FFT-BP), 
(ii) one-shot neural decoders (MLP, CNN, Transformer, GNN, NBP, Tanner-Attention), and 
(iii) generic masked diffusion (MDD). 
All learning-based models are trained with AdamW on the same train/validation/test splits. 
At inference, one-shot baselines decode in a single forward pass, while MDD and \cider run \(T\) refinement steps. 
Full baseline definitions, permutation-invariant training, hyperparameters, and training schedules are provided in Appendix~\Cref{app:arch_details,app:exp_train_infer}.

Because the decoded message set is unordered, predicted and ground-truth grids are matched by a minimum-Hamming-distance Hungarian assignment before evaluation. 
Let \(\pi^\star\) denote this optimal row permutation. We report symbol error rate (SER) and codeword error rate (CER). For one test example, we compute
\begin{equation}
\mathrm{SER}(X^\star,\hat X)
=
\frac{1}{KL}\sum_{k,\ell}
\mathbf{1}\!\left[\hat X_{k,\ell}\neq X^\star_{\pi^\star(k),\ell}\right],
\mathrm{CER}(X^\star,\hat X)
=
\frac{1}{K}\sum_{k}
\mathbf{1}\!\left[\hat X_{k,:}\neq X^\star_{\pi^\star(k),:}\right].
\end{equation}
Reported SER/CER are empirical averages over the test set.
Thus, SER evaluates token-level accuracy after row matching, whereas CER evaluates row-level recovery: a decoded codeword is counted as correct only if all \(L\) symbols in that row are recovered exactly.
The Hungarian matching definition and metric implementation details are given in Appendix~\Cref{app:metric_details}.
\subsection{Main results}
\label{sec:exp_results}

\begin{figure}[t]
  \centering
  \includegraphics[width=1.0\columnwidth]{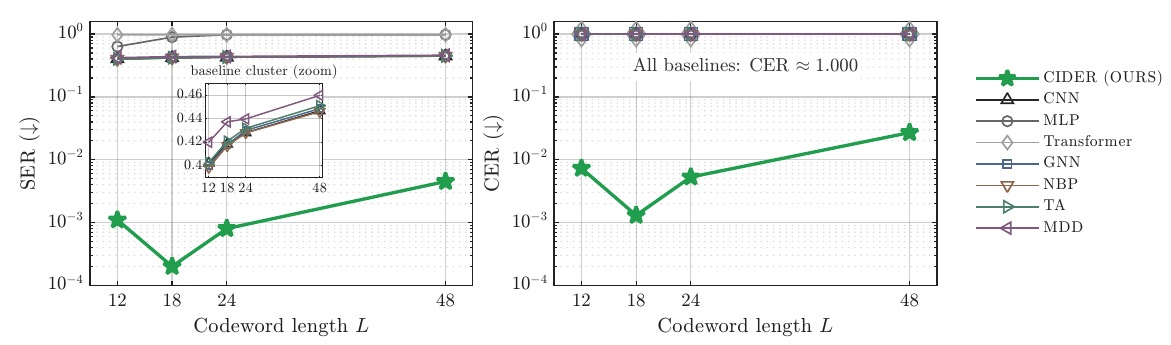}
  
  \captionsetup{font=footnotesize}
\caption{%
\textbf{\cider vs.\ neural baselines across code lengths ($K=2$).}
\textbf{Left:} SER vs.\ code length $L$ at $K=2$; \cider achieves low SER, whereas the other neural decoders largely fail and collapse to nearly overlapping total-failure curves.
\textbf{Right:} CER vs.\ code length $L$ at $K=2$. Numerical values are reported in Appendix~\Cref{app:neural_baseline_table}.}
  \label{fig:main_neural_mechanism}
  
\end{figure}

\paragraph{Neural baseline comparison.}
Figure~\ref{fig:main_neural_mechanism} compares \cider with one-shot neural decoders and generic masked diffusion (MDD) at $K=2$. Across the four LDPC code lengths, the generic baselines show nearly overlapping high-error curves, while \cider remains in the low-error regime. The difference is consistent with the need to resolve row ownership and enforce parity when refining a set of codewords from slot-local evidence. Full numerical results are in Appendix~\Cref{app:full_results}.

\paragraph{Classical decoder comparison.}
Table~\ref{tab:main_results_classical} compares \cider with SIC-BP, FFT-BP, and Top-$J$ search under the same AMP evidence interface. The table combines SER/CER with optimized same-GPU latency measurements. \cider has the lowest SER and CER at all four code lengths and the shortest measured latency among the compared methods. Top-$J$ search becomes costly at larger lengths.

\begin{table*}[h]

\centering
\caption{Classical decoder accuracy and same-GPU latency under shared AMP evidence ($K=2$, $Q=64$). Time is ms/sample on an RTX 3090: BP and Top-$J$ run in FP32 (BP with syndrome early exit) and deployed \cider uses 16-bit mixed precision. Top-$J$ uses two candidates per slot; at $L=48$ it exceeded the 60-second cap.}
\label{tab:main_results_classical}
\scriptsize
\setlength{\tabcolsep}{2.8pt}
\renewcommand{\arraystretch}{0.92}
\resizebox{\textwidth}{!}{%
\begin{tabular}{@{}l|ccc|ccc|ccc|ccc@{}}
\toprule
\multirow{2}{*}{\textbf{Method}}
& \multicolumn{3}{c|}{$L=12$}
& \multicolumn{3}{c|}{$L=18$}
& \multicolumn{3}{c|}{$L=24$}
& \multicolumn{3}{c}{$L=48$} \\
\cmidrule(lr){2-4}
\cmidrule(lr){5-7}
\cmidrule(lr){8-10}
\cmidrule(lr){11-13}
& SER & CER & Time
& SER & CER & Time
& SER & CER & Time
& SER & CER & Time \\
\midrule

\rowcolor{CIDERGreenFill!55}
\textbf{\cider}
& \textbf{0.0011} & \textbf{0.0073} & \textbf{1.51}
& \textbf{0.0002} & \textbf{0.0013} & \textbf{3.12}
& \textbf{0.0008} & \textbf{0.0053} & \textbf{4.31}
& \textbf{0.0045} & \textbf{0.0270} & \textbf{10.08} \\

SIC-BP
& 0.0015 & 0.0078 & 7.29
& 0.0025 & 0.0081 & 14.95
& 0.0031 & 0.0076 & 89.81
& 0.1144 & 0.2680 & 914.15 \\

FFT-BP
& 0.0015 & 0.0078 & 12.38
& 0.0025 & 0.0081 & 27.18
& 0.0031 & 0.0076 & 162.83
& 0.1144 & 0.2680 & 1687.95 \\

Top-$J$ (Top 2)
& 0.0476 & 0.0504 & 2.52
& 0.0711 & 0.0706 & 4.83
& 0.0911 & 0.0975 & 77.61
& -- & -- & $>60{,}000$ \\

\bottomrule
\end{tabular}%
}

\end{table*}

\begin{figure*}[t]
  \centering
  \includegraphics[width=0.85\textwidth]{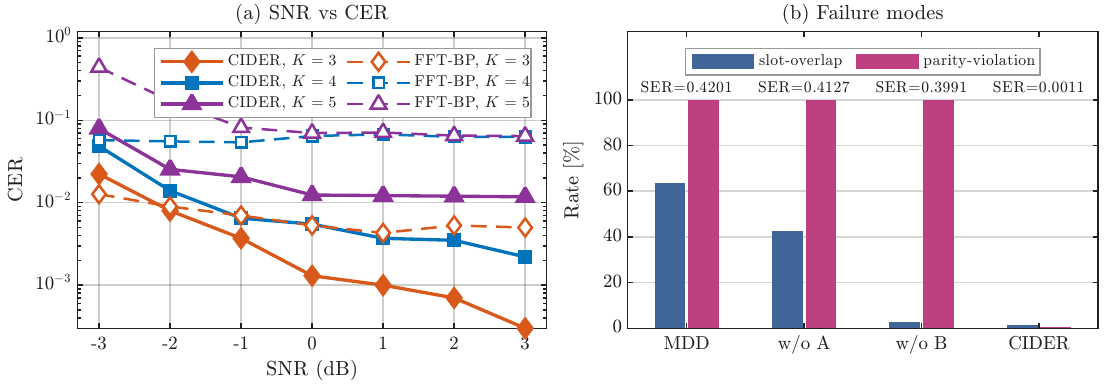}
  
  \caption{
  \textbf{Classical scaling and \cider mechanism.}
  \textbf{Left:} CER (= PUPE) versus per-user per-channel-use SNR. \cider is competitive at $K=3$ and shows a growing advantage at higher per-bin loads ($K=4,5$); the model is trained at a single operating point (SNR $=-0.79$ dB ($E_b/N_0=10$ dB)) and evaluated across the sweep without retraining. Runtime is in Appendix~\Cref{app:runtime_k3}.
  \textbf{Right:} Failure-mode diagnostics on Tiny $(Q,L)=(64,12)$, $K=2$. Module A primarily resolves duplicate-row overlap; Module B restores code consistency; only the full \cider achieves both.
  }
  \label{fig:main_classical_mechanism}
  
\end{figure*}

\paragraph{Module ablation.}
The ablation in Table~\ref{tab:app_full_results_ablation_main} separates the contributions of masked refinement and the two structural modules. Removing either Module~A or Module~B drives CER close to one, while generic MDD and non-diffusion controls using the same modules also fall short of full \cider. Figure~\ref{fig:main_classical_mechanism}(b) diagnoses the corresponding failure modes: Module~A reduces duplicate-row overlap, and Module~B reduces parity violations. Figure~\ref{fig:main_classical_mechanism}(a) additionally reports CER across SNRs and per-bin loads.

\paragraph{Runtime.}
For the same-GPU comparison in Table~\ref{tab:main_results_classical}, we vectorized and batched the BP baselines with synchronized FP32 timing and syndrome early exit, and measured deployed 16-bit mixed-precision \cider on the same RTX 3090 (Appendix~\Cref{app:response_fairness}). At $L=12/48$, \cider takes 1.51/10.08 ms/sample, versus 7.29/914.15 for SIC-BP and 12.38/1687.95 for FFT-BP. At the CER $\leq .05$ operating point, $K=4/7$ \cider takes 10.19/31.90 ms/sample versus 90.9/525 for SIC-BP; at $K=8$, SIC-BP does not reach that accuracy target in the tested setting (Appendix~\Cref{tab:response_gpu_operating}). These are implementation-specific wall-clock results. Batch throughput and memory remain in Appendix~\Cref{app:response_fairness}.

\paragraph{Generalization and scaling.}
We next evaluate \cider beyond the $K=2$ LDPC benchmark. \Cref{tab:main_prism_k_scaling} reports single-bin results for per-bin loads up to $K=8$; the low-load error rates are not monotonic across the separately trained $K$-specific models. At higher loads, PRISM-style quality-guided remasking improves reliability by selectively re-decoding low-confidence rows while clamping high-confidence rows. At $K=8$, it reduces CER from .2576 to .0403 while increasing end-to-end latency from 30.77 to 63.00 ms/sample, including quality scoring and triggered decoding passes (Appendix~\Cref{tab:response_remasking}).

On the tree-code instance, \cider achieves substantially lower SER/CER than the tree-code stitching decoder and Top-$J$ search, while retaining millisecond-scale runtime (\Cref{tab:main_tree_code_summary}). This suggests that the gains are not merely due to a particular code construction: even when compared against a code-specific stitching baseline, structured masked refinement provides more reliable accuracy.

We then wrap \cider in a stochastic-binning protocol \citep{marshakov2019polar,vem2019sic}: each active user randomly selects one of $\zeta$ preambles, which assigns its payload to a bin. Following Appendix~\Cref{app:protocol_scaling}, let $K_\chi$ be the load of bin $\chi\in[\zeta]$. We scale $\zeta$ with $K_{\mathrm{tot}}$ so that $\mathbb{E}[K_\chi] \approx 4$. For this protocol-level evaluation, the receiver is given $K_\chi$ and uses the corresponding $K$-specific \cider decoder (+PRISM for $K_\chi\ge 6$); bins with $K_\chi>8$ are counted as erasures. We test estimated loads separately in \Cref{tab:estimated_load}. 
Table~\ref{tab:ura_scaling} reports system-level SER/CER (= PUPE) for $K_{\mathrm{tot}}$ up to 100. Further code-family, load-scaling, and protocol details are in Appendix~\Cref{app:other_codes,app:larger_k_results,app:protocol_scaling}.

\begin{table*}[h]
\centering
\caption{
Compact summary of ablations, load scaling, tree-code baseline comparison, and protocol scaling ($\zeta = \lceil K_{\mathrm{tot}}/4 \rceil$, $\mathbb{E}[K_\chi] \approx 4$).
Panels (a)--(c) use Tiny $(Q,L)=(64,12)$; (d) shows selected protocol loads. The full protocol sweep is in Appendix~\Cref{tab:protocol_scalability_main2}.
}
\label{tab:compact_summary}
\small
\captionsetup[subtable]{justification=centering,singlelinecheck=false,font=small}
\setlength{\tabcolsep}{4pt}
\renewcommand{\arraystretch}{1.05}

\begin{subtable}[t]{0.575\textwidth}
\centering
\caption{Module ablation.}
\label{tab:app_full_results_ablation_main}
\begin{tabular}{@{}lccccc@{}}
\toprule
\textbf{Model} & \textbf{Diff.} & \textbf{A} & \textbf{B} & \textbf{SER} & \textbf{CER} \\
\midrule
\rowcolor{CIDERGreenFill!55}
\cider & $\checkmark$ & $\checkmark$ & $\checkmark$ & \textbf{.0011} & \textbf{.0073} \\
w/o A & $\checkmark$ & $\times$ & $\checkmark$ & .4127 & .9993 \\
w/o B & $\checkmark$ & $\checkmark$ & $\times$ & .3991 & .9993 \\
MDD & $\checkmark$ & $\times$ & $\times$ & .4201 & .9996 \\
One-shot A+B & $\times$ & $\checkmark$ & $\checkmark$ & .1013 & .3979 \\
Iter. A+B, \(N_{\rm iter}{=}1\) & $\times$ & $\checkmark$ & $\checkmark$ & .1039 & .4093 \\
Iter. A+B, \(N_{\rm iter}{=}12\) & $\times$ & $\checkmark$ & $\checkmark$ & .3863 & .9989 \\
\bottomrule
\end{tabular}
\end{subtable}
\hfill
\begin{subtable}[t]{0.395\textwidth}
\centering
\caption{Load scaling (+PRISM).}
\label{tab:main_prism_k_scaling}
\begin{tabular}{@{}rcccc@{}}
\toprule
\multirow{2}{*}{$K$}
& \multicolumn{2}{c}{\cider}
& \multicolumn{2}{c}{+\textsc{PRISM}} \\
\cmidrule(lr){2-3}\cmidrule(l){4-5}
& SER & CER & SER & CER \\
\midrule
2 & .0011 & .0073 & -- & -- \\
3 & .0006 & .0044 & -- & -- \\
4 & .0015 & .0058 & -- & -- \\
5 & .0048 & .0141 & -- & -- \\
6 & .0149 & .0349 & \cellcolor{CIDERGreenFill!55}\textbf{.0064} & \cellcolor{CIDERGreenFill!55}\textbf{.0163} \\
7 & .0441 & .1006 & \cellcolor{CIDERGreenFill!55}\textbf{.0096} & \cellcolor{CIDERGreenFill!55}\textbf{.0247} \\
8 & .1339 & .2576 & \cellcolor{CIDERGreenFill!55}\textbf{.0166} & \cellcolor{CIDERGreenFill!55}\textbf{.0403} \\
\bottomrule
\end{tabular}
\end{subtable}
\par\medskip
\noindent
\begin{subtable}[t]{0.505\textwidth}
\centering
\caption{Tree-code baselines.}
\label{tab:main_tree_code_summary}
\begin{tabular}{@{}lccc@{}}
\toprule
Method & SER & CER & ms \\
\midrule
\rowcolor{CIDERGreenFill!55}
\cider & \textbf{.0001} & \textbf{.0007} & \textbf{1.08} \\
SIC-BP & .0017 & .0065 & 6.82 \\
FFT-BP & .0017 & .0065 & 11.56 \\
Stitch & .0474 & .0490 & 1.20 \\
Top-2 & .0474 & .0490 & 2.55 \\
\bottomrule
\end{tabular}
\end{subtable}
\hfill
\begin{subtable}[t]{0.465\textwidth}
\centering
\caption{Protocol scaling (selected loads).}
\label{tab:ura_scaling}
\begin{tabular}{@{}rcc@{}}
\toprule
$K_{\rm tot}$ & SER & CER \\
\midrule
10  & .0027 & .0108 \\
20  & .0225 & .0322 \\
40  & .0423 & .0517 \\
60  & .0493 & .0590 \\
80  & .0502 & .0598 \\
100 & .0525 & .0619 \\
\bottomrule
\end{tabular}
\end{subtable}

\end{table*}

\paragraph{Transfer without weight updates.}
To test whether \cider depends on retraining for each code, we apply the $L=12,K=2$ checkpoint to different sparse graphs and longer codewords, changing only the masked-refinement step count $T$ (\Cref{tab:response_transfer}). PEG-LDPC denotes a graph constructed by progressive edge growth. Transfer CER remains at most .0110 through $L=24$, then rises to .0762 at $L=96$; dedicated target-trained models perform better on the other tested settings. The $L=72,96$ rows each use 256 test examples. The Training LDPC row uses the Tiny LDPC setting of Table~\ref{tab:main_results_classical} on separately generated on-the-fly test frames. \par\begin{table}[!htbp]\centering\footnotesize\setlength{\tabcolsep}{5pt}\caption{One-checkpoint graph and length transfer at $K=2$ without weight updates. Only the refinement count $T$ changes. Target-trained results use a dedicated model; $L=72,96$ use 256 examples each. Lower CER is better.}\label{tab:response_transfer}\begin{tabular}{@{}lcccc@{}}\toprule Target code & $L$ & $T$ & Transfer CER & Target-trained CER \\ \midrule Training LDPC & 12 & 12 & .00807 & .00807 \\ PEG-LDPC & 12 & 12 & .00783 & .00230 \\ Tree code & 12 & 12 & .00820 & .00070 \\ LDPC & 18 & 16 & .0078 & .0013 \\ LDPC & 24 & 20 & .0110 & .0053 \\ LDPC & 48 & 28 & .0352 & .0270 \\ LDPC & 72 & 40 & .0410 & --- \\ LDPC & 96 & 60 & .0762 & --- \\ \bottomrule\end{tabular}\end{table}

\paragraph{Decoding with estimated per-bin load.}
In the stochastic-binning setting above, each bin has a variable number of active users $K_\chi$, which the receiver does not observe directly. Writing $K$ for a bin's true load, we replace the known-$K$ assumption in the preceding single-bin results with an external energy-based estimate $\hat K$. Bank uses $\hat K$ for AMP and checkpoint selection. Single instead runs one $K=8$ checkpoint for every bin, deduplicates and ranks its candidate rows by summed slot-wise log-evidence, and retains the top $\hat K$ rows. Both modes use the same estimator. Table~\ref{tab:estimated_load} compares them for $K=2$--8. After matching the reported and transmitted codeword sets, missed-detection rate (MDR) is the fraction of true codewords missed, and false-alarm rate (FAR) is the fraction of reported codewords that are spurious. The $K=2$ Bank row uses the Tiny LDPC setting of Table~\ref{tab:main_results_classical} on separately generated on-the-fly test frames; the Single columns evaluate the separate $K=8$ checkpoint on this load-estimation task.

\begin{table}[!htbp]
\centering
\caption{Decoding with estimated per-bin load $\hat K$ for true load $K$. The Single oracle column uses true $K$ only for final row pruning. Lower MDR and FAR are better.}
\label{tab:estimated_load}
\footnotesize
\setlength{\tabcolsep}{7pt}
\renewcommand{\arraystretch}{1.02}
\begin{tabular}{@{}ccccccc@{}}
\toprule
\multirow{2}{*}{True $K$} & \multirow{2}{*}{$\Pr(\hat K=K)$}
& \multicolumn{2}{c}{Bank} & \multicolumn{2}{c}{Single}
& \multicolumn{1}{c}{Single oracle} \\
\cmidrule(lr){3-4}\cmidrule(lr){5-6}\cmidrule(lr){7-7}
& & MDR & FAR & MDR & FAR & MDR \\
\midrule
2 & .997 & .0067 & .0080 & .0003 & .0017 & .0003 \\
3 & .989 & .0067 & .0056 & .0031 & .0016 & .0007 \\
4 & .959 & .0095 & .0130 & .0045 & .0025 & .0015 \\
5 & .939 & .0195 & .0197 & .0076 & .0023 & .0016 \\
6 & .895 & .0224 & .0254 & .0116 & .0084 & .0043 \\
7 & .853 & .0388 & .0389 & .0203 & .0179 & .0105 \\
8 & .842 & .0563 & .0458 & .0524 & .0415 & .0435 \\
\midrule
\rowcolor{CIDERGreenFill!45}
\textbf{Mean} & .925 & .0293 & .0278 & \textbf{.0199} & \textbf{.0153} & .0133 \\
\bottomrule
\end{tabular}
\end{table}

Across $K=2$--8, the estimator has 92.5\% mean exact-load accuracy. Single lowers mean MDR/FAR from .0293/.0278 (Bank) to .0199/.0153 but always incurs $K=8$ computation; oracle pruning lowers its MDR to .0133. A one-user underestimate causes misses, whereas an overestimate adds a false alarm (Appendix~\Cref{tab:load_mismatch}).

\paragraph{AMP and SNR mismatch.}
To assess sensitivity to the fixed AMP-generated evidence interface, we evaluate \cider on the Tiny scale without retraining under two types of mismatch: fewer AMP iterations and SNR shifts.
\Cref{tab:detector_mismatch} shows little CER change when AMP iterations change from the matched 10 to 5 or 20 at matched SNR, but CER reaches .3087 under the largest negative SNR shift.

\FloatBarrier

\section{Conclusion}

We introduced \cider for within-bin shared-codebook URA outer decoding, where the receiver recovers an unordered set of codewords from fixed slot-wise soft evidence.
The key is to refine the unordered codeword set jointly: row competition discourages duplicate rows, while Tanner-graph propagation steers each row toward a valid codeword. The ablations show that both mechanisms are needed. In the evaluated short-block settings, \cider achieves lower CER than the tested neural baselines; at $K=2$, it also improves CER and measured outer-decoder latency over the optimized classical baselines on the same GPU. \par The results extend beyond one fixed code and load. A single checkpoint transfers across sparse-graph constructions and moderately longer codewords by changing only the refinement count, although target-trained models are usually more accurate. Quality-guided remasking improves reliability at higher per-bin loads at additional runtime cost, and estimated-load decoding shows the importance of load inference. Improving load estimation, demixing under heavy codeword overlap, and robustness to evidence mismatch are the main next steps toward an end-to-end receiver.

\section*{Acknowledgements and Disclosures}
This work was partly supported by ARO Award W911NF2310062, ONR Award N000142412542, NSF CAREER Award 2443857, NSF RINGS Award CNS-2148141, and the 6G@UT center within the Wireless Networking and Communications Group (WNCG) at the University of Texas at Austin. The authors declare no competing interests.

\renewcommand{\bibfont}{\small}
\setlength{\bibsep}{0pt}
\bibliographystyle{plainnat}
\bibliography{camera_ready}

\newpage
\appendix

\section{Notation glossary}
\label{app:notation}

This appendix summarizes the main notation used for the receiver, decoder, and experiments; see Table~\ref{tab:notation_glossary_ext}.

{\small
\setlength{\tabcolsep}{3pt}
\renewcommand{\arraystretch}{1.0}

\begin{longtable}{@{}p{3.4cm}p{9.9cm}@{}}
\caption{Notation glossary.}
\label{tab:notation_glossary_ext}\\
\toprule
Symbol & Meaning \\
\midrule
\endfirsthead

\caption[]{Notation glossary (continued).}\\
\toprule
Symbol & Meaning \\
\midrule
\endhead

\midrule
\multicolumn{2}{r}{\emph{Continued on next page}}\\
\endfoot

\bottomrule
\endlastfoot

\multicolumn{2}{l}{\textbf{Indices and sizes}}\\
$[M]$ & Message index set $\{0,1,\ldots,M-1\}$; $M\triangleq 2^B$. \\
$[Q]$ & Alphabet/signature index set $\{0,1,\ldots,Q-1\}$. \\
$[K]$ & Row index set $\{0,1,\ldots,K-1\}$. \\
$[L]$ & Slot index set $\{0,1,\ldots,L-1\}$. \\
$[\zeta]$ & Preamble-bin index set $\{1,\ldots,\zeta\}$ in the protocol wrapper. \\
$k\in[K]$ & Row (hypothesis/user) index in the $K\times L$ grid. \\
$\ell\in[L]$ & Slot index; also codeword coordinate index. \\
$a\in[Q]$ & Symbol/signature index within a slot. \\
$\chi\in[\zeta]$ & Bin index in the protocol wrapper. \\
$B,\,R$ & Payload length in bits per message and code rate, respectively. \\
$K$ & Number of active users/messages in one bin; $K_\chi$ is the bin's true load and $\hat K$ its estimate. \\
$K_{\mathrm{tot}}$ & Total number of active users in a frame (protocol wrapper). \\
$K_{\max}$ & Maximum supported per-bin load (decoder bank is trained up to $K_{\max}$). \\
$\zeta$ & Number of preamble/payload bins in the protocol wrapper. \\
$L$ & Codeword length, or number of payload slots in one bin. \\
$Q$ & Alphabet size. \\
$D$ & Latent/embedding dimension. \\
$P$ & Number of parity-check equations (rows of $H$). \\
$N_{\mathrm{seq}}$ & Sequence length in the generic masked-diffusion preliminaries; in the URA grid instantiation, $N_{\mathrm{seq}}=KL$. \\

\midrule
\multicolumn{2}{l}{\textbf{Codebook, grids, and decoding}}\\
$\mathcal{C}=\{c_m\}_{m\in[M]}$ & Shared codebook. \\
$c_m\in[Q]^L$ & Codeword for message $m$. \\
$\mathcal{W}^\star\subseteq[M]$ & Ground-truth transmitted message set, $|\mathcal{W}^\star|=K$. \\
$\hat{\mathcal{W}}\subseteq[M]$ & Decoded message set. \\
$X^\star\in[Q]^{K\times L}$ & Ground-truth codeword grid (row order immaterial). \\
$\hat X\in[Q]^{K\times L}$ & Decoded codeword grid (row order immaterial). \\
$X^{(t)}\in([Q]\cup\{\mask\})^{K\times L}$ & Discrete masked grid at \cider refinement step $t$ (the only persistent state across steps). \\
$\gamma_t$ & Mask ratio at step $t$; $\gamma_0=1$, $\gamma_T=0$. \\
$\mask$ & Mask token. \\

\midrule
\multicolumn{2}{l}{\textbf{Symbol-wise soft detector evidence}}\\
$Y^{(\ell)}\in\mathbb{C}^{n_s}$ & Random received vector in slot $\ell$; $n_s$ is its dimension, $A^{(\ell)}\in\mathbb{C}^{n_s\times Q}$ is the slot signature dictionary, and $y^{(\ell)}$ denotes a realization. \\
$U^{(\ell)}\in\mathbb{C}^{Q}$ & Slot-$\ell$ sparse activity vector (random variable). \\
$b_q$ & $q$-th standard basis vector in $\mathbb{C}^Q$ (used to form the activity vector $U^{(\ell)}$). \\
$S_{\ell,a}$ & Entry of $S\in\mathbb{R}^{L\times Q}$: log-posterior score that symbol $a$ is active in slot $\ell$. \\
$S^{(\chi)}\in\mathbb{R}^{L\times Q}$ & Evidence matrix for bin $\chi$ (protocol wrapper). \\

\midrule
\multicolumn{2}{l}{\textbf{Diffusion / \cider}}\\
$T$ & Number of refinement steps. \\
$Z^{(t)}\in\mathbb{R}^{K\times L\times D}$ & Embedded latent constructed from the current discrete grid at step $t$ (recomputed from $X^{(t)}$ each step). \\
\mbox{$\tilde Z^{(t)}\in\mathbb{R}^{K\times L\times D}$} & Demixed latent after Module~A at step $t$. \\
\mbox{$\hat Z^{(t)}\in\mathbb{R}^{K\times L\times D}$} & Refined latent after Module~B at step $t$ (used to form logits). \\
$\Lambda^{(t)}\in\mathbb{R}^{K\times L\times Q}$ & Logits output by the denoiser at step $t$. \\
$E\in\mathbb{R}^{(Q+1)\times D}$ & Embedding table (includes \mask\ token). \\
$W\in\mathbb{R}^{Q\times D}$ & Output projection matrix (rows are $w_a^\top$). \\
Module A, Module B & Demixing / parity propagation modules. \\
$r^{(t)}_{k,\ell,a}$ & Responsibility (soft assignment) used in demixing. \\
$g_{\psi}$ & Per-token quality head used for quality-guided remasking. \\
$\omega^{(t)}_{k,\ell}$ & Per-token correctness score predicted by $g_{\psi}$. \\
$\bar \omega_{k}$ & Row-level confidence for row $k$ (average of $\{\omega^{(T)}_{k,\ell}\}_{\ell\in[L]}$ over slots). \\
$\mathcal{K}_{\mathrm{low}}\subseteq[K]$ & Set of remasked (low-confidence) rows in quality-guided remasking at stage $s$. \\
$\varphi_s$ & Quality threshold at remasking stage $s\in\{1,\dots,\varrho\}$. \\
$\varrho$ & Number of remasking stages (thresholds) in the multi-stage strategy. \\
$X_{\mathrm{rm}}^{(t)}\in([Q]\cup\{\mask\})^{K\times L}$ & Discrete masked grid during the remasking pass (rows in $\mathcal{K}_{\mathrm{low}}$ are remasked; other rows are clamped). \\

\midrule
\multicolumn{2}{l}{\textbf{LDPC / Tanner graph}}\\
$\mathbb{F}_Q$ & Finite field of size $Q$. \\
$H\in\mathbb{F}_Q^{P\times L}$ & Parity-check matrix. \\
$\mathcal{N}_{\mathrm{var}}(j)$ & Variable-node neighbors of check $j$: $\{\ell : H_{j,\ell}\neq 0\}$. \\
$\mathcal{N}_{\mathrm{chk}}(\ell)$ & Check-node neighbors of variable $\ell$: $\{j : H_{j,\ell}\neq 0\}$. \\
$\alpha\in\mathbb{F}_Q^\times$ & Nonzero GF$(Q)$ edge coefficient; $\alpha = H_{j,\ell}$ for edge $(j,\ell)$. \\
$\Pi_{\alpha}$ & Symbol permutation induced by multiplication by $\alpha\in\mathbb{F}_Q^\times$. \\
$\tau_{\mathrm{demix}}$ & Temperature for demixing softmax (Module A). \\

\midrule
\multicolumn{2}{l}{\textbf{Matching and metrics}}\\
$\mathfrak{S}_K$ & Set of permutations of $[K]$ (row matching). \\
$\mathrm{SER}$ & Symbol error rate (after optimal row matching). \\
$\mathrm{CER}$ & Codeword error rate (after optimal row matching). \\

\midrule
\multicolumn{2}{l}{\textbf{Complexity notation}}\\
$J$ & Per-slot candidate list size retained from $S$ (top-$J$). \\
$|E_H|$ & Number of nonzeros in $H$ (Tanner-graph edges). \\
$I_{\mathrm{BP}}$ & Total Tanner-graph iterations in SIC-BP. \\

\end{longtable}
}

\newpage

\section{Wireless communication context and motivation: scalable random access}
\label{app:wireless_context}

\paragraph{Grant-based random access and contention.}
In contemporary cellular systems, uplink transmissions are typically organized around grant-based access. 
A device first performs random access to establish timing and request an uplink grant, and the base station then schedules that grant for payload transmission.
This design works well when the number of simultaneously requesting devices is moderate and payloads are large, because the access overhead can be amortized.

\paragraph{Limitations of contention-based access for low-latency communication.}
A key challenge of contention-based random access is that it incurs non-negligible access latency due to its reliance on multi-step coordination prior to data exchange.
For example, the four-step handshaking procedure in 5G NR requires two round-trip signaling exchanges between a device and the BS to obtain an uplink grant, leading to a baseline access delay that is difficult to amortize for short, latency-critical messages.
This latency overhead becomes significantly more severe as the number of simultaneously requesting devices increases: contention opportunities are limited, collision probability on the finite preamble pool rises, and access efficiency degrades due to repeated failures and retransmissions.
In latency-stringent regimes, the need to resolve \emph{who transmitted what} through contention resolution and subsequent retransmissions can therefore dominate the end-to-end delay, rendering coordination-first access ill-suited for scalable low-latency communication.

\paragraph{Two-step random access with reduced latency.}
To reduce access latency and signaling overhead, contemporary wireless systems also support a coordination-based random access procedure with fewer round-trip exchanges in certain regimes.
In this design, a device transmits a preamble together with a short message on preconfigured access opportunities using a commonly known coding scheme, and the BS responds only if this concatenated transmission is successfully decoded.
While this approach reduces baseline access latency compared to four-step random access, it remains effective primarily under light contention.
When multiple devices transmit initial access payloads simultaneously, collisions occur and decoding typically fails, except for occasional capture of the strongest transmission, since current standards do not perform joint multiuser decoding.
As a result, this reduced-latency coordination-based access improves latency only when the number of simultaneous access attempts is small, and does not scale to latency-stringent scenarios with many concurrently active devices.

\paragraph{Grant-free access with fixed user--access mapping.}
Beyond contention-based random access, some systems attempt to reduce handshake overhead by preconfiguring communication resources through device-specific assignments or pilot sequences \citep{ke2023next}.
Under these designs, each potential device is associated with a distinct assignment, signature, or codebook ahead of time, so decoding proceeds under a fixed mapping between received observations and users.
While effective in relatively static settings, this approach scales poorly with a large population of potential users and unpredictable activity.
Maintaining per-user signatures, codebooks, or scheduling state can incur substantial overhead, since preallocated resources are wasted whenever only a fraction of devices are active.
Moreover, rapid changes in the device set require continual reconfiguration, rendering fixed user–resource mappings ill-suited for scalable low-latency communication.

\paragraph{Motivation for URA.}
Latency-stringent systems with many potential users often cannot afford the overhead of explicit handshaking, scheduling, or repeated access attempts prior to data transmission.
In such regimes, transmitting payloads without prior coordination provides a scalable operating point, allowing many users to access the channel simultaneously while avoiding per-user access overhead. This operating principle fundamentally changes how random access should be modeled and decoded.
The unsourced random access (URA) abstraction captures this regime by formulating random access as a joint coding and inference problem rather than a coordination-first protocol.
When payloads are transmitted without prior coordination, per-user allocation and transmitter identification cannot be assumed during decoding.
As a result, URA defines the receiver’s objective as recovering the set of transmitted messages directly from the superposed uplink signal. 
By avoiding per-user preallocation and coordination dependencies, URA decouples system performance from the total number of potential users, making it particularly well suited for large, dynamic systems with sporadic activity and strict latency constraints.
From a physical-layer perspective, this amounts to using the random access channel itself as a data transmission channel, rather than merely as a means for access coordination.

\paragraph{Decoding implications of uncoordinated access.}
Operating without prior coordination has two key implications for the decoding problem. 
First, since the receiver cannot rely on device-specific configuration or scheduling and does not know which devices are active at the time of decoding, all devices must use a commonly known codebook. This shared encoding rule ensures that the joint decoding is well defined even when the active set is unknown.
Second, since the receiver lacks prior knowledge of which devices are active before data exchange, decoding is naturally formulated in terms of message-set recovery rather than message-to-transmitter association. The transmitter identity is therefore unknown prior to transmission and, if necessary, may be embedded within the payload and recovered only after successful message decoding.

\section{Additional related work details}
\label{app:rw_appendix}

\paragraph{URA problem context.}
Unsourced random access (URA) is commonly studied as a model for massive sporadic uplink access, including formulations based on the Gaussian many-access channel (GMAC) \citep{polyanskiy2017perspective}. Within this line of work, a range of URA constructions and analyses have been developed, including coded compressed sensing pipelines \citep{amalladinne2020coded}, SPARCS-type schemes \citep{fengler2021sparcs}, and information-theoretic characterizations \citep{kowshik2021fundamental}.

\subsection{URA architectural families}
\label{app:ura_families}

The recent survey \citep{liva2024unsourced} classifies URA receivers into four architectural families:

\begin{itemize}[leftmargin=1.2em,itemsep=2pt,topsep=2pt]
  \item \textbf{Multipacket-reception (MPR) slotted-Aloha schemes.} Each frame is divided into slots, and within each slot a small group of colliding users is jointly decoded rather than treated as destructive collisions. Examples include T-fold polar Aloha \citep{marshakov2019polar} and T-fold LDPC schemes \citep{vem2019sic}, often combined with SIC across slots.

  \item \textbf{Coded compressed sensing (CCS).} Messages are split into fragments, each compressed via a per-slot signature dictionary, and recovered fragments are assembled into valid codewords using a tree code or list-recoverable code \citep{amalladinne2020coded,fengler2021sparcs,andreev2022list}. Joint AMP and multiuser-code message passing has also been integrated within this family \citep{amalladinne2022integrating,ebert2022coded}.

  \item \textbf{Preamble-based architectures.} The user message is split into two parts: the first selects a preamble that determines a per-user resource or access pattern, and the second is encoded with a channel code. Decoding is then performed within each bin, using either joint multiuser decoding (e.g., sparse IDMA \citep{pradhan2022sparse}) or single-user decoders aided by SIC \citep{ozates2024unsourced}.

  \item \textbf{Spreading-based architectures.} Message bits select spreading sequences from a dictionary, and recovered codewords are extracted by energy detection followed by per-user channel decoding with SIC \citep{pradhan2020polar}.
\end{itemize}

\paragraph{Learning for decoding: connections to channel decoding and URA.}
Beyond these rule-based multiuser decoders, learning-based decoders have been studied for classical channel decoding by augmenting iterative message passing, including neural belief propagation and learned Tanner-graph updates \citep{nachmani2016learning,nachmani2017improved}, and by using attention-based architectures for soft decoding \citep{choukroun2022ecct}.
End-to-end learned coding has also been explored, where encoder and decoder are jointly trained (e.g., TurboAE) \citep{jiang2019turboae}. These approaches are typically formulated for decoding a single codeword with a fixed observation-to-codeword association.
In URA, the decoding target is an \emph{unordered list} (set) of messages (unsourced decoding), and the receiver must resolve ownership ambiguity across users in addition to enforcing code constraints \citep{polyanskiy2017perspective,amalladinne2020coded,kowshik2021fundamental}.

\paragraph{Discrete diffusion and masked  denoising models.}
Discrete diffusion models have been developed for categorical variables.
D3PM defines a Markov noising process that gradually corrupts a discrete token by mixing it with a base categorical distribution (often uniform over the vocabulary), and learns a reverse-time model that predicts the corresponding denoising transitions \citep{austin2021d3pm}.
In contrast, masked denoising models (MDM), also described as masked-token diffusion, use an absorbing mask token and train with a masked prediction objective, enabling reconstruction via iterative unmasking procedures \citep{chang2022maskgit,lou2024sedd,sahoo2024mdlm,prism2025}.
Both formulations provide a principled way to perform \emph{iterative refinement} of discrete hypotheses under uncertainty, which matches the role of shared-codebook multiuser decoding: the slot-wise soft detector produces noisy, incomplete per-slot symbol evidence, and the decoding stage refines it into a globally consistent discrete output under code constraints.
We build our decoder on the masked denoising (MDM) formulation; additional uniform-corruption (D3PM-style) comparisons are deferred to Appendix~\Cref{app:full_results_d3pm}.

\section{Formal URA model and two-stage decoding interface}
\label{app:ura_formal}

This appendix provides a formal description of the URA two-stage receiver abstraction used throughout the paper.
We define the slot-wise codebook interface, the per-slot observation model, the evidence matrix produced by the fixed symbol-level soft detector, and the multiuser decoder objective as set recovery under permutation invariance.

\subsection{URA system model (frame-level set recovery)}
\label{app:ura_formal_system}

\paragraph{Messages and unsourced objective.}
Let $B$ be the payload length in bits and $M\triangleq 2^B$ be the number of possible messages.
We index messages by $[M]\triangleq\{0,1,\dots,M-1\}$.
In one decoding window (frame), an unknown subset of users becomes active and transmits a subset of messages
\[
\mathcal{W}^\star \subseteq [M],\qquad |\mathcal{W}^\star|=K,
\]
where $K$ is the number of active messages in the frame (often treated as known in evaluation settings).
In \emph{unsourced} random access, user identities are not part of the decoding objective; the receiver outputs only the unordered set $\hat{\mathcal{W}}$.
In particular, the receiver is not required to determine \emph{which} device sent a decoded message.
(If two devices transmit the same payload, the observation is indistinguishable from a single transmission of that payload under the set-valued objective; when $M$ is large this event is typically negligible.)

\subsection{Slot grid and shared codebook}
\label{app:ura_formal_codebook}

\paragraph{Slot decomposition.}
We represent each message by a length-$L$ sequence over a $Q$-ary alphabet $[Q]\triangleq\{0,1,\dots,Q-1\}$.
The shared codebook is
\[
\mathcal{C}=\{c_m\}_{m\in[M]},\qquad c_m \in [Q]^L.
\]
The $\ell$-th slot symbol of message $m$ is denoted by $c_{m,\ell}\in[Q]$.
Equivalently, $c_{m,\ell}$ is an \emph{index} selecting one of $Q$ pre-defined per-slot signatures/waveforms in slot $\ell$.

\paragraph{Set-valued codeword representation.}
Let $(w_0,\dots,w_{K-1})$ be an arbitrary ordering of $\mathcal{W}^\star$.
The corresponding transmitted codeword set is
\[
\mathcal{C}^\star \triangleq \{c_m : m\in\mathcal{W}^\star\}.
\]
For notational convenience, we represent the same unordered set $\mathcal{C}^\star$ by a grid
$X^\star\in[Q]^{K\times L}$ whose rows are the $K$ codewords in any order:
\[
X^\star_{k,\ell} \triangleq c_{w_k,\ell}.
\]
Any row permutation of $X^\star$ represents the same solution.

\subsection{Per-slot channel model and symbol-level soft evidence}
\label{app:ura_formal_inner}

\paragraph{Signature dictionary per slot.}
For each slot $\ell\in[L]$, each symbol $q\in[Q]$ corresponds to a known waveform (signature)
$a^{(\ell)}_{q}\in\mathbb{C}^{n_s}$.
Collecting them forms a dictionary
\[
A^{(\ell)} \triangleq \big[ a^{(\ell)}_{0}\ \cdots\ a^{(\ell)}_{Q-1}\big]\in\mathbb{C}^{n_s\times Q}.
\]

\paragraph{Superposition observation.}
In slot $\ell$, the receiver observes a superposition of the $K$ active signatures plus noise:
\begin{equation}
\label{eq:app_ura_slot_obs}
Y^{(\ell)}
\;=\;
\sum_{k=0}^{K-1} a^{(\ell)}_{c_{w_k,\ell}}
\;+\;
\epsilon^{(\ell)},
\qquad
\epsilon^{(\ell)}\sim \mathcal{CN}(0,\sigma^2 I).
\end{equation}
This is a simplified (yet standard) URA abstraction in which each active transmission contributes one unit-amplitude signature per slot.
More general models include per-user (or per-slot) complex gains multiplying each selected signature; our experiments use the above normalization for simplicity.
Equivalently, define a sparse activity vector $U^{(\ell)}\in\mathbb{C}^{Q}$ by
\begin{equation}
\label{eq:app_ura_activity_vec}
U^{(\ell)} \triangleq \sum_{k=0}^{K-1} b_{c_{w_k,\ell}},
\end{equation}
where $b_q$ is the $q$-th standard basis vector. Then \eqref{eq:app_ura_slot_obs} can be rewritten as
\begin{equation}
\label{eq:app_ura_slot_obs_dict}
Y^{(\ell)} = A^{(\ell)}U^{(\ell)} + \epsilon^{(\ell)}.
\end{equation}
To see \eqref{eq:app_ura_slot_obs_dict} directly, note that
$A^{(\ell)}b_{c_{w_k,\ell}} = a^{(\ell)}_{c_{w_k,\ell}}$ by definition of the dictionary columns, hence
\[
A^{(\ell)}U^{(\ell)}
=
A^{(\ell)}\sum_{k=0}^{K-1} b_{c_{w_k,\ell}}
=
\sum_{k=0}^{K-1} A^{(\ell)}b_{c_{w_k,\ell}}
=
\sum_{k=0}^{K-1} a^{(\ell)}_{c_{w_k,\ell}}.
\]
Collisions correspond to multiple users selecting the same index and summing at the same coordinate of $U^{(\ell)}$.
When collisions occur, the corresponding entry $U^{(\ell)}_a$ becomes an integer count (or, under more general fading models, an aggregated complex amplitude).

\paragraph{Evidence interface $S$.}
The symbol-level soft detector is fixed throughout this paper.
For each slot $\ell$, it maps $(Y^{(\ell)},A^{(\ell)})$ to a real-valued score vector over $[Q]$.
We define log-posterior evidence (logits)
\[
S_{\ell,a}
\triangleq
\log \Pr\!\left(U^{(\ell)}_{a}\neq 0 \,\middle|\, Y^{(\ell)},A^{(\ell)}\right),
\qquad a\in[Q],\ \ell\in[L].
\]
Equivalently, normalizing over $a$ yields per-slot soft beliefs
\[
p_{\ell,a} \;\propto\; \exp(S_{\ell,a}),\qquad a\in[Q],
\]
which we use as a normalized per-slot categorical belief for the downstream decoder. Since multiple symbols can be active in one slot, this softmax normalization should not be interpreted as a calibrated multi-hot activity probability.
Stacking all slots yields an evidence matrix $S\in\mathbb{R}^{L\times Q}$.
All multiuser decoders compared in this paper operate only on $S$ (the symbol-level soft detector is not modified).
Details of the AMP--MMSE detector used in our implementation are given in Appendix~\Cref{app:amp_mmse}.

\subsection{Multiuser decoding objective and set-valued output}
\label{app:ura_formal_outer}

\paragraph{Multiuser decoder as a mapping from evidence to a $K\times L$ grid.}
The multiuser decoding stage is modeled as a mapping
\[
\hat X = \mathcal{G}(S),
\qquad
\hat X\in[Q]^{K\times L},
\]
where $\hat X$ is defined up to row permutation.
Many implementations of $\mathcal{G}$ produce per-site logits
$\Lambda\in\mathbb{R}^{K\times L\times Q}$ and then decode by argmax
\[
\hat X_{k,\ell} = \arg\max_{a\in[Q]} \Lambda_{k,\ell,a}.
\]
In our learned decoders (MDD/\cider), the logits $\Lambda$ arise from a masked-denoising refinement loop conditioned on $S$.
In rule-based baselines, $\Lambda$ may be produced implicitly (e.g., via BP marginals) or explicitly as per-slot posterior scores.

\paragraph{Mapping a decoded grid to a message set.}
Given $\hat X$, the decoded message set is
\[
\hat{\mathcal W}
\;\triangleq\;
\left\{
m \in [M]\;:\;
\exists\, k\in[K]\ \text{s.t. }\ \hat X_{k,:}=c_m
\right\}.
\]
That is, we keep the message indices whose codewords appear among the decoded rows.
(When $\mathcal{C}$ is an LDPC code, this corresponds to keeping only parity-valid rows under the shared codebook.)
In practice, the ``row $\mapsto$ message'' step can be implemented either by explicit lookup in $\mathcal{C}$ (when the full codebook is enumerated) or by checking code membership/validity (e.g., $H\hat X_{k,:}=0$ for an LDPC code) together with a payload-to-codeword convention.

\paragraph{Permutation invariance and row matching.}
Because URA is unsourced, row order is meaningless.
When comparing $\hat X$ and $X^\star$, we first align rows by an optimal permutation (Hungarian matching).
Let $\mathfrak{S}_K$ denote the set of permutations of $[K]$.
We define
\begin{equation}
\label{eq:app_ura_hungarian}
\pi^\star
\triangleq
\arg\min_{\pi\in\mathfrak{S}_K}
\sum_{k=0}^{K-1}\sum_{\ell=0}^{L-1}
\mathbf{1}\!\left[\hat X_{k,\ell}\neq X^\star_{\pi(k),\ell}\right].
\end{equation}
All symbol- and codeword-level metrics in the main text are computed after this alignment.
Equivalently, \eqref{eq:app_ura_hungarian} maximizes the total number of matched symbol positions under a one-to-one assignment between predicted and true rows.

\paragraph{What this paper replaces.}
Classical URA pipelines implement $\mathcal{G}$ using rule-based stitching and sequential SIC-style peeling under code constraints.
In this paper, we keep the symbol-level soft detector and evidence interface $S$ unchanged and replace the mapping $\mathcal{G}$ with our masked-diffusion-based decoder (\cider).

\section{Slot-wise AMP--MMSE detector}
\label{app:amp_mmse}

We use a fixed detector that operates independently for each slot $\ell\in[L]$.
Within a slot, the receiver faces a sparse inverse problem: among $Q$ candidate signatures, only a small subset actually appears, because each of the $K$ active devices selects (at most) one signature index in that slot.
We adopt Approximate Message Passing (AMP) with an MMSE denoiser to produce \emph{soft} activity evidence \citep{donoho2009messagepassing,bayati2011dynamics}.

\paragraph{Slot observation model (why it is sparse recovery).}
In slot $\ell$, we observe a length-$n_s$ received vector $Y^{(\ell)}\in\mathbb{C}^{n_s}$:
\[
Y^{(\ell)} = A^{(\ell)} U^{(\ell)} + \epsilon^{(\ell)},\qquad
\epsilon^{(\ell)}\sim\mathcal{CN}(0,\sigma^2 I).
\]
Here $A^{(\ell)}\in\mathbb{C}^{n_s\times Q}$ is a known sensing/dictionary matrix whose $Q$ columns correspond to the $Q$ possible signatures in slot $\ell$ (Partial-DFT in our experiments).
The unknown vector $U^{(\ell)}\in\mathbb{C}^{Q}$ encodes which signatures participated in the slot:
the coordinate $U^{(\ell)}_a$ is \emph{active} if signature index $a\in[Q]$ was used by at least one active device in slot $\ell$, and \emph{inactive} if that signature was not used in the slot.
Equivalently, ``active'' means $U^{(\ell)}_a\neq 0$ and ``inactive'' means $U^{(\ell)}_a=0$.
Since typically $K\ll Q$, only a small fraction of the $Q$ coordinates are active, so $U^{(\ell)}$ is sparse.
When $n_s<Q$, recovering $U^{(\ell)}$ from $Y^{(\ell)}$ is underdetermined unless we exploit this sparsity. In the simplified URA model in \Cref{eq:app_ura_activity_vec}, collisions make some active coordinates take integer values larger than $1$; AMP remains applicable because it performs approximate Bayesian inference under a continuous-valued sparse prior.

\paragraph{Probabilistic sparsity model.}
To produce calibrated \emph{soft} evidence (rather than only a hard support estimate), we use a Bernoulli--Gaussian prior on each coordinate:
\[
U^{(\ell)}_{a}\ \sim\ (1-\rho_{\mathrm{BG}})\,\delta_0 \;+\; \rho_{\mathrm{BG}}\,\mathcal{CN}(0,\sigma_u^2),
\qquad a\in[Q].
\]
Under this model, $\rho_{\mathrm{BG}}$ controls the probability that a given signature coordinate is active (i.e., nonzero) in a slot, and $\sigma_u^2$ sets the typical magnitude scale of a nonzero (active) coordinate.
In a URA slot, $\rho_{\mathrm{BG}}$ is conceptually tied to the expected sparsity level: if there are $K$ active devices choosing among $Q$ signatures, then only on the order of $K$ coordinates are expected to be active (with collisions potentially reducing the number of distinct active coordinates). We treat $(\rho_{\mathrm{BG}},\sigma_u^2)$ as fixed symbol-wise soft detector hyperparameters; the multiuser decoder sees only the resulting evidence matrix $S$.

\paragraph{Why AMP and what the iterations represent.}
AMP is an iterative method designed for large sensing matrices where repeated linear updates can be approximated by a ``signal + effective Gaussian noise'' scalar channel.
At a high level, AMP alternates between:
($i$) forming a per-coordinate pseudo-observation by backprojecting the current residual through $(A^{(\ell)})^H$, and
($ii$) applying a scalar denoiser coordinate-wise to update the estimate.
Crucially, AMP includes an additional correction (the Onsager correction) in the residual update, which cancels the dominant self-interference created by reusing the same sensing matrix across iterations; this stabilization is what makes the scalar ``effective noise'' view accurate in practice. When $A^{(\ell)}$ has approximately orthogonal columns (as with partial-DFT) and the problem dimensions are moderate-to-large, this approximation is accurate enough that a simple per-coordinate denoiser yields useful posterior estimates.

\paragraph{AMP--MMSE: denoiser and iterations.}
AMP is built around a simple scalar view of each coordinate: it treats a \emph{pseudo-observation} as if it were a noisy scalar measurement
\[
R \;=\; U + W,\qquad W\sim\mathcal{CN}(0,\nu),
\]
and then applies the \emph{scalar MMSE denoiser}
\begin{equation}
\label{eq:amp_mmse_denoiser_def}
\eta_{\mathrm{MMSE}}(r;\nu)\ \triangleq\ \mathbb{E}\!\left[U \,\middle|\, R=r\right].
\end{equation}
AMP also uses a companion \emph{sensitivity} term $\eta'_{\mathrm{MMSE}}(r;\nu)$ to scale the Onsager correction; in real-valued notation, one may view it as the derivative of the denoiser,
\begin{equation}
\label{eq:amp_mmse_sensitivity_def}
\eta'_{\mathrm{MMSE}}(r;\nu)\ \triangleq\ \frac{\partial}{\partial r}\,\eta_{\mathrm{MMSE}}(r;\nu),
\end{equation}
and in the complex setting we use the standard AMP analogue of this scalar sensitivity (see \citet{bayati2011dynamics}).

In our symbol-wise soft detector, the prior on $U$ is Bernoulli--Gaussian:
$U \sim (1-\rho_{\mathrm{BG}})\delta_0 + \rho_{\mathrm{BG}}\mathcal{CN}(0,\sigma_u^2)$.
Here $U\neq 0$ (``present'') means that the corresponding signature index actually participated in the slot, i.e., $U^{(\ell)}_a\neq 0$ for that coordinate $a$.
Under this prior and the scalar model above, the posterior activity probability $\hat p(r)=
\Pr(U\neq 0\mid R=r)
$ and the MMSE denoiser \eqref{eq:amp_mmse_denoiser_def} admit closed forms:
\begin{align}
\label{eq:bg_activity_prob}
\hat p(r)
&=
\frac{\rho_{\mathrm{BG}}\,\phi(r;0,\nu+\sigma_u^2)}
{(1-\rho_{\mathrm{BG}})\,\phi(r;0,\nu)\;+\;\rho_{\mathrm{BG}}\,\phi(r;0,\nu+\sigma_u^2)},\\
\label{eq:bg_mmse_mean}
\eta_{\mathrm{MMSE}}(r;\nu)
&=
\hat p(r)\cdot \frac{\sigma_u^2}{\nu+\sigma_u^2}\,r,
\end{align}
where $\phi(\cdot;0,\cdot)$ denotes the circular complex Gaussian density. 
For completeness, the activity posterior in \eqref{eq:bg_activity_prob} follows from Bayes' rule using the mixture prior:
\[
\Pr(U\neq 0\mid R=r)
=
\frac{\rho_{\mathrm{BG}}\,p(r\mid U\neq 0)}{(1-\rho_{\mathrm{BG}})\,p(r\mid U=0)+\rho_{\mathrm{BG}}\,p(r\mid U\neq 0)},
\]
with $p(r\mid U=0)=\phi(r;0,\nu)$ and $p(r\mid U\neq 0)=\phi(r;0,\nu+\sigma_u^2)$ after marginalizing $U\sim\mathcal{CN}(0,\sigma_u^2)$.

The MMSE mean \eqref{eq:bg_mmse_mean} then combines the posterior mean under the Gaussian component,
$\mathbb{E}[U\mid R=r, U\neq 0]=\frac{\sigma_u^2}{\nu+\sigma_u^2}r$, with the activity probability $\hat p(r)$.
In practice, AMP evaluates \eqref{eq:bg_mmse_mean} coordinate-wise, and uses the corresponding sensitivity term \eqref{eq:amp_mmse_sensitivity_def} (or its standard AMP complex counterpart) inside the Onsager correction.

With these definitions, for each slot $\ell$ AMP maintains an estimate $\hat U^{(i)}\in\mathbb{C}^{Q}$ of $U^{(\ell)}$ and a measurement-domain residual $\tilde Y^{(i)}\in\mathbb{C}^{n_s}$ (the part of $Y^{(\ell)}$ not yet explained by the current estimate). We initialize $\hat U^{(0)}=0$ and $\tilde Y^{(0)}=Y^{(\ell)}$, and for $i=0,1,\ldots,I_{\mathrm{AMP}}-1$ we iterate:
\begin{align}
\label{eq:amp_iter_r}
R^{(i)} &= \hat U^{(i)} + (A^{(\ell)})^{\!H}\tilde Y^{(i)},\\
\label{eq:amp_iter_u}
\hat U^{(i+1)}_a &= \eta_{\mathrm{MMSE}}\!\left(R^{(i)}_a;\,\nu^{(i)}\right)\qquad (a\in[Q]),\\
\label{eq:amp_iter_res}
\tilde Y^{(i+1)} &= Y^{(\ell)} - A^{(\ell)}\hat U^{(i+1)} \;+\;
\frac{\tilde Y^{(i)}}{n_s}\sum_{a\in[Q]}\eta'_{\mathrm{MMSE}}\!\left(R^{(i)}_a;\,\nu^{(i)}\right),
\end{align}
where $\nu^{(i)}$ is an effective noise-variance estimate for the pseudo-observation (e.g., $\nu^{(i)}\approx \|\tilde Y^{(i)}\|_2^2/n_s$). The final term in \eqref{eq:amp_iter_res} is the Onsager correction, which uses the denoiser sensitivity $\eta'_{\mathrm{MMSE}}(\cdot)$ to compensate for iteration-to-iteration correlations when reusing $A^{(\ell)}$ \citep{bayati2011dynamics}.

\paragraph{Evidence interface.}
After $I_{\mathrm{AMP}}$ iterations, AMP provides a final pseudo-observation
$R^{(I_{\mathrm{AMP}})}\in\mathbb{C}^{Q}$ (and an associated effective variance estimate $\nu^{(I_{\mathrm{AMP}})}$).
We then convert these into a slot-wise evidence heatmap by evaluating the posterior activity probability under the Bernoulli--Gaussian
scalar model:
\[
S_{\ell,a}\ \triangleq\ \log \Pr\!\big(U^{(\ell)}_{a}\neq 0 \mid Y^{(\ell)},A^{(\ell)}\big)
\;\approx\; \log \hat p\!\left(R^{(I_{\mathrm{AMP}})}_a;\nu^{(I_{\mathrm{AMP}})}\right),
\]
where $\hat p(\cdot;\cdot)$ is given in \eqref{eq:bg_activity_prob}. Stacking over slots yields $S\in\mathbb{R}^{L\times Q}$.
Throughout the paper, the symbol-wise soft detector (and hence $S$) is fixed, and we train/compare the multiuser decoder.

\section{\texorpdfstring{$Q$-ary LDPC code}{Q-ary LDPC code}}
\label{app:ldpc_details}

Our main experiments use a \emph{$Q$-ary} low-density parity-check (LDPC) code over the finite field $\mathbb{F}_Q$ (here $Q=64$) to impose \emph{global consistency across slots}.
In the main text, we denote the shared URA codebook by $\mathcal{C}=\{c_m\}_{m\in[M]}$ with codewords $c_m\in[Q]^L$.
In our LDPC setting, we fix a bijection $\phi:[Q]\to\mathbb{F}_Q$ and use it to interpret each discrete symbol index as a field element. With a slight abuse of notation, we write $c_m\in\mathbb{F}_Q^L$ (equivalently $\mathbf{c}_m=\phi(c_m)\in\mathbb{F}_Q^L$) when discussing parity constraints.
In this regime, the message index set $[M]$ can be viewed as an enumeration of the LDPC codebook; when $H$ has full row rank, $|\mathcal{C}_{\mathrm{LDPC}}|=Q^{L-P}$ and thus $B=\log_2 |\mathcal{C}_{\mathrm{LDPC}}|$ (so $M=2^B$).

\paragraph{Linear code definition via a sparse parity-check matrix.}
A $Q$-ary LDPC code is a linear block code specified by a sparse parity-check matrix
$H\in\mathbb{F}_Q^{P\times L}$.
The set of valid codewords is
\begin{equation}
\label{eq:app_ldpc_code_def}
\mathcal{C}_{\mathrm{LDPC}} \;\triangleq\; \{\,\mathbf{c}\in\mathbb{F}_Q^L : H\mathbf{c} = \mathbf{0} \text{ in } \mathbb{F}_Q\,\}.
\end{equation}
In our experiments, the codebook $\mathcal{C}$ is chosen to be (or to be identified with) this LDPC code:
\begin{equation}
\label{eq:app_codebook_equals_ldpc}
\mathcal{C}=\{\mathbf{c}_m\}_{m\in[M]}=\mathcal{C}_{\mathrm{LDPC}},
\qquad\text{so each }\mathbf{c}_m\text{ satisfies }H\mathbf{c}_m=\mathbf{0}.
\end{equation}

Equivalently, each check row $j\in[P]$ enforces a single parity constraint
\begin{equation}
\label{eq:app_ldpc_check_eq}
\sum_{\ell=0}^{L-1} H_{j,\ell}\,(\mathbf{c}_m)_\ell \;=\; 0
\qquad\text{in }\mathbb{F}_Q,
\end{equation}
where only a small number of coefficients $H_{j,\ell}$ are nonzero. The defining feature of LDPC is that each constraint involves only a few symbol positions, enabling scalable constraint enforcement.

\paragraph{Tanner graph viewpoint (why sparsity matters).}
The matrix $H$ induces a bipartite Tanner graph with variable nodes $\ell\in[L]$ (one per slot position) and check nodes $j\in[P]$ (one per parity equation).
Define the edge set
\[
E_H \triangleq \{(j,\ell)\in[P]\times[L] : H_{j,\ell}\neq 0\}, \qquad |E_H|=\#\{H_{j,\ell}\neq 0\}.
\]
We also define neighbor sets
\[
\mathcal{N}_{\mathrm{var}}(j)\triangleq\{\ell\in[L]:(j,\ell)\in E_H\},\qquad
\mathcal{N}_{\mathrm{chk}}(\ell)\triangleq\{j\in[P]:(j,\ell)\in E_H\}.
\]
Then \eqref{eq:app_ldpc_check_eq} can be written more explicitly as
\begin{equation}
\label{eq:app_ldpc_check_eq_neighbors}
\sum_{\ell'\in\mathcal{N}_{\mathrm{var}}(j)} H_{j,\ell'}\,(\mathbf{c}_m)_{\ell'} \;=\; 0
\qquad\text{in }\mathbb{F}_Q,
\end{equation}
highlighting that each check touches only the few variables in $\mathcal{N}_{\mathrm{var}}(j)$.

\paragraph{Why we use LDPC in shared-codebook multiuser decoding.}
In URA, the symbol-wise soft detector outputs only \emph{slot-wise} evidence for each slot $\ell$ and each symbol index $a\in[Q]$, but it does not specify how to assemble these local hypotheses into globally consistent messages.
The LDPC constraint $Hc_m=0$ provides strong global structure: among all $Q^L$ possible length-$L$ sequences, only those in $\mathcal{C}_{\mathrm{LDPC}}$ satisfy the parity checks.
Thus, the multiuser decoder can use $H$ to rule out spurious stitched sequences while still allowing efficient constraint propagation because $H$ is sparse.

\paragraph{Soft information from the detector.}
Let $S\in\mathbb{R}^{L\times Q}$ be the evidence matrix produced by the fixed detector.
A standard probabilistic form is the per-slot categorical belief
\begin{equation}
\label{eq:app_ldpc_local_belief_from_S}
\lambda_\ell(a)
\;\propto\;
\exp(S_{\ell,a}),
\qquad a\in[Q],\ \ell\in[L],
\end{equation}
which can be interpreted (after normalization over $a$) as a soft belief that symbol $a$ appeared in slot $\ell$.
LDPC decoding can be viewed as combining these local beliefs with the global parity constraints \eqref{eq:app_ldpc_code_def}.

\paragraph{Standard LDPC decoding principle: belief propagation on the Tanner graph.}
A classical way to enforce \eqref{eq:app_ldpc_code_def} given soft beliefs \eqref{eq:app_ldpc_local_belief_from_S} is belief propagation (BP) on the Tanner graph.
BP passes $Q$-dimensional messages along edges $(j,\ell)\in E_H$.
Denote variable-to-check messages by $m_{\ell\to j}(a)$ and check-to-variable messages by $m_{j\to \ell}(a)$, both defined for $a\in[Q]$. Here and below, the symbol arguments $a$ and $\{a_{\ell'}\}$ are interpreted as elements of $\mathbb{F}_Q$ via the fixed labeling $\phi$.

The variable update multiplies the local belief with incoming check messages:
\begin{equation}
\label{eq:app_ldpc_v2c}
m_{\ell\to j}(a)
\;\propto\;
\lambda_\ell(a)\prod_{j'\in\mathcal{N}_{\mathrm{chk}}(\ell)\setminus\{j\}} m_{j'\to \ell}(a).
\end{equation}
The check update enforces the parity constraint by summing over assignments of neighboring variables that satisfy \eqref{eq:app_ldpc_check_eq_neighbors}:
\begin{equation}
\label{eq:app_ldpc_c2v}
m_{j\to \ell}(a)
\;\propto\;
\sum_{\{a_{\ell'}\}_{\ell'\in\mathcal{N}_{\mathrm{var}}(j)\setminus\{\ell\}}}
\mathbf{1}\!\left[
H_{j,\ell}\,a \;+\!\!\!\sum_{\ell'\in\mathcal{N}_{\mathrm{var}}(j)\setminus\{\ell\}}\!\!\! H_{j,\ell'}\,a_{\ell'} = 0
\right]
\prod_{\ell'\in\mathcal{N}_{\mathrm{var}}(j)\setminus\{\ell\}} m_{\ell'\to j}(a_{\ell'}).
\end{equation}
After a fixed number of BP iterations, the approximate marginal at each variable node is
\begin{equation}
\label{eq:app_ldpc_marginal}
b_\ell(a)
\;\propto\;
\lambda_\ell(a)\prod_{j\in\mathcal{N}_{\mathrm{chk}}(\ell)} m_{j\to \ell}(a),
\qquad a\in[Q],
\end{equation}
and a hard decision can be made by $\hat c_\ell=\arg\max_{a\in[Q]} b_\ell(a)$.
Our SIC-BP baseline uses this classical principle (with additional multiuser heuristics) to decode multiple rows.

\paragraph{How \cider uses $H$.}
\cider leverages the same parity structure but integrates it into the learned denoiser (Module B) through Tanner-graph propagation.
In particular, the nonzero coefficients $\alpha = H_{j,\ell}\in\mathbb{F}_Q^\times$ define edge-dependent actions in $\mathbb{F}_Q$ under the fixed $[Q]\leftrightarrow \mathbb{F}_Q$ convention described above, and \cider applies these coefficients consistently during propagation.
In our implementation this is realized through fixed, field-consistent coefficient actions (e.g., $T_\alpha$) inside the propagation module.
This injects LDPC constraints while preserving the standard URA modularity (the symbol-wise soft detector and evidence interface $S$ remain fixed).

\paragraph{Construction of $H$ in our experiments.}
For each evaluation scale, we build the Tanner-graph connectivity pattern (i.e., the locations of nonzeros in $H$) from a structured regular graph that avoids short cycles, and then assign i.i.d.\ nonzero coefficients $\alpha_{j,\ell}$ on each edge:
\[
\alpha_{j,\ell} = H_{j,\ell}\in\mathbb{F}_Q^\times \ \text{i.i.d.\ for } (j,\ell)\in E_H,
\qquad
H_{j,\ell}=0 \ \text{for } (j,\ell)\notin E_H.
\]
The nominal rate is approximately $R \approx (L-P)/L$ (and equals $(L-\mathrm{rank}(H))/L$ in general). For background on LDPC codes and belief-propagation decoding, see \citet{gallager1962ldpc}.

\section{Additional background on masked discrete diffusion}
\label{app:maskdiff_details}

This appendix collects standard identities for masked (absorbing-state) categorical diffusion, and summarizes the practical inference procedure used by masked denoising models (MDM).

\subsection{Masked diffusion: reverse transitions}
\label{app:maskdiff_posterior}

\paragraph{Time indexing and direction.}
We index refinement steps by $t\in\{0,1,\ldots,T\}$, where $t=0$ denotes the most corrupted initialization (all-\mask) and $t=T$ denotes a fully revealed state.
If one prefers the standard diffusion convention in which time increases with corruption, define $s\triangleq T-t$.
Under this reparameterization, the forward noising direction is $s:0\to T$ (clean $\to$ corrupted) and the reverse denoising direction is $s:T\to 0$.

\paragraph{Discrete-step reverse model.}
At each refinement step $t$, the reverse process predicts clean tokens from the current masked state.
Given denoiser prediction $\hat x_{\theta,j}(\tilde X^{(t)},\xi,t)\in\Delta^Q$, unmasked positions are selected based on confidence ranking (see \Cref{app:maskdiff_infer}).

\paragraph{Reverse transition.}
The reverse transition is parameterized by a denoiser that predicts a clean-token distribution from a noisy input and conditioning context $\xi$
(in our URA instantiation, $\xi\equiv S$, where $S\in\mathbb{R}^{L\times Q}$ is the slot-wise evidence matrix from the fixed symbol-wise soft detector).
Given a denoiser prediction $\hat x_{\theta,j}(\tilde X^{(t)},\xi,t)\in\Delta^Q$, we define the reverse transition by substituting $\hat x_\theta$ into the analytic single-site posterior:
\begin{equation}
\label{eq:app_reverse_transition}
p_\theta\!\left(\tilde x^{(s)}_{j}\mid \tilde x^{(t)}_{j},\xi\right)
\triangleq
q\!\left(\tilde x^{(s)}_{j}\mid \tilde x^{(t)}_{j},\hat x_{\theta,j}(\tilde X^{(t)},\xi,t)\right).
\end{equation}
Here $s>t$ denotes a less-corrupted step than $t$ under our $t=0\to T$ refinement convention.
In practice, we use this identity only to motivate the standard masked-denoising objective and the iterative unmasking sampler; the implementation directly computes logits $\Lambda^{(t)}$ and updates the discrete masked state.

\paragraph{Uniform categorical corruption (D3PM-style).}
As a comparison point, instead of masked diffusion, D3PM replaces the absorbing mask with \emph{uniform categorical} corruption \citep{austin2021d3pm}.
Let $\bar X^{(t)}\in[Q]^{N_{\mathrm{seq}}}$ denote the corrupted state (no mask token) and let $\bar\gamma_t\in[0,1]$ be the corruption rate.
For each position $j\in[N_{\mathrm{seq}}]$ and symbol $a\in[Q]$,
\begin{equation}
\label{eq:app_uniform_forward}
q_{\mathrm{unif}}\!\left(\bar X^{(t)}_{j}=a \,\middle|\, X_j=x_j\right)
= (1-\bar\gamma_t)\,\mathbf{1}[a=x_j] \;+\; \bar\gamma_t\cdot\frac{1}{Q}.
\end{equation}
Equivalently, $\bar X^{(t)}_{j}=x_j$ with probability $(1-\bar\gamma_t)$, and otherwise $\bar X^{(t)}_{j}$ is resampled uniformly from $[Q]$.

\subsection{Inference: confidence ranking and cosine reveal schedule}
\label{app:maskdiff_infer}

\paragraph{Unmasking order.}
While the masked diffusion model defines a family of reverse-time transitions, practical masked denoising models (MDM) must additionally specify \emph{which sites are revealed at each iteration}.
In particular, decoding maintains a discrete masked state $\tilde X^{(t)}\in([Q]\cup\{\mask\})^{K\times L}$ (with $K$ rows and $L$ slots) and iteratively replaces a subset of $\mask$ entries with predicted symbols over $T$ refinement steps. Thus, inference is determined not only by the denoiser outputs but also by an \emph{unmasking schedule}. 

We adopt a cosine reveal schedule \citep{chang2022maskgit} to determine how many sites should be unmasked by step $t$:
\begin{equation}
\label{eq:cosine_reveal_fraction}
\rho(t) \triangleq \frac{1}{2}\left(1-\cos\!\left(\pi\frac{t}{T}\right)\right), \qquad t=0,1,\dots,T.
\end{equation}
During inference, the scalar mask-ratio input to the denoiser is set to the remaining masked fraction,
\[
\gamma_t^{\mathrm{infer}} = 1-\rho(t),
\]
so the denoiser conditioning is consistent with the fraction of tokens that remain masked at each refinement step.

Concretely, at step $t$ we compute logits
\[
\Lambda^{(t)} = f_\theta(\tilde X^{(t-1)}, S, \gamma_t^{\mathrm{infer}})
\]
and probabilities
\[
p^{(t)}_{k,\ell,:}
=
\mathrm{softmax}\!\left(\Lambda^{(t)}_{k,\ell,:}/\tau_{\mathrm{infer},t}\right).
\]
We define the per-site \emph{confidence} as
\[
\kappa^{(t)}_{k,\ell}\;\triangleq\;\max_{a\in[Q]} p^{(t)}_{k,\ell,a},
\]
to avoid clashing with the codebook notation $c_m$ in the main text.

At step $t$, we reveal the highest-confidence entries among the currently masked sites so that the \emph{cumulative} fraction of revealed sites is approximately $\rho(t)$.

\paragraph{First-reveal stabilization.}
In our default \cider inference procedure, we use a first-reveal stabilization rule that gives priority to row hypotheses without any revealed site. For $K\ge 3$, whenever a top-ranked masked site belongs to such a row $k$, we reveal only that single site at that step, which anchors row $k$. Once every row is anchored (and after the remasking in PRISM passes), decoding follows the cosine schedule in \eqref{eq:cosine_reveal_fraction}. This keeps early reveals from concentrating on a few rows while others remain fully masked, which matters most at higher per-bin loads. For $K\le 2$, the rule reduces to first revealing the most confident site of one randomly chosen row.

\subsection{PRISM-style quality head and remasking plug-in}
\label{app:prism_remasking}

This section describes how we plug a PRISM-style confidence head into \cider and use it for quality-guided remasking at inference~\citep{prism2025}. The key property is that the head is trained post-hoc and does not modify the backbone denoiser; it only reads the backbone latent representations and outputs per-token correctness probabilities that we use to trigger selective remasking.

\paragraph{Backbone outputs and notation.}
\cider produces token logits
$\Lambda^{(t)}\in\mathbb{R}^{K\times L\times Q}$ at each refinement step $t\in\{1,\dots,T\}$.
We denote the refined per-step latent used by the confidence head by $\hat Z^{(t)}\in\mathbb{R}^{K\times L\times D}$.
We write
\[
p^{(t)}_{k,\ell,:} \triangleq \mathrm{softmax}\!\left(\Lambda^{(t)}_{k,\ell,:}/\tau_t\right)\in\Delta^{Q}
\]
for the categorical distribution at site $(k,\ell)$ (with temperature $\tau_t>0$ if used).
Let $X^\star\in[Q]^{K\times L}$ be the ground-truth codeword grid (row order immaterial). As in the main text, we use Hungarian matching to align predicted rows to ground-truth rows; we denote the resulting permutation by $\pi^\star\in\mathfrak{S}_K$.

\paragraph{Per-token quality head.}
We attach a lightweight two-layer MLP $g_{\psi}$ to each site latent and predict a per-token correctness probability
\begin{equation}
\label{eq:app_prism_head}
\omega^{(t)}_{k,\ell}
\;=\;
\sigma\!\left(g_{\psi}\!\left(\hat Z^{(t)}_{k,\ell}\right)\right)
\in[0,1],
\end{equation}
where $\sigma(\cdot)$ is the sigmoid. Concretely, $g_{\psi}$ is a 2-layer MLP (e.g., LayerNorm/GELU/Dropout between layers) mapping $\mathbb{R}^{D}\to\mathbb{R}$.

\paragraph{PRISM-style supervision for the head.}
We train the head using PRISM-style labels constructed from the backbone’s own sampled predictions.
For each training example, we sample a diffusion step (mask ratio) and form a partially-masked grid input $\tilde X^{(t)}$,
run the frozen backbone to obtain logits $\Lambda^{(t)}$ and probabilities $p^{(t)}_{k,\ell,:}$ at masked sites,
and compute the row-matching permutation $\pi^\star$ (as in the main text) using a discrete grid derived from $\Lambda^{(t)}$ (e.g., per-site argmax).
For a selected subset of masked sites $\mathcal{S}\subseteq\{(k,\ell):\tilde X^{(t)}_{k,\ell}=\mask\}$, we sample a token prediction
\begin{equation}
\label{eq:app_prism_sample_pred}
\hat x^{(t)}_{k,\ell} \sim \mathrm{Cat}\!\left(p^{(t)}_{k,\ell,:}\right),
\qquad (k,\ell)\in\mathcal{S}.
\end{equation}
We define the binary correctness label (using the matched ground-truth row)
\begin{equation}
\label{eq:app_prism_label}
b^{(t)}_{k,\ell}
\;\triangleq\;
\mathbf{1}\!\left[\hat x^{(t)}_{k,\ell} = X^\star_{\pi^\star(k),\ell}\right],
\end{equation}
and minimize binary cross-entropy on $\mathcal{S}$:
\begin{equation}
\label{eq:app_prism_bce}
\mathcal{L}_{\mathrm{qual}}(\psi)
\;=\;
\sum_{(k,\ell)\in\mathcal{S}}
\Big(
- b^{(t)}_{k,\ell}\log \omega^{(t)}_{k,\ell}
- (1-b^{(t)}_{k,\ell})\log(1-\omega^{(t)}_{k,\ell})
\Big).
\end{equation}
In all experiments, the backbone denoiser is frozen when training $g_{\psi}$; only the head parameters $\psi$ are updated.

\paragraph{Sequence-level confidence for multiuser decoding.}
We make remasking decisions at the \emph{sequence} (row) level.
After a decoding pass completes (all tokens revealed at the final step $t=T$), we compute a sequence-level confidence score by averaging per-token quality predictions across slots:
\begin{equation}
\label{eq:app_row_conf}
\bar \omega_{k}
\;\triangleq\;
\frac{1}{L}\sum_{\ell=0}^{L-1} \omega^{(T)}_{k,\ell}.
\end{equation}
We interpret smaller $\bar \omega_k$ as indicating a more error-prone sequence (row) that is likely to contain incorrect symbols across rows.

\paragraph{Quality-guided remasking and second pass.}
Let $\hat X\in[Q]^{K\times L}$ be the discrete decoded grid after the first pass.
We use a multi-stage threshold-based remasking strategy indexed by $s\in\{1,\dots,\varrho\}$, with quality thresholds $\{\varphi_s\}_{s=1}^{\varrho}$.

After each decoding pass, we identify low-confidence sequences as those falling below the threshold: 
\[
\mathcal{K}_{\mathrm{low}}
\;\triangleq\;
\{k \in [K] : \bar \omega_k < \varphi_s\}.
\]

We initialize the second-pass discrete state by remasking exactly these sequences:
\begin{equation}
\label{eq:app_remask_init_threshold}
X_{\mathrm{rm}}^{(0)}{}_{k,\ell}
=
\begin{cases}
\mask, & k\in\mathcal{K}_{\mathrm{low}},\\
\hat X_{k,\ell}, & k\notin\mathcal{K}_{\mathrm{low}}.
\end{cases}
\end{equation}
We then rerun the masked-diffusion inference procedure starting from $X_{\mathrm{rm}}^{(0)}$,
while clamping the high-confidence codewords throughout the second pass: for all refinement steps $t$ and all $k\notin\mathcal{K}_{\mathrm{low}}$,
we enforce $X_{\mathrm{rm}}^{(t)}{}_{k,\ell}=\hat X_{k,\ell}$ (i.e., token updates are disallowed outside $\mathcal{K}_{\mathrm{low}}$).
The number of refinement steps for each remasking pass scales with the number of sequences being remasked:
$$T_{\mathrm{rm}} = \max\left(1,\ \left\lfloor T \cdot \frac{|\mathcal{K}_{\mathrm{low}}|}{K} \right\rfloor\right)$$
where $T$ is the number of steps used in the initial decoding pass.
This process repeats (increasing $s$) until $\mathcal{K}_{\mathrm{low}}=\emptyset$ or all $\varrho$ thresholds are exhausted.

\paragraph{Algorithmic summary.}
Algorithm~\ref{alg:prism_remask} summarizes the procedure for one remasking round.

\begin{algorithm}[H]
\caption{PRISM-style quality-guided remasking for \cider}
\label{alg:prism_remask}
\begin{enumerate}[leftmargin=1.2em,itemsep=1pt,topsep=1pt]
\item Run masked-diffusion decoding for $T$ steps to obtain $\hat X$ and final refined latents $\hat Z^{(T)}$.

\item Compute per-token qualities $\omega^{(T)}_{k,\ell}=\sigma(g_{\psi}(\hat Z^{(T)}_{k,\ell}))$ and sequence confidences $\bar \omega_k=\frac{1}{L}\sum_{\ell} \omega^{(T)}_{k,\ell}$.
\item Select $\mathcal{K}_{\mathrm{low}}$ as the sequences with $\bar \omega_k < \varphi_s$.
\item Initialize $X_{\mathrm{rm}}^{(0)}$ by remasking sequences (rows) in $\mathcal{K}_{\mathrm{low}}$ (Eq.~\eqref{eq:app_remask_init_threshold}).
\item Rerun masked-diffusion decoding from $X_{\mathrm{rm}}^{(0)}$ for $T_{\mathrm{rm}}$ steps while clamping $k\notin\mathcal{K}_{\mathrm{low}}$; output the refined grid.

\end{enumerate}
\end{algorithm}

\paragraph{Practical notes.}
($i$) The remasking head adds negligible runtime because it is a small per-token MLP applied to already-computed latents.
($ii$) The number of thresholds $\varrho$ and their values trade compute for reliability; more stages with stricter (higher) thresholds trigger additional remasking passes but improve accuracy.
($iii$) We use $\varrho=1,2,3$ remasking rounds for $K=6,7,8$, respectively, in our reported results.

\section{Experimental setup and implementation details}
\label{app:exp_details}

This appendix collects implementation details omitted from \Cref{sec:experiments}: ($i$) data generation and channel model, ($ii$) sensing matrices and symbol-wise soft detector evidence construction, ($iii$) multiuser code families, ($iv$) training and inference schedules, and ($v$) per-scale model hyperparameters.

\subsection{Simulation pipeline and channel model}
\label{app:exp_data}

For each scale in \Cref{tab:app_exact_dataset_scales}, we generate synthetic frames by sampling $K$ active messages uniformly from $[M]$,
mapping them through the shared codebook to obtain the ground-truth grid $X^\star\in[Q]^{K\times L}$,
and transmitting one signature index per slot.
For each slot $\ell\in[L]$, we form the sparse activity vector $U^{(\ell)}$ as in \Cref{eq:app_ura_activity_vec} and generate
\begin{equation}
\label{eq:app_channel_model}
Y^{(\ell)} = A^{(\ell)} U^{(\ell)} + \epsilon^{(\ell)},
\qquad
\epsilon^{(\ell)}\sim\mathcal{CN}(0,\sigma^2 I).
\end{equation}

\paragraph{Simulation constants.}

Unless stated otherwise, we use $K=2$ active users.
We generate $Y^{(\ell)} = A^{(\ell)}U^{(\ell)} + \epsilon^{(\ell)}$ with $\epsilon^{(\ell)}\sim\mathcal{CN}(0,\sigma^2 I)$ and fixed $\sigma^2=1.0$.
We control the effective SNR by scaling the transmit power in the symbol encoder as
$P_{\mathrm{sym}} = \frac{B\cdot 10^{E_b/10}}{L}$ (with $E_b$ in dB). 

\paragraph{Dataset sizes and sensing dimensions.}
We use $n_s = 24$ in our main implementation unless otherwise specified.
Additionally, we use a $70\mathrm{K}/15\mathrm{K}/15\mathrm{K}$ train/validation/test split per scale.

\subsection{Sensing-matrix construction}
  \label{app:sensing_matrices}

  All experiments use the partial-DFT sensing matrix.
  Let $\mathbf{F}\in\mathbb{C}^{Q\times Q}$ be the unnormalized DFT matrix
  ($\mathbf{F}_{k,n} = e^{-2\pi i kn/Q}$) and let $\mathcal{R}\subset\{0,\dots,Q-1\}$
  be $n_s$ row indices drawn uniformly at random:
  \begin{equation}
  \mathbf{A}_{\mathrm{DFT}} = \frac{1}{\sqrt{n_s}}\, \mathbf{F}_{\mathcal{R},\,:}
  \;\in\;\mathbb{C}^{n_s\times Q}.
  \end{equation}
  
This normalization ensures each column $a_q$ satisfies $\lVert a_q\rVert_2^2=1$.

\subsection{Per-user per-channel-use SNR}
Since each column of the sensing matrix is normalized, one transmitted symbol carries energy $P_{\mathrm{sym}}$ spread uniformly across its $n_s$ complex channel uses. Thus the per-user per-channel-use signal energy is $P_{\mathrm{sym}}/n_s$, and
\begin{align}
      \mathrm{SNR}
      &= \frac{P_{\mathrm{sym}}}{n_s\,\sigma^2}
       = \frac{B\cdot 10^{E_b/10}}{L\,n_s\,\sigma^2}, \\
      \mathrm{SNR}\,[\mathrm{dB}]
      &= E_b - 10\log_{10}\!\bigl(L n_s / B\bigr) \quad (\sigma^2 = 1).
\end{align}

\subsection{Fixed symbol-wise soft detector and evidence interface}
\label{app:exp_evidence}

For each slot $\ell$, the fixed symbol-wise soft detector (AMP--MMSE) takes $(Y^{(\ell)},A^{(\ell)})$ and produces log-posterior evidence scores
$s_{\ell,a}$ for every $a\in[Q]$.
We aggregate these into the evidence heatmap $S\in\mathbb{R}^{L\times Q}$ with entries $S_{\ell,a}=s_{\ell,a}$.
The matrices $S$ are pre-computed offline and stored in the dataset; all multiuser decoders operate only on $S$.

\subsection{Multiuser-code families and parity-check metadata}
\label{app:exp_codes}

In the experiments reported in this paper, we use LDPC codes (with rate $R$ as specified per experiment).
We store the parity-check matrix $H\in\mathbb{F}_Q^{P\times L}$ and provide it to parity-aware decoders.

\subsection{Training and inference schedules}
\label{app:exp_train_infer}

\paragraph{Optimization.}
All learned models are trained with AdamW under identical dataset splits (shared across model families unless otherwise stated).

\paragraph{Permutation-invariant training.}
Because the \(K\) decoded rows are unordered, we align predicted rows to target rows during training. For a masked training example, let \(\Omega_t\subseteq[K]\times[L]\) be the masked grid sites and let \(p^{(t)}_{k,\ell,:}\) be the predicted categorical distribution at site \((k,\ell)\). We form the row-matching cost
\[
D_{k,k'}=
\sum_{\ell:(k,\ell)\in\Omega_t}
\mathrm{CE}\!\left(
X^\star_{k',\ell},
p^{(t)}_{k,\ell,:}
\right),
\]
compute
\[
\pi^\star=\arg\min_{\pi\in\mathfrak{S}_K}\sum_{k=0}^{K-1}D_{k,\pi(k)},
\]
and minimize the matched masked cross-entropy
\[
\mathcal{L}_{\mathrm{train}}
=
\frac{1}{|\Omega_t|}
\sum_{(k,\ell)\in\Omega_t}
\mathrm{CE}\!\left(
X^\star_{\pi^\star(k),\ell},
p^{(t)}_{k,\ell,:}
\right).
\]
This matching removes the arbitrary row order from the supervision while preserving the standard masked-diffusion denoising objective.

\paragraph{MDM mask-ratio sampling (training).}
We sample a discrete timestep $t$ uniformly from $\{0,1,\ldots,T_{\mathrm{train}}\}$ with $T_{\mathrm{train}}$ to each model's step count (10/12/20/24 for Tiny/Small/Moderate/Large) and set
\begin{equation}
\gamma_t \triangleq \gamma_{\max} - (\gamma_{\max}-\gamma_{\min})\frac{t}{T_{\mathrm{train}}},
\qquad
(\gamma_{\min},\gamma_{\max})=(0.1,1.0).
\end{equation}
We use a short warmup: for the first $4$ epochs we exclude $t=0$ (fully masked), and afterwards we include $t=0$ (so fully masked examples occur with probability $1/(T_{\mathrm{train}}+1)$ under uniform $t$ sampling). 

\paragraph{MDM inference schedule (cosine reveal).}
We run $T$ refinement steps (scale-dependent; see Table~\ref{tab:app_exact_model_sizing}) and reveal tokens by confidence ranking according to the cosine schedule in \eqref{eq:cosine_reveal_fraction}.

\paragraph{Inference temperature annealing.}
We use cosine annealing for the inference temperature from \(\tau_{\max}\) at the first refinement step to \(\tau_{\min}\) at the final refinement step:
\begin{equation}
\tau_{\mathrm{infer},t} \triangleq
\tau_{\max}\,w(t) + \tau_{\min}\,\big(1-w(t)\big),
\qquad
w(t)\triangleq \frac{1}{2}\!\left(1+\cos\!\left(\pi\frac{t-1}{T-1}\right)\right).
\end{equation}
The denoiser probabilities are computed as
\(
p^{(t)}_{k,\ell,:}=\mathrm{softmax}(\Lambda^{(t)}_{k,\ell,:}/\tau_{\mathrm{infer},t}).
\)

\paragraph{Final refinement pass.}
After all tokens are revealed, we run one additional forward pass at $\gamma=0$ and output by per-site argmax.

\subsection{Model sizes by evaluation scale}
\label{app:per_scale_hparams}

We report four evaluation scales (Tiny/Small/Moderate/Large) in  \Cref{tab:app_exact_model_sizing} with a fixed alphabet size $Q=64$ and varying codeword lengths $L\in\{12,18,24,48\}$.
For each scale, we scale the denoiser capacity ($D$, layers/heads) and compare the same set of baselines under an identical evidence interface $S$ and Hungarian-matched supervision.

\section{Complexity derivation}
\label{app:complexity_derivation}

Let $J$ be the per-slot candidate list size retained from $S$ (top-$J$ indices per slot).
Let $H\in\mathbb{F}_Q^{P\times L}$ be the parity-check matrix and let $|E_H|$ be the number of nonzeros in $H$
(i.e., Tanner-graph edges).
We write $I_{\mathrm{BP}}$ for the 
number of Tanner-graph message-passing iterations performed by SIC-BP.
For \cider, we write $T$ for the number of refinement steps and $N_{\mathrm{layer}}$ for the number of stacked denoiser
blocks inside one refinement step.

\paragraph{Top-$J$ exhaustive search.}
Top-$J$ Exhaustive search forms candidate codewords by selecting one of $J$ candidates per slot, yielding $J^L$ candidates in the worst case.
Checking parity validity for one candidate can be implemented by aggregating along Tanner edges, costing $\mathcal{O}(|E_H|)$.
Thus the worst-case complexity is
\[
\mathcal{O}(J^L)\times \mathcal{O}(|E_H|)=\mathcal{O}(J^L|E_H|),
\]
which is exponential in $L$ (over the reduced Top-$J$ search space).

\paragraph{SIC-BP.}
SIC-BP performs belief propagation (BP) on the non-binary Tanner graph.
The dominant operation in non-binary BP is the check-to-variable update.
For a check node of degree $d_c$, computing one outgoing check-to-variable message amounts to convolving
$(d_c-1)$ distributions over $\mathbb{F}_Q$.
With direct GF$(Q)$ convolution, one convolution costs $\mathcal{O}(Q^2)$, so one outgoing message costs
$\mathcal{O}(d_c\,Q^2)$.
Since LDPC degrees are constant in our setting, we treat $d_c$ as a constant factor and write the per-edge BP update cost as
$\Theta(Q^2)$.
Therefore, $I_{\mathrm{BP}}$ BP iterations over $K$ decoded rows cost
\[
\mathcal{O}\!\left(K\,I_{\mathrm{BP}}\,|E_H|\,Q^2\right).
\]

\paragraph{FFT-BP.}
FFT-BP is algorithmically identical to SIC-BP but accelerates the check-to-variable update by exploiting the additive structure of $\mathbb{F}_{2^m}$. Since field addition coincides with bitwise XOR, the convolution of $(d_c{-}1)$ distributions over $\mathbb{F}_Q$ becomes pointwise multiplication in the Walsh--Hadamard transform (WHT) domain. Each WHT and its inverse costs $\mathcal{O}(Q\log Q)$, so the per-edge BP update cost drops from $\Theta(Q^2)$ to $\Theta(Q\log Q)$. This WHT-based formulation is the one used by recent coded random-access work \citep{ebert2022coded}. At $Q=64$ this reduces the per-edge operation count by about $10\times$, which matches the wall-clock speedup in per-sample CPU implementation.

\paragraph{\cider.}
\cider runs $T$ refinement steps on the discrete grid $X^{(t)}$ and constructs a latent grid of size $K\times L\times D$
at each step.
We summarize the dominant costs per refinement step.

\emph{($i$) Logit projection and demixing/evidence fusion.}
Computing $Q$ logits at each of $KL$ sites from $D$-dimensional latents costs $\mathcal{O}(KLQD)$, and the demixing/evidence
fusion has the same $Q$-summation structure per site.
Applying $N_{\mathrm{layer}}$ denoiser blocks multiplies this term by $N_{\mathrm{layer}}$.

\emph{($ii$) Parity-aware propagation.}
Parity propagation processes Tanner edges for each of the $K$ rows.
The main per-edge operation is the coefficient action used for normalization and denormalization inside Module~B.
In our implementation, a coefficient action is applied as
\[
T_{\alpha}(x) \triangleq W^\top \Pi_{\alpha} W x,
\]
which can be implemented by projecting to symbol space, applying the fixed permutation, and projecting back.
This costs $\mathcal{O}(QD)$ per application.
Since normalization and denormalization are applied on each edge (constant number of times per edge per block),
the parity-propagation cost scales as
\[
\mathcal{O}\!\left(N_{\mathrm{layer}}\,K\,|E_H|\,QD\right),
\]
up to constant factors from local aggregation over a constant check degree.

\emph{($iii$) Stabilization (optional).}
If enabled, applying $\Phi_U$ over all $KL$ sites costs $\mathcal{O}(KL\cdot \mathrm{cost}(\Phi_U))$ per refinement step.

Putting these together, the total cost over $T$ refinement steps is
\[
\mathcal{O}\!\left(
T\,N_{\mathrm{layer}}
\bigl[
KL(QD+D^2)
+
K|E_H|(QD+D^2)
\bigr]
+
T\,KL\cdot \mathrm{cost}(\Phi_U)
\right).
\]

\paragraph{Reading the comparison.}
This asymptotic accounting should be interpreted together with hardware execution. 
\cider is not claimed to have a smaller scalar operation count than every optimized BP variant in all regimes; for example, FFT-BP reduces non-binary check updates to \(\mathcal{O}(Q\log Q)\). 
The empirical speedup comes from fixed-depth, highly parallel dense tensor operations on GPU, avoidance of sequential SIC-style peeling, and avoidance of Top-\(J\) combinatorial enumeration.

\section{Exact configurations for reproducibility}
\label{app:exact_configs}

This appendix summarizes the exact configuration values used in \cider experiments (dataset scales, model sizing, and training/inference hyperparameters).

\subsection{Dataset scales (LDPC over GF(64))}
All scales use $Q=64$ (GF(64)) and $K=2$ active users. We vary the codeword length $L$ while keeping the code rate fixed at $R=1/3$.
To avoid notational conflicts with the main text (where $k$ denotes a row index), we denote the number of parity checks by $P$ and the number of information symbols by $L_{\mathrm{info}}\triangleq L-P$.

\begin{table}[H]
\centering
\footnotesize
\setlength{\tabcolsep}{4pt}
\renewcommand{\arraystretch}{1.05}
\caption{LDPC dataset scales used in all experiments (GF(64), $K=2$).}
\label{tab:app_exact_dataset_scales}
\begin{tabular}{l c c c c c}
\toprule
Scale & $Q$ & $L$ & $P$ & $L_{\mathrm{info}}$ & $R=L_{\mathrm{info}}/L$ \\
\midrule
Tiny     & 64 & 12 &  8 &  4 & 1/3 \\
Small    & 64 & 18 & 12 &  6 & 1/3 \\
Moderate & 64 & 24 & 16 &  8 & 1/3 \\
Large    & 64 & 48 & 32 & 16 & 1/3 \\
\bottomrule
\end{tabular}
\end{table}

\paragraph{Information payload.}
Each symbol carries $\log_2(64)=6$ bits, so the payload size is $6L_{\mathrm{info}}$ bits per message.

\subsection{Model sizing and inference steps}
Diffusion models run in discrete masked-token mode (\texttt{use\_soft\_input=false}) and use scale-dependent model sizing and inference steps.

\begin{table}[H]
\centering
\footnotesize
\setlength{\tabcolsep}{4pt}
\renewcommand{\arraystretch}{1.05}
\caption{Diffusion-model sizing by scale.}
\label{tab:app_exact_model_sizing}
\begin{tabular}{l c c c c}
\toprule
Scale & $D$ & $N_{\mathrm{layer}}$ & \#Heads & Inference steps $T$ \\
\midrule
Tiny     & 128 & 4 & 4 & 12 \\
Small    & 128 & 6 & 4 & 16 \\
Moderate & 128 & 6 & 4 & 20 \\
Large    & 128 & 8 & 4 & 28 \\
\bottomrule
\end{tabular}
\end{table}

\subsection{\texorpdfstring{Per-$K$ model sizing}{Per-K model sizing}}
Diffusion models run in discrete masked-token mode (\texttt{use\_soft\_input=false}) and use scale-dependent model sizing, inference steps, and remasking threshold for different $K$.

\begin{table}[H]
  \centering
  \footnotesize
  \setlength{\tabcolsep}{4pt}
  \renewcommand{\arraystretch}{1.05}
  \caption{Diffusion-model sizing per $K$. For $K\ge 6$, the PRISM-head
  pipeline performs $\varrho$ remasking rounds on top of the initial
  $T_{\mathrm{init}}$-step diffusion pass, gated by the threshold
  schedule $(\varphi_1,\dots,\varphi_\varrho)$.}
  \label{tab:app_exact_K_sizing}
  \begin{tabular}{c c c c c c c}
  \toprule
  $K$ & $D$ & $N_{\mathrm{layer}}$ & \#Heads & $T_{\mathrm{init}}$ &
  Remask rounds $\varrho$ & Thresholds $(\varphi_1,\varphi_2,\varphi_3)$ \\
  \midrule
  2 & 128 & 4 & 4 & 12 & -- & -- \\
  3 & 128 & 4 & 4 & 24 & -- & -- \\
  4 & 128 & 4 & 4 & 42 & -- & -- \\
  5 & 128 & 4 & 4 & 60 & -- & -- \\
  6 & 128 & 5 & 4 & 50 & 1 & $(0.97)$ \\
  7 & 128 & 5 & 4 & 62 & 2 & $(0.96,\;0.90)$ \\
  8 & 128 & 5 & 4 & 74 & 3 & $(0.96,\;0.90,\;0.85)$ \\
  \bottomrule
  \end{tabular}
  \end{table}

\subsection{Training hyperparameters}

\begin{table}[H]
\centering
\footnotesize
\setlength{\tabcolsep}{4pt}
\renewcommand{\arraystretch}{1.05}
\caption{Training hyperparameters.}
\label{tab:app_exact_training}
\begin{tabular}{l c}
\toprule
Parameter & Value \\
\midrule
Batch size & 128 \\
Epochs & 100 \\
Optimizer & AdamW (weight decay $10^{-4}$ for CNN/MLP/Transformer/GNN/NBP \\
& weight decay $10^{-2}$ for \cider/TA/MDD) \\
Learning rate & $10^{-3}$ \\
LR schedule & cosine, warmup 10 epochs, $\eta_{\min}=10^{-6}$ \\
EMA decay & 0.9999 \\
Precision & 16-mixed \\
Mask ratio $\gamma$ & $\gamma_t$ with $t\sim\mathrm{Unif}\{0,\ldots,T_{\mathrm{train}}\}$, $\gamma_t\in[0.1,1.0]$ after warmup \\
Mask warmup & 4 epochs (exclude $\gamma=1$) \\
$\Pr[\gamma=1.0]$ & $1/(T_{\mathrm{train}}+1)$ after warmup \\
AMP iterations \(I_{\mathrm{AMP}}\) & 10 \\
\bottomrule
\end{tabular}
\end{table}

\subsection{Compute resources}
\label{app:compute}

Training and inference were run on separate machines.

\paragraph{Training environment.}
All model training used a single NVIDIA A100-SXM4-80GB GPU operated in MIG mode with a 40GB compute slice. Each training run used a single GPU slice; no multi-GPU or distributed training was used.

\paragraph{Training cost.}
Backbone training at $K=2$ (100 epochs, batch size 128, single GPU) took $\sim$4.75\,h at the tiny scale, $\sim$4.9\,h at small, $\sim$5.0\,h at moderate and $\sim$6.4\,h at large.
Backbone training becomes more expensive at higher $K$ because the Hungarian-matching step in the loss is $O(K^3)$ per sample. 

\paragraph{Inference / evaluation environment.}
All evaluation results reported in this paper—including neural and classical baselines—were generated on a workstation with a single NVIDIA GeForce RTX 3090 GPU with 24GB memory and an Intel Core i5-14500 CPU.

\section{Architectures of compared multiuser decoders}
\label{app:arch_details}

\subsection{Shared input/output interfaces}
\label{app:arch_shared}

\paragraph{Evidence and outputs.}
All compared decoders consume the same symbol-wise soft detector evidence $S\in\mathbb{R}^{L\times Q}$ (batched as $\mathbb{R}^{B_{\mathrm{sz}}\times L\times Q}$) and output token logits $\Lambda\in\mathbb{R}^{K\times L\times Q}$ (or batched as $\mathbb{R}^{B_{\mathrm{sz}}\times K\times L\times Q}$).
The decoded grid is obtained by per-site $\arg\max$ on $\Lambda$.
When using a batch dimension, we denote the batch size by $B_{\mathrm{sz}}$ to avoid confusion with the payload length $B$ in the main text. Architecture diagrams are provided in \Cref{fig:arch_set_1,fig:arch_mdd}. All accuracy metrics are computed after Hungarian row matching to account for permutation invariance.

\paragraph{K-head outputs.}

For one-shot baselines, a shared length-$L$ hidden sequence is produced and then mapped to $K$ output rows using $K$ independent heads, each predicting a distribution over $[Q]$ per slot.

\subsection{MLP baseline: direct flattening of evidence}
\label{app:arch_mlp}

\Cref{fig:arch_set_1} illustrates this baseline.

\paragraph{Tokenization/embedding.}
The MLP baseline applies no learned token embedding.
It flattens the evidence as $x=\mathrm{vec}(S)\in\mathbb{R}^{B_{\mathrm{sz}}\times (LQ)}$ and feeds $x$ into an MLP.

\paragraph{Backbone and output.}

The backbone is an $n_{\mathrm{mlp}}$-layer MLP of hidden width \texttt{hidden\_dim}. Each hidden layer uses an affine transform followed by LayerNorm, GELU, and dropout (rate $p$), and the final layer projects to logits $\Lambda\in\mathbb{R}^{B_{\mathrm{sz}}\times K\times L\times Q}$.

\subsection{CNN baseline: Conv1d along the slot axis}
\label{app:arch_cnn}

\Cref{fig:arch_set_1} illustrates this baseline.

\paragraph{Input projection.}

The CNN baseline first projects each slot evidence vector $S_{\ell,:}\in\mathbb{R}^{Q}$ to a $C$-dimensional feature via
Linear($Q\!\to\!C$)$\rightarrow$LayerNorm$\rightarrow$GELU. Convolutions are applied only along the slot axis $L$ (no convolution across $Q$). 

\paragraph{Convolution axis (key point).}

The backbone is a stack of residual 1D convolution blocks operating over the slot dimension $L$ (padding preserves length).

\paragraph{Residual blocks.}
Each residual block uses \emph{two} Conv1d layers:
\[
\begin{aligned}
\text{res}&\leftarrow x,\\
x&\leftarrow \mathrm{Conv1d}\rightarrow\mathrm{BN}\rightarrow\mathrm{GELU}\rightarrow\mathrm{Dropout},\\
x&\leftarrow \mathrm{Conv1d}\rightarrow\mathrm{BN},\\
x&\leftarrow \mathrm{GELU}(x+\text{res}).
\end{aligned}
\]

\paragraph{Output heads.}

After the backbone, $K$ independent slot-wise heads map features to logits via
Linear($C\!\to\!C$)$\rightarrow$GELU$\rightarrow$Dropout$\rightarrow$Linear($C\!\to\!Q$),
yielding $\Lambda\in\mathbb{R}^{B_{\mathrm{sz}}\times K\times L\times Q}$.

\subsection{Transformer baseline: linear projection + sinusoidal positional encoding}
\label{app:arch_transformer}

\Cref{fig:arch_set_1} illustrates this baseline.

\paragraph{Evidence tokenization.}

The transformer treats each slot as a token by applying a learned linear projection $Q\!\to\! D_{\text{tr}}$, producing $x\in\mathbb{R}^{B_{\mathrm{sz}}\times L\times D_{\text{tr}}}$.

\paragraph{Positional encoding and backbone.}

Sinusoidal positional encodings are added (with dropout rate $p$), followed by a standard Transformer encoder over the $L$ slot tokens. Each block uses multi-head self-attention and a position-wise FFN with GELU, both wrapped with residual connections and LayerNorm.

\paragraph{Output.}

The model uses $K$ independent slot-wise output heads, each implemented as Linear($D_{\text{tr}}\!\to\! D_{\text{tr}}$)$\rightarrow$GELU$\rightarrow$Dropout$\rightarrow$Linear($D_{\text{tr}}\!\to\!Q$), and stacks the results to obtain $\Lambda\in\mathbb{R}^{B_{\mathrm{sz}}\times K\times L\times Q}$.

\subsection{GNN baseline}
\label{app:arch_gnn}

\Cref{fig:arch_set_1} illustrates this baseline.

\paragraph{Evidence tokenization.}
The GNN baseline treats each variable node as a token by applying a learned linear projection $Q \to D$, producing $x \in \mathbb{R}^{B_{\text{sz}} \times L \times D_{\text{gnn}}}$. 
Learnable row embeddings $s_k \in \mathbb{R}^D_{\text{gnn}}$ are added to create $K$ parallel copies.

\paragraph{Message passing on $H$.}
Each GNN layer performs bidirectional message passing on the Tanner graph defined by $H$. 
Variable nodes aggregate to check nodes via $H$, check nodes update their state, then send messages back via $H^\top$. 
Check node states are initialized from a learnable parameter.
All transformations use two-layer MLPs (LN$\to$Linear$\to$GELU$\to$Dropout$\to$Linear) and residual connections. 

\paragraph{Output heads.}
The model uses $K$ independent output heads, each implemented as Linear($D_{\text{gnn}} \to D_{\text{gnn}}$)$\to$GELU$\to$Dropout$\to$Linear($D_{\text{gnn}} \to Q$), producing $\Lambda \in \mathbb{R}^{B_{\text{sz}} \times K \times L \times Q}$.

\subsection{Neural Belief Propagation (NBP) baseline}
\label{app:arch_nbp}

\Cref{fig:arch_set_1} illustrates this baseline.

\paragraph{Evidence tokenization.}
A learned projection $Q \to D_{\text{nbp}}$ encodes channel evidence to $x \in \mathbb{R}^{B_{\text{sz}} \times L \times D_{\text{nbp}}}$, with row embeddings added for $K$ copies.

\paragraph{Message passing on $H$.}
Each layer unfolds one BP iteration on the Tanner graph. 
Messages flow through VN$\to$CN through $H$ and CN$\to$VN through $H^\top$, with 2-layer MLP at each step.
Two key differences from GNN: (1) \emph{channel skip}---the encoded evidence $x$ is re-injected at every layer rather than only at initialization; (2) \emph{learnable damping}---beliefs are updated as $b^{(\ell)} = \alpha b^{(\ell-1)} + (1-\alpha)\tilde{b}$ where $\alpha = \sigma(\theta)$ is learned.

\paragraph{Output heads.}
The model uses $K$ independent output heads, each implemented as Linear($D_{\text{nbp}} \to D_{\text{nbp}}$)$\to$GELU$\to$Dropout$\to$Linear($D_{\text{nbp}} \to Q$), producing $\Lambda \in \mathbb{R}^{B_{\text{sz}} \times K \times L \times Q}$.

\subsection{Tanner-Attention (TA) baseline}
\label{app:arch_mpa}

\Cref{fig:arch_set_1} illustrates this baseline.

\paragraph{Overview.}
TA is a one-shot, parity-aware baseline that injects code constraints via attention/message passing on the Tanner graph.
Given evidence $S\in\mathbb{R}^{L\times Q}$ (mini-batched as $\mathbb{R}^{B_{\mathrm{sz}}\times L\times Q}$) and a sparse parity-check matrix
$H\in\mathbb{F}_Q^{P\times L}$, TA constructs variable-node tokens (one per slot) and check-node tokens (one per parity equation),
then performs neighbor-restricted attention updates along Tanner edges. TA explicitly exploits Tanner sparsity by restricting interactions to graph neighborhoods. It is not belief propagation: it performs a single forward pass of Tanner-structured attention/message passing.

\paragraph{Tokenization and shared shapes.}
TA represents ($i$) variable nodes as a length-$L$ sequence of embeddings in $\mathbb{R}^{B_{\mathrm{sz}}\times L\times D}$
and ($ii$) check nodes as a length-$P$ sequence in $\mathbb{R}^{B_{\mathrm{sz}}\times P\times D}$.
The output is a logit tensor $\Lambda\in\mathbb{R}^{B_{\mathrm{sz}}\times K\times L\times Q}$, obtained by projecting the final variable-node embeddings
through $K$ independent output heads.

\paragraph{Evidence encoder.}
Evidence vectors $S_{\ell,:}\in\mathbb{R}^{Q}$ are converted into variable-node inputs by a small MLP encoder:
LayerNorm $\rightarrow$ Linear $\rightarrow$ GELU $\rightarrow$ Dropout $\rightarrow$ Linear, producing
\[
\mathbf{T}^{\mathrm{var},(0)} \in \mathbb{R}^{B_{\mathrm{sz}}\times L\times D}.
\]
A positional embedding over slots is added to break slot symmetry (followed by dropout), yielding the initial variable-node tokens.

\paragraph{Check-node initialization.}
Check-node tokens are initialized by a learnable embedding indexed by check equation $j\in[P]$ (and broadcast across the batch),
producing an initial sequence
\[
\mathbf{T}^{\mathrm{chk},(0)} \in \mathbb{R}^{B_{\mathrm{sz}}\times P\times D}.
\]
This makes TA ``structure-aware'' in the sense that the graph topology comes from $H$, while the check-node states are learned.

\paragraph{Tanner-structured message passing.}
Let $\mathcal{N}_{\mathrm{var}}(j)=\{\ell\in[L]\;:\;H_{j,\ell}\neq 0\}$ denote the variable neighbors of check $j$,
and $\mathcal{N}_{\mathrm{chk}}(\ell)=\{j\in[P]\;:\;H_{j,\ell}\neq 0\}$ denote check neighbors of variable $\ell$.
TA alternates two neighbor-restricted updates:

\emph{Variable $\rightarrow$ check (VN$\to$CN).}
Each check token aggregates messages from its neighboring variables using multi-head attention where
queries are $\mathbf{T}^{\mathrm{chk}}$ and keys/values are $\mathbf{T}^{\mathrm{var}}$ restricted to $\mathcal{N}_{\mathrm{var}}(j)$.
The update uses residual connections and LayerNorm, followed by a position-wise FFN (GELU, dropout).

\emph{Check $\rightarrow$ variable (CN$\to$VN).}
Each variable token aggregates messages from its neighboring checks using attention where
queries are variable tokens and keys/values are the check tokens restricted to $\mathcal{N}_{\mathrm{chk}}(\ell)$,
again followed by residual + LayerNorm and an FFN (GELU, dropout).

\paragraph{Non-binary coefficient handling.}
For non-binary LDPC codes, each edge $(j,\ell)$ carries a coefficient $H_{j,\ell}\in\mathbb{F}_Q^\times$.
TA accounts for coefficients by applying a coefficient-conditioned transform on edge messages before aggregation,
so that check-node updates depend on coefficient-normalized neighbor information and variable-node updates receive denormalized messages.
(Implementation-wise, this is realized by fixed, coefficient-indexed mixing/permutation operators consistent with the $\mathbb{F}_Q$ convention of $H$.)

\paragraph{Output heads.}
After $L_{\mathrm{mp}}$ message-passing layers, the final variable tokens $v\in\mathbb{R}^{B_{\mathrm{sz}}\times L\times D}$ are mapped to logits via
$K$ independent heads (each a small MLP head with GELU and dropout, ending in a Linear to $Q$ logits),
stacked to form $\Lambda\in\mathbb{R}^{B_{\mathrm{sz}}\times K\times L\times Q}$.
Decoding uses per-site argmax.

\paragraph{Algorithm (forward pass).}
Given a mini-batch evidence tensor $S\in\mathbb{R}^{B_{\mathrm{sz}}\times L\times Q}$ and parity matrix $H\in\mathbb{F}_Q^{P\times L}$,
TA computes logits $\Lambda\in\mathbb{R}^{B_{\mathrm{sz}}\times K\times L\times Q}$ as:
\begin{enumerate}[leftmargin=1.4em,itemsep=1pt,topsep=2pt]
  \item \textbf{Encode evidence.} Apply an evidence MLP (LayerNorm, GELU, dropout) to obtain initial variable tokens $\mathbf{T}^{\mathrm{var},(0)}$,
        and add slot positional embeddings.
  \item \textbf{Initialize checks.} Create check tokens $\mathbf{T}^{\mathrm{chk},(0)}$ from a learnable check-index embedding.
  \item \textbf{Repeat for $L_{\mathrm{mp}}$ layers:}
    \begin{enumerate}[leftmargin=1.2em,itemsep=1pt,topsep=1pt]
      \item \emph{VN$\to$CN update:} update each check token by neighbor-restricted attention over $\{\mathbf{T}^{\mathrm{var}}_\ell:\ell\in\mathcal{N}_{\mathrm{var}}(j)\}$,
            applying coefficient conditioning on edges $(j,\ell)$; apply residual + LayerNorm and an FFN.
      \item \emph{CN$\to$VN update:} update each variable token by neighbor-restricted attention over $\{\mathbf{T}^{\mathrm{chk}}_j:j\in\mathcal{N}_{\mathrm{chk}}(\ell)\}$,
            again with coefficient conditioning; apply residual + LayerNorm and an FFN.
    \end{enumerate}
  \item \textbf{Project to logits.} Apply $K$ independent output heads on variable tokens and stack outputs to obtain $\Lambda$.
\end{enumerate}

\paragraph{Implementation notes.}
($i$) All attention/FFN blocks follow the standard Transformer pattern (multi-head attention, residual connections, LayerNorm, dropout; FFN uses GELU).
($ii$) This baseline is \emph{one-shot}: unlike MDD/\cider, it does not perform iterative diffusion refinement; all parity conditioning is injected in a single forward pass via Tanner-structured updates.
\subsection{Masked Diffusion Decoder (MDD) baseline: concatenation fusion of discrete tokens and evidence}
\label{app:arch_mdd_dit}

\Cref{fig:arch_mdd} compares the MDD denoiser with \cider under the same masked-refinement loop.

\paragraph{Overview.}
MDD instantiates masked discrete diffusion with a generic Transformer denoiser.
At each refinement step, the denoiser consumes a partially masked grid $\tilde X^{(t)}\in([Q]\cup\{\mask\})^{K\times L}$
together with the evidence heatmap $S\in\mathbb{R}^{L\times Q}$ and the mask ratio $\gamma_t$,
and outputs logits $\Lambda^{(t)}\in\mathbb{R}^{K\times L\times Q}$ for masked-site prediction.
(For interface consistency, the implementation also accepts the parity-check matrix $H\in\mathbb{F}_Q^{P\times L}$ via an embedding interface; however, MDD does not perform explicit parity propagation and treats any parity metadata only as generic context.)

\paragraph{Discrete token embeddings.}
For a mini-batch, the masked grid is represented as
$\tilde X^{(t)}\in([Q]\cup\{\mask\})^{B_{\mathrm{sz}}\times K\times L}$, where $\mask$ is an absorbing mask token.
Each site $(k,\ell)$ is embedded as the sum of ($i$) a learnable symbol embedding for $\tilde X^{(t)}_{k,\ell}$
and ($ii$) a learnable positional embedding for the slot index $\ell$, yielding token embeddings
$\mathbf{T}^{\mathrm{tok}}\in\mathbb{R}^{B_{\mathrm{sz}}\times K\times L\times D}$.
Dropout is applied after embedding and after each residual sublayer.

\paragraph{Evidence (magnitude) encoder.}
The evidence tensor is $S\in\mathbb{R}^{B_{\mathrm{sz}}\times L\times Q}$.
We standardize each slot over the alphabet dimension and project it to $D$ dimensions via a two-layer MLP with GELU:
\[
\tilde S_{b,\ell,:}=\frac{S_{b,\ell,:}-\mu(S_{b,\ell,:})}{\sigma(S_{b,\ell,:})+\varepsilon},
\qquad
\mathbf{T}^{\mathrm{ev}}_{b,\ell}\in\mathbb{R}^{D}.
\]
The resulting $\mathbf{T}^{\mathrm{ev}}\in\mathbb{R}^{B_{\mathrm{sz}}\times L\times D}$ is broadcast across $k$ and combined with $\mathbf{T}^{\mathrm{tok}}$.

\paragraph{Syndrome context and gating.}
We compute a syndrome tensor
$\mathrm{syn}\in[Q]^{B_{\mathrm{sz}}\times K\times P}$ from the current grid and parity matrix (same interface as parity-aware models).
Each syndrome entry is embedded, pooled across parity checks, and projected to a context vector
$\mathbf{T}^{\mathrm{syn}}\in\mathbb{R}^{B_{\mathrm{sz}}\times K\times D}$.
To emphasize structural cues late in denoising, the syndrome context is gated by $(1-\gamma_t)$ and broadcast across the slot dimension. This is a generic conditioning mechanism and should not be interpreted as enforcing parity constraints; MDD does not perform Tanner-graph message passing.

\paragraph{Time conditioning via adaLN.}
The mask ratio $\gamma_t\in[0,1]$ is embedded with a sinusoidal embedding followed by an MLP to produce $\mathbf{t}(\gamma_t)\in\mathbb{R}^{B_{\mathrm{sz}}\times D}$. We condition on time through adaptive LayerNorm (adaLN) in each Transformer block. The remaining conditioning signals are concatenated and fused:
\[ \mathbf{T}^{\mathrm{fuse}} = \phi\Big(\big[\mathbf{T}^{\mathrm{tok}},\ \mathbf{T}^{\mathrm{ev}},\ \mathbf{T}^{\mathrm{syn}}\big]\Big) \in\mathbb{R}^{B_{\mathrm{sz}}\times K\times L\times D},
\]
where $\phi$ is a fusion MLP with LayerNorm, GELU, and dropout. Each Transformer block modulates its LayerNorm statistics using learned affine parameters predicted from $\mathbf{t}(\gamma_t)$.

\paragraph{Backbone and output.}
The fused grid is flattened into a length-$K\!\cdot\!L$ token sequence and processed by $N_{\mathrm{blk}}$ Transformer encoder blocks with adaLN conditioning. Each block uses multi-head self-attention and a position-wise FFN with GELU, with dropout and residual connections. LayerNorm statistics are modulated by the time embedding $\mathbf{t}(\gamma_t)$. A final adaLN-modulated LayerNorm and linear projection produce logits $\Lambda^{(t)}\in\mathbb{R}^{B_{\mathrm{sz}}\times K\times L\times Q}$.

\paragraph{Algorithm (per-step denoiser forward).}
Given $(\tilde X^{(t)}, S, \gamma_t)$, MDD computes logits $\Lambda^{(t)}$ as:
\begin{enumerate}[leftmargin=1.4em,itemsep=1pt,topsep=2pt]
  \item Embed $\tilde X^{(t)}$ with symbol + slot-position embeddings to form $\mathbf{T}^{\mathrm{tok}}$.
  \item Encode evidence $S$ into $\mathbf{T}^{\mathrm{ev}}$ (slot-wise standardization $\rightarrow$ MLP with GELU $\rightarrow$ LayerNorm),
        then broadcast across $k$.
  \item Encode syndrome into $\mathbf{T}^{\mathrm{syn}}$, gate by $(1-\gamma_t)$, and broadcast across $\ell$.
  \item Embed time $\gamma_t$ into $\mathbf{t}(\gamma_t)$ via sinusoidal embedding and MLP.
  \item Concatenate $[\mathbf{T}^{\mathrm{tok}}, \mathbf{T}^{\mathrm{ev}}, \mathbf{T}^{\mathrm{syn}}]$ and fuse to obtain $\mathbf{T}^{\mathrm{fuse}}$.
  \item Flatten to length-$K\!\cdot\!L$ and apply $N_{\mathrm{blk}}$ adaLN Transformer blocks (MHA + GELU-FFN, with dropout, residual, adaLN), each conditioned on $\mathbf{t}(\gamma_t)$.
  \item Apply an adaLN-modulated final layer to produce logits $\Lambda^{(t)}$ over $[Q]$ for each site.
\end{enumerate}

\paragraph{Implementation notes.}
($i$) MDD is deliberately generic: it does not exploit Tanner sparsity or perform explicit parity propagation in the denoiser.
($ii$) All MLPs and FFNs use GELU; dropout is applied after embeddings and within each residual sublayer; LayerNorm follows standard Transformer practice.
($iii$) MDD is used inside an iterative refinement loop (the diffusion process), but the denoiser itself is a single forward pass per step.

\subsection{\texorpdfstring{\cider: constraint-aware iterative diffusion decoding for error-correcting refinement}{CIDER: constraint-aware iterative diffusion decoding for error-correcting refinement}}
\label{app:arch_cider}

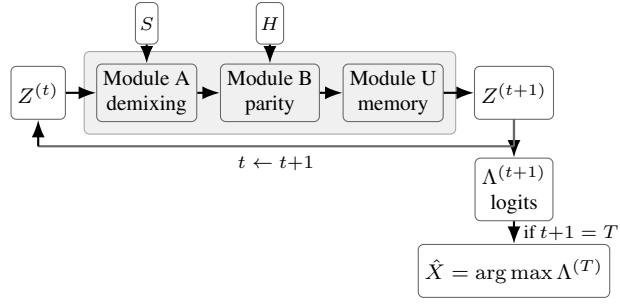
\begin{figure}[t]
\centering
\captionsetup{font=footnotesize}
  \begin{tikzpicture}[archfont]
\tikzset{
  crfivebox/.style={draw=black, line width=0.95pt, fill=white,
    align=center, inner sep=3pt, minimum height=10mm,
    font=\footnotesize\sffamily\bfseries},
  crfivemodule/.style={crfivebox, fill=black!11, text width=22mm},
  crfiveoptional/.style={crfivebox, fill=black!5, dashed, text width=23mm},
  crfiveinput/.style={crfivebox, minimum height=7mm, inner sep=2pt},
  crfivearrow/.style={-{Latex[length=2.4mm,width=1.8mm]}, draw=black, line width=1pt},
  crfivenote/.style={font=\footnotesize\sffamily\bfseries, fill=white, inner sep=1pt}
}

\node[crfivebox, minimum width=16mm] (Zt) {$Z^{(t)}$};
\node[crfivemodule, right=5mm of Zt] (A) {Module A\\demixing};
\node[crfivemodule, right=5mm of A] (B) {Module B\\parity};
\node[crfiveoptional, right=5mm of B] (U) {Module U\\memory (optional)};
\node[crfivebox, minimum width=16mm, right=5mm of U] (Znext) {$Z^{(t+1)}$};

\node[crfiveinput, above=5mm of A] (S) {$S$};
\node[crfiveinput, above=5mm of B] (H) {$H$};
\draw[crfivearrow] (Zt) -- (A);
\draw[crfivearrow] (A) -- (B);
\draw[crfivearrow] (B) -- (U);
\draw[crfivearrow] (U) -- (Znext);
\draw[crfivearrow] (S) -- (A);
\draw[crfivearrow] (H) -- (B);

\coordinate (loopR) at ($(Znext.north)+(0,16mm)$);
\coordinate (loopL) at ($(Zt.north |- loopR)$);
\draw[draw=black, line width=1pt] (Znext.north) -- (loopR) -- (loopL);
\draw[crfivearrow] (loopL) -- (Zt.north);
\node[crfivenote, above=0mm] at ($(loopL)!0.5!(loopR)$) {repeat: $t\leftarrow t+1$};

\node[crfivebox, below=8mm of Znext] (proj) {$\Lambda^{(t+1)}$\\logits};
\node[crfivebox, below=5mm of proj] (Xhat) {$\hat X=\arg\max\Lambda^{(T)}$};
\draw[crfivearrow] (Znext) -- (proj);
\draw[crfivearrow] (proj) -- node[right, crfivenote] {if $t+1=T$} (Xhat);
  \end{tikzpicture}%

\caption{\cider refinement loop with the optional memory/stabilization module (U).}
\label{fig:cider_overview_U}
\end{figure}

\paragraph{Overview.}
We use \cider to denote the default variant used in the main paper, which consists of
Module A for demixing and Module B for parity-aware propagation. We also consider an \emph{optional} memory/stabilization Module U; when enabled, we denote the variant as \textbf{\cider+U}. See \Cref{fig:arch_cider_main} for the default refinement loop (main paper) and \Cref{fig:cider_overview_U} for the variant with Module U (appendix).

At each refinement step $t$, \cider maintains a discrete masked grid $X^{(t)}\in([Q]\cup\{\mask\})^{K\times L}$ and
constructs a latent grid $Z^{(t)}=\mathrm{Embed}(X^{(t)})\in\mathbb{R}^{K\times L\times D}$.
The denoiser processes $Z^{(t)}$ and produces token logits $\Lambda^{(t)}\in\mathbb{R}^{K\times L\times Q}$,
which are used to reveal additional entries of $X^{(t)}$.
The denoiser combines ($i$) \emph{slot competition} (demixing) driven by evidence $S\in\mathbb{R}^{L\times Q}$ and
($ii$) \emph{parity-aware propagation} on the Tanner graph using the parity-check matrix $H\in\mathbb{F}_Q^{P\times L}$.
\cider uses standard neural components (MLPs with GELU, dropout, LayerNorm, residual connections), but applies code constraints via fixed field-consistent permutations rather than learning finite-field arithmetic.

\paragraph{Latent and symbol parameterization.}
We write $\hat Z^{(t)}_{k,\ell}\in\mathbb{R}^{D}$ for the refined latent vector at site $(k,\ell)$.
Symbol logits are obtained by a shared projection
\[
\Lambda^{(t)}_{k,\ell,a} = w_a^\top \hat Z^{(t)}_{k,\ell}, \qquad a\in[Q],
\]
where $\{w_a\}_{a\in[Q]}\subset\mathbb{R}^{D}$ are learnable symbol vectors (equivalently, a matrix $W\in\mathbb{R}^{Q\times D}$).
A learnable embedding table $E\in\mathbb{R}^{(Q+1)\times D}$ maps discrete symbols and the mask token to $D$ dimensions when embedding a masked grid for initialization/conditioning.

\paragraph{Module A: slot-competition demixing.}
Because $S$ is unsourced, multiple hypothesis rows can collapse onto the same high-evidence symbols.
\cider counteracts this by enforcing competition across rows.
Using compatibility scores derived from the current representation 
$\langle Z^{(t)}_{k,\ell}, w_a\rangle  + \beta_S S_{\ell,a}$,
we compute per-slot responsibilities
$r^{(t)}_{k,\ell,a}$ by a softmax over $k$ (temperature $\tau$),
and form row-specific evidence summaries
\[
e^{(t)}_{k,\ell}=\sum_{a\in[Q]} r^{(t)}_{k,\ell,a}\, S_{\ell,a}\, v_a,
\]
where $\{v_a\}_{a\in[Q]}\subset\mathbb{R}^{D}$ are learnable symbol embeddings.
The latent $Z^{(t)}_{k,\ell}$ and summary $e^{(t)}_{k,\ell}$ are combined using a per-site gated update: a gating network (MLP with GELU and dropout) produces a gate $\beta^{(t)}_{\mathrm{demix},k,\ell}\in[0,1]^D$ and updates
\[
\tilde Z^{(t)}_{k,\ell}
= \beta^{(t)}_{\mathrm{demix},k,\ell}\odot Z^{(t)}_{k,\ell}
+ \bigl(1-\beta^{(t)}_{\mathrm{demix},k,\ell}\bigr)\odot e^{(t)}_{k,\ell}.
\]

\paragraph{Module B: parity-aware propagation.}
\cider propagates parity information along Tanner edges using the parity-check matrix $H\in\mathbb{F}_Q^{P\times L}$.
For non-binary LDPC codes, each nonzero coefficient $\alpha\in\mathbb{F}_Q^\times$ induces a permutation $\Pi_{\alpha}$ of the $Q$ symbols.
\cider precomputes these permutations once (using the same $\mathbb{F}_Q$ convention as $H$) and applies them as fixed operators.
Conceptually, coefficient actions are implemented as a linear--permutation--linear map in symbol space:
\[
T_{\alpha}(x) \triangleq W^\top \Pi_{\alpha} W x, \qquad x\in\mathbb{R}^{D}.
\]
For each check node, \cider aggregates neighbor information with an \emph{extrinsic} pattern (excluding the edge being updated),
implemented by attention over the incident edges plus a residual MLP (GELU, dropout, LayerNorm).
The resulting check-to-variable messages are denormalized by the inverse coefficient action and scatter-added back to variable sites, followed by a fusion MLP to update the site latents.
We denote the output of this module by $\hat Z^{(t)}\in\mathbb{R}^{K\times L\times D}$, which is the latent representation used for the logit projection.

\paragraph{\texorpdfstring{Memory and stabilization (Module U).}{Memory and stabilization (U).}}
(\textbf{Optional; \cider+U only.}) To stabilize refinement across diffusion steps, 
\cider applies a GRU cell for recurrent update:
a GRU cell produces an update gate $\beta^{(t)}_{\mathrm{gru},k,\ell}\in[0,1]^D$ and a proposal $\bar Z^{(t)}_{k,\ell}$, yielding
\[
  Z^{(t+1)}_{k,\ell} = \bigl(1-\beta^{(t)}_{\mathrm{gru},k,\ell}\bigr)\odot Z^{(t)}_{k,\ell}+ \beta^{(t)}_{\mathrm{gru},k,\ell}\odot \bar Z^{(t)}_{k,\ell}.
\]
In the main text, we report the default \cider variant without Module U and defer this ablation to Appendix~\Cref{app:ablation_U}.
 
\paragraph{Algorithm (one denoiser step).}
For a mini-batch of size $B_{\mathrm{sz}}$, let $Z^{(t)}\in\mathbb{R}^{B_{\mathrm{sz}}\times K\times L\times D}$,
$S\in\mathbb{R}^{B_{\mathrm{sz}}\times L\times Q}$, and $H\in\mathbb{F}_Q^{P\times L}$.
One denoiser step stacks $N_{\mathrm{layer}}$ structured layers, each with Module~A and two Tanner-graph message-passing passes, before producing logits $\Lambda^{(t)}$ as:
\begin{enumerate}[leftmargin=1.4em,itemsep=1pt,topsep=2pt]
  \item \textbf{Intermediate logit projection.} Compute intermediate logits \(\bar\Lambda^{(t)}\) from \(Z^{(t)}\) using the shared symbol projection \(W\). These logits are used solely for demixing.
  \item \textbf{Demixing.} Compute responsibilities \(r^{(t)}_{k,\ell,a}\) from \(\bar\Lambda^{(t)}\) using a softmax over \(k\), form evidence summaries \(e^{(t)}_{k,\ell}\) from \(S\), and blend them into \(Z^{(t)}\) with a gated interpolation to obtain \(\tilde Z^{(t)}\).
  \item \textbf{Parity propagation.} Within each structured layer, perform two Tanner-graph message-passing passes using \(H\): normalize by fixed permutations from \(H\), aggregate extrinsically with attention and MLPs, inverse-permute, scatter-add, degree-normalize, and fuse. The final layer yields \(\hat Z^{(t)}\).
  \item \textbf{Final logit projection.} Compute final logits \(\Lambda^{(t)}_{k,\ell,a}=w_a^\top \hat Z^{(t)}_{k,\ell}\). These final logits are used for masked-token reveal.
  \item \textbf{Optional stabilization (\cider+U only).} If Module U is enabled, apply the gated memory update to the persistent continuous state. The default \cider variant does not carry a continuous hidden state across steps and re-embeds \(Z^{(t+1)}\) from the updated discrete grid \(X^{(t+1)}\).
\end{enumerate}
In the default masked-diffusion loop, the decoder uses $\Lambda^{(t)}$ to update the discrete grid $X^{(t)}$ and re-embeds the next latent state from that grid. The optional \cider+U variant additionally carries its recurrent continuous state across steps.

\paragraph{Implementation notes.}
($i$) All MLP submodules use GELU and dropout, with LayerNorm and residual connections in standard practice.
($ii$) Coefficient handling in $\mathbb{F}_Q$ is enforced by fixed permutations derived from $H$, ensuring field-consistent parity messaging without learning arithmetic.
($iii$) \cider differs from generic denoisers primarily through (a) row-wise competition for demixing and (b) Tanner-structured parity propagation.

\begin{figure}[t]
\centering
\captionsetup{font=footnotesize}
\resizebox{\linewidth}{!}{%
  \begin{tikzpicture}[archfont]
\tikzset{
  crarchpanel/.style={draw=black, line width=0.95pt, fill=white},
  crarchheader/.style={draw=black, line width=0.95pt, fill=black!15},
  crarchstage/.style={draw=black, line width=0.85pt, fill=white,
    align=center, inner sep=1.3pt, minimum height=12mm,
    font=\scriptsize\sffamily\bfseries},
  crarchcore/.style={crarchstage, line width=1pt, fill=black!9},
  crarcharrow/.style={-{Latex[length=1.8mm,width=1.3mm]}, draw=black, line width=0.9pt}
}
\newcommand{\CRArchPanel}[7]{%
  \begin{scope}[shift={(#2,#3)}]
    \draw[crarchpanel] (0,0) rectangle (72mm,30mm);
    \draw[crarchheader] (0,24mm) rectangle (72mm,30mm);
    \node[font=\footnotesize\sffamily\bfseries] at (36mm,27mm) {#4};
    \node[crarchstage, text width=15mm] (#1a) at (12mm,12mm) {#5};
    \node[crarchcore, text width=19mm] (#1b) at (36mm,12mm) {#6};
    \node[crarchstage, text width=15mm] (#1c) at (60mm,12mm) {#7};
    \draw[crarcharrow] (#1a.east) -- (#1b.west);
    \draw[crarcharrow] (#1b.east) -- (#1c.west);
  \end{scope}%
}

\CRArchPanel{mlp}{0mm}{68mm}{MLP: dense map}
  {flatten\\$S$}{feed-forward\\layers}{reshape\\$\Lambda$}
\CRArchPanel{cnn}{77mm}{68mm}{CNN: local slot context}
  {slot\\projection}{residual\\Conv1D on $L$}{$K$ heads\\$\Lambda$}
\CRArchPanel{trf}{0mm}{34mm}{Transformer: global attention}
  {slot + position\\tokens}{Transformer\\encoder}{$K$ heads\\$\Lambda$}
\CRArchPanel{gnn}{77mm}{34mm}{GNN: Tanner graph}
  {variable + check\\embeddings}{Tanner MP on $H$\\residual update}{$K$ heads\\$\Lambda$}
\CRArchPanel{nbp}{0mm}{0mm}{NBP: Tanner with damping}
  {variable + check\\embeddings}{Tanner MP on $H$\\gated/damped}{$K$ heads\\$\Lambda$}
\CRArchPanel{ta}{77mm}{0mm}{TA: coefficient-aware attention}
  {variable + check\\tokens}{Tanner attn.\\$H$ + GF$(Q)$\\edges}{$K$ heads\\$\Lambda$}
  \end{tikzpicture}%
}
\caption{Compared multiuser decoder architectures. Each panel maps the shared evidence $S\in\mathbb{R}^{L\times Q}$ to $K\times L\times Q$ logits $\Lambda$. GNN, NBP, and TA additionally use the parity-check matrix $H$ for Tanner connectivity; TA also uses its GF$(Q)$ edge coefficients. Implementation details are given in this section.}
\label{fig:arch_set_1}
\end{figure}

\begin{figure}[t]
\centering
\captionsetup{font=footnotesize}
\resizebox{\linewidth}{!}{%
\begin{tikzpicture}[
  font=\footnotesize\sffamily\bfseries,
  crbox/.style={draw=black, line width=0.9pt, fill=white, align=center, inner sep=3pt, minimum height=10mm},
  crlabel/.style={crbox, fill=black!13, line width=1.1pt},
  crkey/.style={crbox, fill=black!7, line width=1.1pt},
  crarrow/.style={-{Latex[length=2.3mm,width=1.7mm]}, line width=0.9pt, draw=black}
]
\node[crbox, text width=20mm] (grid) {Masked grid\\$X^{(t)}$};
\node[crbox, text width=23mm, right=5mm of grid] (denoiser) {Denoiser $f_\theta$\\time $\gamma_t$};
\node[crbox, text width=20mm, right=5mm of denoiser] (logits) {Logits\\$\Lambda^{(t)}$};
\node[crbox, text width=23mm, right=5mm of logits] (reveal) {Reveal\\schedule $U_t$};
\node[crbox, text width=20mm, right=5mm of reveal] (next) {Updated grid\\$X^{(t+1)}$};
\draw[crarrow] (grid) -- (denoiser);
\draw[crarrow] (denoiser) -- (logits);
\draw[crarrow] (logits) -- (reveal);
\draw[crarrow] (reveal) -- (next);
\node[font=\scriptsize\sffamily, above=2mm of logits] {Shared masked refinement (repeat for $T$ steps)};

\node[crlabel, text width=15mm, below=12mm of grid] (mdd) {MDD};
\node[crbox, text width=23mm, right=5mm of mdd] (membed) {Encode tokens\\and evidence $S$};
\node[crbox, text width=23mm, right=5mm of membed] (mfuse) {Concatenate\\and fuse};
\node[crbox, text width=27mm, right=5mm of mfuse] (mcore) {Transformer\\blocks};
\node[crbox, text width=19mm, right=5mm of mcore] (mout) {Linear\\$D\!\to\!Q$};
\draw[crarrow] (mdd) -- (membed);
\draw[crarrow] (membed) -- (mfuse);
\draw[crarrow] (mfuse) -- (mcore);
\draw[crarrow] (mcore) -- (mout);

\node[crlabel, text width=15mm, below=8mm of mdd] (ciderlab) {CIDER};
\node[crbox, text width=23mm, right=5mm of ciderlab] (cembed) {Encode tokens\\and evidence $S$};
\node[crkey, text width=23mm, right=5mm of cembed] (modulea) {Module A\\row demixing};
\node[crkey, text width=27mm, right=5mm of modulea] (moduleb) {Module B\\Tanner propagation ($H$)};
\node[crbox, text width=19mm, right=5mm of moduleb] (cout) {Linear\\$D\!\to\!Q$};
\draw[crarrow] (ciderlab) -- (cembed);
\draw[crarrow] (cembed) -- (modulea);
\draw[crarrow] (modulea) -- (moduleb);
\draw[crarrow] (moduleb) -- (cout);
\end{tikzpicture}%
}
\caption{Shared masked-refinement loop and denoiser comparison. Both methods convert a partially masked grid and slot evidence $S$ to logits, then reveal tokens over $T$ steps. MDD fuses token and evidence features in a Transformer denoiser (with optional syndrome context); default \cider adds row-wise demixing (Module A) and parity-aware Tanner propagation using $H$ (Module B). The optional memory Module U is omitted.}
\label{fig:arch_mdd}
\label{fig:arch_cider_appendix}
\end{figure}

\subsection{Metric details}
\label{app:metric_details}

Because the decoded set is unordered, \(\hat X\in[Q]^{K\times L}\) and \(X^\star\in[Q]^{K\times L}\) are defined only up to row permutation. 
We choose the permutation
\[
\pi^\star
\triangleq
\arg\min_{\pi\in\mathfrak{S}_K}
\sum_{k=0}^{K-1}\sum_{\ell=0}^{L-1}
\mathbf{1}\!\left[\hat X_{k,\ell}\neq X^\star_{\pi(k),\ell}\right],
\]
where \(\mathfrak{S}_K\) is the set of permutations of \([K]\). 
All SER and CER values in the paper are computed after this matching.

\section{Additional results}
\label{app:full_results}

\subsection[Numerical values underlying the neural-baseline figure]{Numerical values underlying \Cref{fig:main_neural_mechanism}}
\label{app:neural_baseline_table}

The neural baselines in \Cref{fig:main_neural_mechanism} largely overlap because most one-shot neural decoders fail in this regime, producing near-identical SER and CER values. \Cref{tab:main_results_ser_cer} reports the exact numerical values across all four LDPC scales for direct comparison. \cider is the only learned decoder that achieves substantially below-floor SER/CER.

\begin{table}[h]
\centering
\caption{SER/CER ($K=2$) for \cider and neural baselines.}
\label{tab:main_results_ser_cer}

\footnotesize
\resizebox{0.46\columnwidth}{!}{%
\begin{tabular}{l l c|c c}
\toprule
{\textbf{Scenario $(Q,L)$}} & \textbf{Model} & {\textbf{Params}} & \textbf{SER ($\downarrow$)} & \textbf{CER ($\downarrow$)} \\
\midrule

\rowcolor{CIDERGreenFill!55}
Tiny: $(64,12)$  & \textbf{\cider (MDM)}  & 3.25M & \textbf{0.0011} & \textbf{0.0073} \\
                 & CNN                   & 3.69M & 0.4024          & 0.9996 \\
                 & MLP                   & 3.67M & 0.6330          & 1.0000 \\
                 & Transformer           & 3.63M & 0.9731          & 1.0000 \\
                 & GNN           & 3.54M & 0.4003          & 0.9996 \\
                 & NBP           & 3.81M & 0.3985          & 0.9996  \\ 
                 & TA                   & 3.61M & 0.4028          & 0.9996 \\
                 & MDD                   & 3.57M & 0.4201          & 0.9996 \\
\midrule

\rowcolor{CIDERGreenFill!55}
Small: $(64,18)$ & \textbf{\cider (MDM)}  & 4.80M & \textbf{0.0002} & \textbf{0.0013} \\
                 & CNN                   & 5.42M & 0.4186          & 1.0000 \\
                 & MLP                   & 5.42M & 0.8942          & 1.0000 \\
                 & Transformer           & 5.35M & 0.9741          & 1.0000 \\
                 & GNN           & 4.87M & 0.4180          & 1.0000 \\
                 & NBP           & 4.75M & 0.4170          & 1.0000  \\ 
                 & TA                   & 5.20M & 0.4210          & 1.0000 \\
                 & MDD                   & 5.36M & 0.4375          & 1.0000 \\
\midrule

\rowcolor{CIDERGreenFill!55}
Moderate: $(64,24)$ & \textbf{\cider (MDM)}  & 4.80M & \textbf{0.0008} & \textbf{0.0053} \\
                    & CNN                   & 5.42M & 0.4282          & 1.0000 \\
                    & MLP                   & 5.42M & 0.9752          & 1.0000 \\
                    & Transformer           & 5.35M & 0.9754          & 1.0000 \\
                    & GNN           & 4.87M & 0.4302          & 1.0000 \\
                    & NBP           & 4.75M & 0.4281          & 1.0000  \\ 
                    & TA                   & 5.20M & 0.4320          & 1.0000 \\
                    & MDD                   & 5.36M & 0.4400          & 1.0000 \\
\midrule

\rowcolor{CIDERGreenFill!55}
Large: $(64,48)$ & \textbf{\cider (MDM)}  & 6.35M & \textbf{0.0045} & \textbf{0.0270} \\
                 & CNN                   & 7.15M & 0.4470          & 1.0000 \\
                 & MLP                   & 7.13M & 0.9774          & 1.0000 \\
                 & Transformer           & 7.11M & 0.9775          & 1.0000 \\
                 & GNN           & 6.27M & 0.4488          & 1.0000 \\
                 & NBP           & 6.19M & 0.4462          & 1.0000  \\ 
                 & TA                   & 7.59M & 0.4513          & 1.0000 \\
                 & MDD                   & 7.24M & 0.4601          & 1.0000 \\
\bottomrule
\end{tabular}
}
\end{table}

\subsection{\texorpdfstring{Effect of the optional stabilization Module U}{Effect of the optional stabilization Module U}}
\label{app:ablation_U}
We also evaluate an optional memory/stabilization Module U implemented as a GRU-style update. \Cref{tab:app_u_ablation} summarizes the effect of adding Module U across LDPC scales.
\begin{table}[h]
\centering
\footnotesize
\setlength{\tabcolsep}{4pt}
\renewcommand{\arraystretch}{1.05}
\caption{Effect of adding the optional stabilization module (U) on LDPC codes. Lower is better.}
\label{tab:app_u_ablation}
\begin{tabular}{l l c c c}
\toprule
Scenario $(Q,L)$ & Model & Params & SER~(\textbf{$\downarrow$}) & CER~(\textbf{$\downarrow$}) \\
\midrule
Tiny $(64,12)$ & \cider (MDM; A+B) & 3.25M & 0.0011 & 0.0073 \\
              & \cider (MDM+U; A+B+U) & 3.65M & 0.0014 & 0.0060 \\
\midrule
Small $(64,18)$ & \cider (MDM; A+B) & 4.80M & 0.0002 & 0.0013 \\
               & \cider (MDM+U; A+B+U) & 5.40M & 0.0031 & 0.0325 \\
\midrule
Moderate $(64,24)$ & \cider (MDM; A+B) & 4.80M & 0.0008 & 0.0053 \\
                  & \cider (MDM+U; A+B+U) & 5.40M & 0.0028 & 0.0352 \\
\midrule
Large $(64,48)$ & \cider (MDM; A+B) & 6.35M & 0.0045 & 0.0270 \\
               & \cider (MDM+U; A+B+U) & 7.14M & 0.0102 & 0.0409 \\
\bottomrule
\end{tabular}
\end{table}

\subsection{Effect of using D3PM for \cider}
\label{app:full_results_d3pm}

This subsection reports additional results for \cider under \emph{uniform categorical corruption} (D3PM-style noising) \citep{austin2021d3pm}. Unless noted otherwise, the D3PM results use the same optional stabilization Module U (i.e., \cider (D3PM+U)), and all other experimental settings match \Cref{sec:experiments}.

\paragraph{D3PM-style uniform-corruption baseline (D3PD).}
We also report D3PD, a generic diffusion baseline that mirrors MDD but replaces the absorbing-mask corruption with \emph{uniform categorical} corruption, following the D3PM-style noising process \citep{austin2021d3pm}. Concretely, D3PD uses the same denoiser architecture and refinement procedure as MDD, but the forward corruption mixes each token with a uniform distribution over $[Q]$ instead of masking.

\paragraph{Scaling across LDPC problem sizes (Tiny/Small/Moderate/Large).}
\Cref{tab:app_d3pm_d3pd_scales} reports the same four LDPC scales used in the main paper and compares \cider (D3PM+U) against D3PD.

\begin{table}[h]
\centering
\footnotesize
\setlength{\tabcolsep}{4pt}
\renewcommand{\arraystretch}{1.05}
\caption{Scaling results for D3PM-style decoding on LDPC codes ($K=2$). Lower is better.}
\label{tab:app_d3pm_d3pd_scales}
\begin{tabular}{l l c c c}
\toprule
Scenario $(Q,L)$ & Model & Params & SER~(\textbf{$\downarrow$}) & CER~(\textbf{$\downarrow$}) \\
\midrule
Tiny $(64,12)$
& \cider (D3PM+U; A+B+U) & 3.65M & 0.0128 & 0.0339 \\
& D3PD (uniform corruption) & 3.57M & 0.4721 & 0.9996 \\
\midrule
Small $(64,18)$
& \cider (D3PM+U; A+B+U) & 5.39M & 0.0057 & 0.0214 \\
& D3PD (uniform corruption) & 5.36M & 0.4761 & 1.0000 \\
\midrule
Moderate $(64,24)$
& \cider (D3PM+U; A+B+U) & 5.39M & 0.0527 & 0.1436 \\
& D3PD (uniform corruption) & 5.36M & 0.4740 & 1.0000 \\
\midrule
Large $(64,48)$
& \cider (D3PM+U; A+B+U) & 7.14M & 0.2026 & 0.6192 \\
& D3PD (uniform corruption) & 7.24M & 0.4853 & 1.0000 \\
\bottomrule
\end{tabular}
\end{table}

\paragraph{Ablations (D3PM).}
\Cref{tab:app_full_results_ablation} reports module ablations under D3PM-style (uniform categorical) corruption at $(Q,L)=(64,12)$ with $K=2$.
Unlike the MDM setting, where the optional stabilization Module U is not always beneficial, under D3PM-style corruption the stabilization module is important for reliable refinement: removing Module U (the Module A+B variant) substantially degrades SER/CER.
\Cref{tab:app_d3pm_by_eb} additionally reports \cider (D3PM+U) performance at two different noise levels (two $\mathrm{SNR}$ values), using the same architecture and training protocol. \Cref{tab:app_d3pm_by_k} reports an additional load check at $K=3$ for the same setting.

\begin{table}[h]
\centering
\footnotesize
\setlength{\tabcolsep}{4pt}
\renewcommand{\arraystretch}{1.05}
\caption{Ablations of \cider (D3PM) on LDPC codes at $(Q,L)=(64,12)$ ($K=2$). Lower is better.}
\label{tab:app_full_results_ablation}
\begin{tabular}{l c c c}
\toprule
Model & Params & SER~(\textbf{$\downarrow$}) & CER~(\textbf{$\downarrow$}) \\
\midrule
\cider (D3PM+U; A+B+U) & 3.65M & \textbf{0.0128} & \textbf{0.0339} \\
No demixing (B+U) & 3.74M & 0.3976 & 0.9995 \\
No parity-aware propagation (A+U) & 1.14M & 0.3867 & 0.9995 \\
No memory (A+B) & 3.25M & 0.4294 & 0.9997 \\
No diffusion (one-shot SC) & 3.65M & 0.4043 & 0.9996 \\
\bottomrule
\end{tabular}
\end{table}

\begin{table}[h]
\centering
\footnotesize
\setlength{\tabcolsep}{4pt}
\renewcommand{\arraystretch}{1.05}
\caption{SER/CER of \cider (D3PM+U) on LDPC codes ($K=2$). Lower is better.}
\label{tab:app_d3pm_by_eb}
\begin{tabular}{l l c c c}
\toprule
Scenario $(E_b)$ & Model & Params & SER~(\textbf{$\downarrow$}) & CER~(\textbf{$\downarrow$}) \\
\midrule
$\mathrm{SNR} = -2.79$~dB  & \cider (D3PM+U) & 3.65M & 0.0243 & 0.0619 \\
$\mathrm{SNR} = -0.79$~dB & \cider (D3PM+U) & 3.65M & 0.0128 & 0.0339 \\
\bottomrule
\end{tabular}
\end{table}

\begin{table}[h]
\centering
\footnotesize
\setlength{\tabcolsep}{4pt}
\renewcommand{\arraystretch}{1.05}
\caption{SER/CER of \cider (D3PM+U) on LDPC codes at $(Q,L)=(64,12)$. Lower is better.}
\label{tab:app_d3pm_by_k}
\begin{tabular}{l l c c c}
\toprule
Scenario $(K)$ & Model & Params & SER~(\textbf{$\downarrow$}) & CER~(\textbf{$\downarrow$}) \\
\midrule
$K=2$ & \cider (D3PM+U) & 3.65M & 0.0128 & 0.0339 \\
$K=3$ & \cider (D3PM+U) & 3.65M & 0.0084 & 0.0204 \\
\bottomrule
\end{tabular}
\end{table}

\begin{table}[h]
\centering
\footnotesize
\setlength{\tabcolsep}{5pt}
\renewcommand{\arraystretch}{1.15}
\caption{Joint sweep over demixing and parity loss weights.
Each entry reports (SER, CER) for the loss-varying MDD baseline.}
\label{tab:app_lambda_sweep}
\begin{tabular}{c c c c c}
\toprule
$\lambda_{\mathrm{d}} \,\backslash\, \lambda_{\mathrm{p}}$
& 0.03 & 0.1 & 0.3 & 1.0 \\
\midrule
0.03 & (0.7011, 1.0000) & (0.7383, 1.0000) & (0.8910, 1.0000) & (0.9505, 1.0000) \\
0.1  & (0.6970, 1.0000) & (0.7344, 1.0000) & (0.7929, 1.0000) & (0.8762, 1.0000) \\
0.3  & (0.6742, 1.0000) & (0.7568, 1.0000) & (0.8656, 1.0000) & (0.9612, 1.0000) \\
1.0  & (0.7499, 1.0000) & (0.8916, 1.0000) & (0.8587, 1.0000) & (0.9445, 1.0000) \\
\bottomrule
\end{tabular}
\end{table}

\begin{table}[h]
\centering
\footnotesize
\setlength{\tabcolsep}{4pt}
\renewcommand{\arraystretch}{1.05}
\caption[SER/CER without first-reveal rule]{SER/CER on LDPC codes ($(Q,L)=(64,12)$), without first-reveal rule. Lower is better.}
\label{tab:larger_K_no_first}
\begin{tabular}{l l c c c}
\toprule
Scenario $(K)$ & Model & Params & SER~(\textbf{$\downarrow$}) & CER~(\textbf{$\downarrow$}) \\
\midrule
$K=3$ & \cider & 3.25M & 0.0209 & 0.2162 \\
$K=4$ & \cider & 3.25M & 0.0845 & 0.2319 \\
$K=5$ & \cider & 3.25M & 0.1697 & 0.3273 \\
$K=6$ & \cider & 4.03M & 0.2778 & 0.4798 \\
\bottomrule
\end{tabular}
\end{table}

\begin{table}[h]
\centering
\caption{Wall-clock times on LDPC codes at $(Q,L)=(64,12)$ across per-bin loads $K\in\{3,4,5\}$. Time is ms/sample. Lower is better. FFT-BP runs are capped at 50 BP iterations with early exit on convergence. Top-$J$ exhaustive search exceeded the 60-second cap for $K\geq 6$.}
\label{tab:app_full_results_time_k3}
\setlength{\tabcolsep}{4pt}
\renewcommand{\arraystretch}{1.0}
{\scriptsize
\begin{tabular}{@{}l|ccc|ccc|ccc@{}}
\toprule
\multirow{2}{*}{\textbf{Method}}
& \multicolumn{3}{c|}{$K=3$}
& \multicolumn{3}{c|}{$K=4$}
& \multicolumn{3}{c}{$K=5$}\\
\cmidrule(lr){2-4}
\cmidrule(lr){5-7}
\cmidrule(lr){8-10}
& SER & CER & Time
& SER & CER & Time
& SER & CER & Time\\
\midrule
\rowcolor{CIDERGreenFill!55}
\textbf{\cider}
& \textbf{0.0006} & \textbf{0.0044} & \textbf{5.47}
& \textbf{0.0015} & \textbf{0.0058} & \textbf{10.19}
& \textbf{0.0048} & \textbf{0.0141} & \textbf{15.29}\\
SIC-BP
& 0.0036 & 0.0063 & 26.46
& 0.0380 & 0.0573 & 90.93
& 0.0393 & 0.0738 & 128.76 \\
FFT-BP
& 0.0036 & 0.0063 & 45.62
& 0.0380 & 0.0573 & 163.19
& 0.0393 & 0.0738 & 230.67 \\
Top-$J$ (Top 3)
& 0.1157 & 0.1173 & 6.36
& --      & --      & --
& --      & --      & --\\
Top-$J$ (Top 4)
& 0.0416 & 0.0420 & 32.98
& 0.2190 & 0.2232 & 36.80
& --  & -- & --\\
Top-$J$ (Top 5)
& 0.0231 & 0.0233 & 423.02
& 0.0983 & 0.1000 & 427.92
& 0.3671 & 0.3736  & 481.71\\
\bottomrule
\end{tabular}%
}
\end{table}
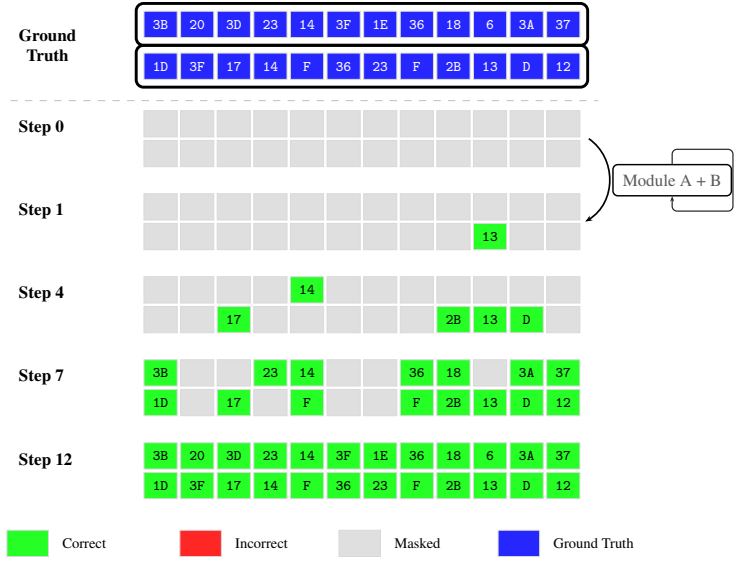
\begin{figure}[!htbp]
\centering
\resizebox{0.9\linewidth}{!}{\usetikzlibrary{backgrounds}

\begin{tikzpicture}[
    cell/.style={
        minimum width=0.48cm, 
        minimum height=0.38cm, 
        font=\scriptsize\sffamily\bfseries,
        draw=black, line width=0.65pt,
        inner sep=1pt,
    },
    revealed/.style={cell, fill=green!85, text=black},
    incorrect/.style={cell, fill=red!85, text=black},
    hidden/.style={cell, fill=gray!25, text=gray!50},
    steplabel/.style={font=\footnotesize\sffamily\bfseries, text=black, anchor=west},
    taulabel/.style={font=\tiny, text=gray, anchor=west},
    acclabel/.style={font=\tiny, text=blue!70, anchor=west},
    gtcell/.style={cell, fill=blue!85, text=white},
]

\node[steplabel, align=center] at (-1.0, 0) {Ground\\Truth};

\matrix (gt) [matrix of nodes, anchor=west,
    row sep=6pt,            
    column sep=1pt,
    nodes={cell}
] at (0.8, 0) {
    |[gtcell]| 3B & |[gtcell]| 20 & |[gtcell]| 3D & |[gtcell]| 23 & |[gtcell]| 14 & |[gtcell]| 3F & |[gtcell]| 1E & |[gtcell]| 36 & |[gtcell]| 18 & |[gtcell]| 6 & |[gtcell]| 3A & |[gtcell]| 37 \\
    |[gtcell]| 1D & |[gtcell]| 3F & |[gtcell]| 17 & |[gtcell]| 14 & |[gtcell]| F & |[gtcell]| 36 & |[gtcell]| 23 & |[gtcell]| F & |[gtcell]| 2B & |[gtcell]| 13 & |[gtcell]| D & |[gtcell]| 12 \\
};

\begin{pgfonlayer}{background}
  \node[
    fit=(gt-1-1)(gt-1-12),
    draw=black,
    line width=1.2pt,
    inner sep=3pt
  ] {};

  \node[
    fit=(gt-2-1)(gt-2-12),
    draw=black,
    line width=1.2pt,
    inner sep=3pt
  ] {};
\end{pgfonlayer}



\draw[black!65, dashed, line width=0.8pt] (-1.0, -0.8) -- (7.5, -0.8);

\node[steplabel] at (-1.0, -1.2) {Step 0};
\matrix (s0) [matrix of nodes, anchor=west,
    row sep=1pt, column sep=1pt,
    nodes={cell}
] at (0.8, -1.35) {
    |[hidden]| \phantom{00} & |[hidden]| \phantom{00} & |[hidden]| \phantom{00} & |[hidden]| \phantom{00} & |[hidden]| \phantom{00} & |[hidden]| \phantom{00} & |[hidden]|\phantom{00}  & |[hidden]| \phantom{00} & |[hidden]| \phantom{00} & |[hidden]|\phantom{00}  & |[hidden]| \phantom{00} & |[hidden]|\phantom{00}  \\
    |[hidden]| \phantom{00} & |[hidden]| \phantom{00} & |[hidden]|\phantom{00}  & |[hidden]| \phantom{00} & |[hidden]| \phantom{00} & |[hidden]|\phantom{00}  & |[hidden]| \phantom{00} & |[hidden]| \phantom{00} & |[hidden]| \phantom{00} & |[hidden]| \phantom{00} & |[hidden]| \phantom{00} & |[hidden]| \phantom{00} \\
};

\node[steplabel] at (-1.0, -2.4) {Step 1};
\matrix (s1) [matrix of nodes, anchor=west,
    row sep=1pt, column sep=1pt,
    nodes={cell}
] at (0.8, -2.55) {
    |[hidden]| \phantom{00} & |[hidden]| \phantom{00} & |[hidden]| \phantom{00} & |[hidden]| \phantom{00} & |[hidden]| \phantom{00} & |[hidden]| \phantom{00} & |[hidden]|\phantom{00}  & |[hidden]| \phantom{00} & |[hidden]| \phantom{00} & |[hidden]|\phantom{00}  & |[hidden]| \phantom{00} & |[hidden]|\phantom{00}  \\
    |[hidden]| \phantom{00} & |[hidden]| \phantom{00} & |[hidden]|\phantom{00}  & |[hidden]| \phantom{00} & |[hidden]| \phantom{00} & |[hidden]|\phantom{00}  & |[hidden]| \phantom{00} & |[hidden]| \phantom{00} & |[hidden]| \phantom{00} & |[revealed]| 13 & |[hidden]| \phantom{00} & |[hidden]| \phantom{00} \\
};

\node[steplabel] at (-1.0, -3.6) {Step 4};
\matrix (s4) [matrix of nodes, anchor=west,
    row sep=1pt, column sep=1pt,
    nodes={cell}
] at (0.8, -3.75) {
    |[hidden]| \phantom{00} & |[hidden]| \phantom{00} & |[hidden]| \phantom{00} & |[hidden]| \phantom{00} & |[revealed]| 14 & |[hidden]| \phantom{00} & |[hidden]| \phantom{00} & |[hidden]| \phantom{00} & |[hidden]| \phantom{00} & |[hidden]| \phantom{00} & |[hidden]| \phantom{00} & |[hidden]| \phantom{00} \\
    |[hidden]| \phantom{00} & |[hidden]| \phantom{00} & |[revealed]| 17 & |[hidden]| \phantom{00} & |[hidden]| \phantom{00} & |[hidden]| \phantom{00} & |[hidden]| \phantom{00} & |[hidden]| \phantom{00} & |[revealed]| 2B & |[revealed]| 13 & |[revealed]| D & |[hidden]| \phantom{00} \\
};

\node[steplabel] at (-1.0, -4.8) {Step 7};
\matrix (s7) [matrix of nodes, anchor=west,
    row sep=1pt, column sep=1pt,
    nodes={cell}
] at (0.8, -4.95) {
    |[revealed]| 3B & |[hidden]| \phantom{00} & |[hidden]| \phantom{00} & |[revealed]| 23 & |[revealed]| 14 & |[hidden]| \phantom{00} & |[hidden]| \phantom{00} & |[revealed]| 36 & |[revealed]| 18 & |[hidden]| \phantom{00} & |[revealed]| 3A & |[revealed]| 37 \\
    |[revealed]| 1D & |[hidden]| \phantom{00} & |[revealed]| 17 & |[hidden]| \phantom{00} & |[revealed]| F & |[hidden]| \phantom{00} & |[hidden]| \phantom{00} & |[revealed]| F & |[revealed]| 2B & |[revealed]| 13 & |[revealed]| D & |[revealed]| 12 \\
};

\node[steplabel] at (-1.0, -6.0) {Step 12};
\matrix (s12) [matrix of nodes, anchor=west,
    row sep=1pt, column sep=1pt,
    nodes={cell}
] at (0.8, -6.15) {
    |[revealed]| 3B & |[revealed]| 20 & |[revealed]| 3D & |[revealed]| 23 & |[revealed]| 14 & |[revealed]| 3F & |[revealed]| 1E & |[revealed]| 36 & |[revealed]| 18 & |[revealed]| 6 & |[revealed]| 3A & |[revealed]| 37 \\
    |[revealed]| 1D & |[revealed]| 3F & |[revealed]| 17 & |[revealed]| 14 & |[revealed]| F & |[revealed]| 36 & |[revealed]| 23 & |[revealed]| F & |[revealed]| 2B & |[revealed]| 13 & |[revealed]| D & |[revealed]| 12 \\
};


\begin{scope}[shift={(-0.75, -7.2)}]
    \node[revealed, draw=black, line width=0.65pt, minimum width=0.6cm] (leg1) at (0, 0) {};
    \node[font=\scriptsize\sffamily\bfseries, right=2pt of leg1] {Correct};

    \node[incorrect, draw=black, line width=0.65pt, minimum width=0.6cm] (legX) at (2.5, 0) {};
    \node[font=\scriptsize\sffamily\bfseries, right=2pt of legX] {Incorrect};

    \node[hidden, draw=black, line width=0.65pt, minimum width=0.6cm] (leg2) at (4.8, 0) {};
    \node[font=\scriptsize\sffamily\bfseries, right=2pt of leg2] {Masked};

    \node[gtcell, draw=black, line width=0.65pt, minimum width=0.6cm] (leg3) at (7.1, 0) {};
    \node[font=\scriptsize\sffamily\bfseries, right=2pt of leg3] {Ground Truth};
\end{scope}

\coordinate (s0-out) at ($ (s0-1-12.east)!0.5!(s0-2-12.east) $);
\coordinate (s1-in) at ($ (s1-1-12.east)!0.5!(s1-2-12.east) $);

\begin{scope}[shift={(7.2, -1.95)}]
  \node[
    draw=black!100,
    line width=0.95pt,
    inner sep=4pt,
    font=\footnotesize\sffamily\bfseries,
    text=black,
    anchor=west,
    align=center
  ] (iterbox) at (0.5,0)
  {Module A + B};
\end{scope}

\coordinate (module-in) at ($(iterbox.north west)!0.28!(iterbox.south west)$);
\coordinate (module-out) at ($(iterbox.north west)!0.72!(iterbox.south west)$);
\draw[black, -{Latex[length=2.3mm,width=1.7mm]}, line width=0.95pt]
  (s0-out) -- (module-in);
\draw[black, -{Latex[length=2.3mm,width=1.7mm]}, line width=0.95pt]
  (module-out) -- (s1-in);

\end{tikzpicture}}
\caption{\cider iterative decoding (full figure)}
\label{fig:tiny_cider_visualization}

\end{figure}

\begin{figure}[!htbp]
\centering
\resizebox{0.9\linewidth}{!}{\usetikzlibrary{backgrounds}

\begin{tikzpicture}[
    cell/.style={
        minimum width=0.48cm, 
        minimum height=0.38cm, 
        font=\scriptsize\sffamily\bfseries,
        draw=black, line width=0.65pt,
        inner sep=1pt,
    },
    revealed/.style={cell, fill=green!85, text=black},
    incorrect/.style={cell, fill=red!85, text=black},
    hidden/.style={cell, fill=gray!25, text=gray!50},
    steplabel/.style={font=\footnotesize\sffamily\bfseries, text=black, anchor=west},
    taulabel/.style={font=\tiny, text=gray, anchor=west},
    acclabel/.style={font=\tiny, text=blue!70, anchor=west},
    gtcell/.style={cell, fill=blue!85, text=white},
]

\node[steplabel, align=center] at (-1.0, 0) {Target\\Codewords};

\matrix (gt) [matrix of nodes, anchor=west,
    row sep=6pt,            
    column sep=1pt,
    nodes={cell}
] at (0.8, 0) {
    |[gtcell]| 3B & |[gtcell]| 20 & |[gtcell]| 3D & |[gtcell]| 23 & |[gtcell]| 14 & |[gtcell]| 3F & |[gtcell]| 1E & |[gtcell]| 36 & |[gtcell]| 18 & |[gtcell]| 6 & |[gtcell]| 3A & |[gtcell]| 37 \\
    |[gtcell]| 1D & |[gtcell]| 3F & |[gtcell]| 17 & |[gtcell]| 14 & |[gtcell]| F & |[gtcell]| 36 & |[gtcell]| 23 & |[gtcell]| F & |[gtcell]| 2B & |[gtcell]| 13 & |[gtcell]| D & |[gtcell]| 12 \\
};

\begin{pgfonlayer}{background}
  \node[
    fit=(gt-1-1)(gt-1-12),
    draw=black,
    line width=1.2pt,
    inner sep=3pt
  ] {};

  \node[
    fit=(gt-2-1)(gt-2-12),
    draw=black,
    line width=1.2pt,
    inner sep=3pt
  ] {};
\end{pgfonlayer}



\draw[black!65, dashed, line width=0.8pt] (-1.0, -0.8) -- (7.5, -0.8);

\node[steplabel] at (-1.0, -1.2) {Step 0};
\matrix (s0) [matrix of nodes, anchor=west,
    row sep=1pt, column sep=1pt,
    nodes={cell}
] at (0.8, -1.35) {
    |[hidden]| \phantom{00} & |[hidden]| \phantom{00} & |[hidden]| \phantom{00} & |[hidden]| \phantom{00} & |[hidden]| \phantom{00} & |[hidden]| \phantom{00} & |[hidden]|\phantom{00}  & |[hidden]| \phantom{00} & |[hidden]| \phantom{00} & |[hidden]|\phantom{00}  & |[hidden]| \phantom{00} & |[hidden]|\phantom{00}  \\
    |[hidden]| \phantom{00} & |[hidden]| \phantom{00} & |[hidden]|\phantom{00}  & |[hidden]| \phantom{00} & |[hidden]| \phantom{00} & |[hidden]|\phantom{00}  & |[hidden]| \phantom{00} & |[hidden]| \phantom{00} & |[hidden]| \phantom{00} & |[hidden]| \phantom{00} & |[hidden]| \phantom{00} & |[hidden]| \phantom{00} \\
};

\node[steplabel] at (-1.0, -2.4) {Step 1};
\matrix (s1) [matrix of nodes, anchor=west,
    row sep=1pt, column sep=1pt,
    nodes={cell}
] at (0.8, -2.55) {
    |[hidden]| \phantom{00} & |[hidden]| \phantom{00} & 
    |[hidden]| \phantom{00} & |[hidden]| \phantom{00} & 
    |[hidden]| \phantom{00} & |[revealed]| 3F & 
    |[hidden]|\phantom{00}  & |[hidden]| \phantom{00} & 
    |[hidden]| \phantom{00} & |[hidden]|\phantom{00}  & 
    |[hidden]| \phantom{00} & |[hidden]|\phantom{00}  \\
    |[hidden]| \phantom{00} & |[hidden]| \phantom{00} & 
    |[hidden]|\phantom{00}  & |[hidden]| \phantom{00} & 
    |[hidden]| \phantom{00} & |[hidden]|\phantom{00}  & 
    |[hidden]| \phantom{00} & |[hidden]| \phantom{00} & 
    |[hidden]| \phantom{00} & |[hidden]| \phantom{00} & 
    |[hidden]| \phantom{00} & |[hidden]| \phantom{00} \\
};

\node[steplabel] at (-1.0, -3.6) {Step 4};
\matrix (s4) [matrix of nodes, anchor=west,
    row sep=1pt, column sep=1pt,
    nodes={cell}
] at (0.8, -3.75) {
    |[revealed]| 3B & |[hidden]| \phantom{00} & 
    |[hidden]| \phantom{00} & |[hidden]| \phantom{00} & 
    |[hidden]| \phantom{00} & |[revealed]| 3F &
    |[hidden]| \phantom{00} & |[hidden]| \phantom{00} & 
    |[hidden]| \phantom{00} & |[hidden]| \phantom{00} & 
    |[hidden]| \phantom{00} & |[incorrect]| 12 \\
    |[revealed]| 1D & |[hidden]| \phantom{00} & 
    |[hidden]| \phantom{00} & |[hidden]| \phantom{00} & 
    |[hidden]| \phantom{00} & |[hidden]| \phantom{00} & 
    |[hidden]| \phantom{00} & |[hidden]| \phantom{00} & 
    |[hidden]| \phantom{00} & |[hidden]| \phantom{00} & 
    |[hidden]| \phantom{00} & |[incorrect]| 37 \\
};

\node[steplabel] at (-1.0, -4.8) {Step 7};
\matrix (s7) [matrix of nodes, anchor=west,
    row sep=1pt, column sep=1pt,
    nodes={cell}
] at (0.8, -4.95) {
    |[revealed]| 3B & |[incorrect]| 3F & 
    |[revealed]| 3D & |[incorrect]| 14  & 
    |[hidden]| \phantom{00} & |[revealed]| 3F &
    |[hidden]| \phantom{00} & |[hidden]| \phantom{00} & 
    |[revealed]| 18 & |[revealed]| 6 & 
    |[hidden]| \phantom{00} & |[incorrect]| 12 \\
    |[revealed]| 1D & |[hidden]| \phantom{00} & 
    |[revealed]| 17 & |[incorrect]| 23 & 
    |[hidden]| \phantom{00} & |[revealed]| 36 & 
    |[hidden]| \phantom{00} & |[hidden]| \phantom{00} & 
    |[incorrect]| 18 & |[incorrect]| 6 & 
    |[hidden]| \phantom{00} & |[incorrect]| 37 \\
};

\node[steplabel] at (-1.0, -6.0) {Step 12};
\matrix (s12) [matrix of nodes, anchor=west,
    row sep=1pt, column sep=1pt,
    nodes={cell}
] at (0.8, -6.15) {
    |[revealed]| 3B & |[incorrect]| 3F & 
    |[revealed]| 3D & |[incorrect]| 14  & 
    |[revealed]| 14 & |[revealed]| 3F &
    |[incorrect]| 23 & |[incorrect]| F & 
    |[revealed]| 18 & |[revealed]| 6 & 
    |[incorrect]| D & |[incorrect]| 12 \\
    |[revealed]| 1D & |[revealed]| 3F & 
    |[revealed]| 17 & |[incorrect]| 23 & 
    |[incorrect]| 14 & |[revealed]| 36 & 
    |[incorrect]| 1E & |[incorrect]| 36 & 
    |[incorrect]| 18 & |[incorrect]| 6 & 
    |[revealed]| D & |[incorrect]| 37 \\
};


\begin{scope}[shift={(-0.75, -7.2)}]
    \node[revealed, draw=black, line width=0.65pt, minimum width=0.6cm] (leg1) at (0, 0) {};
    \node[font=\scriptsize\sffamily\bfseries, right=2pt of leg1] {Correct};

    \node[incorrect, draw=black, line width=0.65pt, minimum width=0.6cm] (legX) at (2.5, 0) {};
    \node[font=\scriptsize\sffamily\bfseries, right=2pt of legX] {Incorrect};

    \node[hidden, draw=black, line width=0.65pt, minimum width=0.6cm] (leg2) at (4.8, 0) {};
    \node[font=\scriptsize\sffamily\bfseries, right=2pt of leg2] {Masked};

    \node[gtcell, draw=black, line width=0.65pt, minimum width=0.6cm] (leg3) at (7.1, 0) {};
    \node[font=\scriptsize\sffamily\bfseries, right=2pt of leg3] {Ground Truth};
\end{scope}

\coordinate (s0-out) at ($ (s0-1-12.east)!0.5!(s0-2-12.east) $);
\coordinate (s1-in) at ($ (s1-1-12.east)!0.5!(s1-2-12.east) $);

\begin{scope}[shift={(7.35, -1.95)}]
  \node[
      draw=black!100,
    line width=0.95pt,
    inner sep=4pt,
    font=\footnotesize\sffamily\bfseries,
    text=black,
    anchor=west,
    align=left
  ] (iterbox) at (0.5,0)
  {DiT denoiser\\$f_\theta(\tilde X,S,\gamma_t)$};

\end{scope}

\coordinate (module-in) at ($(iterbox.north west)!0.28!(iterbox.south west)$);
\coordinate (module-out) at ($(iterbox.north west)!0.72!(iterbox.south west)$);
\draw[black, -{Latex[length=2.3mm,width=1.7mm]}, line width=0.95pt]
  (s0-out) -- (module-in);
\draw[black, -{Latex[length=2.3mm,width=1.7mm]}, line width=0.95pt]
  (module-out) -- (s1-in);

\end{tikzpicture}}
\caption{MDD iterative decoding}
\label{fig:tiny_mdd_visualization}

\end{figure}
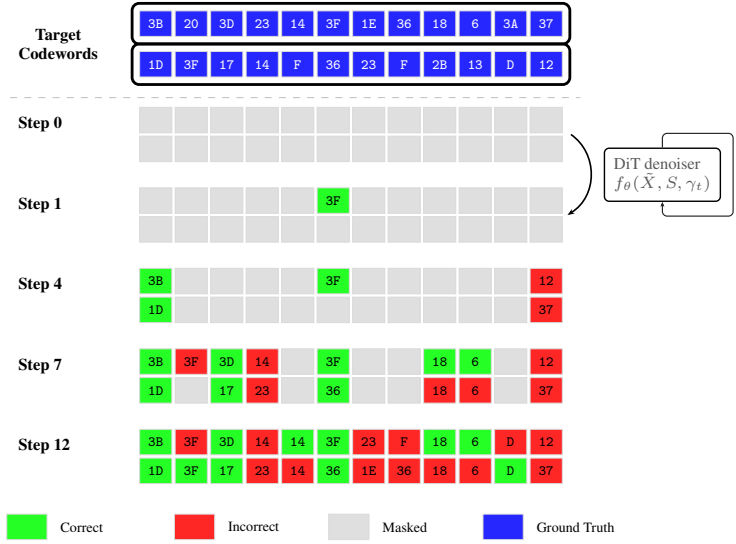

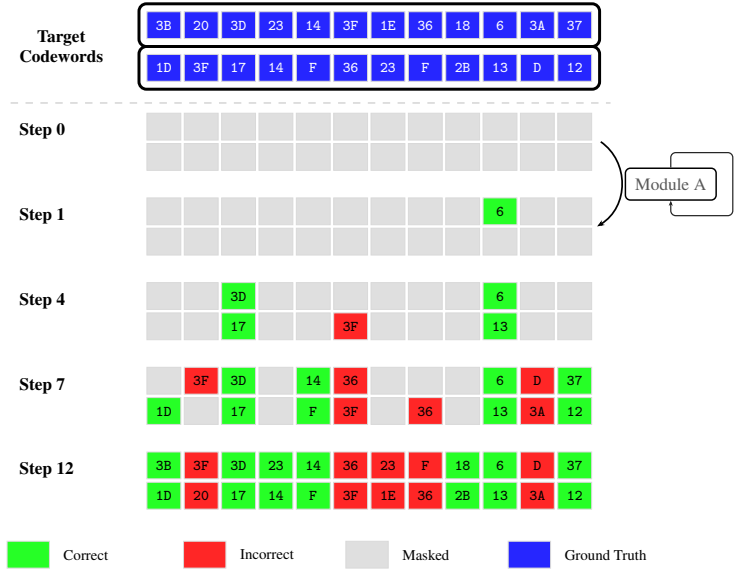
\begin{figure}[!htbp]
\centering
\resizebox{0.9\linewidth}{!}{\usetikzlibrary{backgrounds}

\begin{tikzpicture}[
    cell/.style={
        minimum width=0.48cm, 
        minimum height=0.38cm, 
        font=\scriptsize\sffamily\bfseries,
        draw=black, line width=0.65pt,
        inner sep=1pt,
    },
    revealed/.style={cell, fill=green!85, text=black},
    incorrect/.style={cell, fill=red!85, text=black},
    hidden/.style={cell, fill=gray!25, text=gray!50},
    steplabel/.style={font=\footnotesize\sffamily\bfseries, text=black, anchor=west},
    taulabel/.style={font=\tiny, text=gray, anchor=west},
    acclabel/.style={font=\tiny, text=blue!70, anchor=west},
    gtcell/.style={cell, fill=blue!85, text=white},
]

\node[steplabel, align=center] at (-1.0, 0) {Target\\Codewords};

\matrix (gt) [matrix of nodes, anchor=west,
    row sep=6pt,            
    column sep=1pt,
    nodes={cell}
] at (0.8, 0) {
    |[gtcell]| 3B & |[gtcell]| 20 & |[gtcell]| 3D & |[gtcell]| 23 & |[gtcell]| 14 & |[gtcell]| 3F & |[gtcell]| 1E & |[gtcell]| 36 & |[gtcell]| 18 & |[gtcell]| 6 & |[gtcell]| 3A & |[gtcell]| 37 \\
    |[gtcell]| 1D & |[gtcell]| 3F & |[gtcell]| 17 & |[gtcell]| 14 & |[gtcell]| F & |[gtcell]| 36 & |[gtcell]| 23 & |[gtcell]| F & |[gtcell]| 2B & |[gtcell]| 13 & |[gtcell]| D & |[gtcell]| 12 \\
};

\begin{pgfonlayer}{background}
  \node[
    fit=(gt-1-1)(gt-1-12),
    draw=black,
    line width=1.2pt,
    inner sep=3pt
  ] {};

  \node[
    fit=(gt-2-1)(gt-2-12),
    draw=black,
    line width=1.2pt,
    inner sep=3pt
  ] {};
\end{pgfonlayer}



\draw[black!65, dashed, line width=0.8pt] (-1.0, -0.8) -- (7.5, -0.8);

\node[steplabel] at (-1.0, -1.2) {Step 0};
\matrix (s0) [matrix of nodes, anchor=west,
    row sep=1pt, column sep=1pt,
    nodes={cell}
] at (0.8, -1.35) {
    |[hidden]| \phantom{00} & |[hidden]| \phantom{00} & |[hidden]| \phantom{00} & |[hidden]| \phantom{00} & |[hidden]| \phantom{00} & |[hidden]| \phantom{00} & |[hidden]|\phantom{00}  & |[hidden]| \phantom{00} & |[hidden]| \phantom{00} & |[hidden]|\phantom{00}  & |[hidden]| \phantom{00} & |[hidden]|\phantom{00}  \\
    |[hidden]| \phantom{00} & |[hidden]| \phantom{00} & |[hidden]|\phantom{00}  & |[hidden]| \phantom{00} & |[hidden]| \phantom{00} & |[hidden]|\phantom{00}  & |[hidden]| \phantom{00} & |[hidden]| \phantom{00} & |[hidden]| \phantom{00} & |[hidden]| \phantom{00} & |[hidden]| \phantom{00} & |[hidden]| \phantom{00} \\
};

\node[steplabel] at (-1.0, -2.4) {Step 1};
\matrix (s1) [matrix of nodes, anchor=west,
    row sep=1pt, column sep=1pt,
    nodes={cell}
] at (0.8, -2.55) {
    |[hidden]| \phantom{00} & |[hidden]| \phantom{00} & 
    |[hidden]| \phantom{00} & |[hidden]| \phantom{00} & 
    |[hidden]| \phantom{00} & |[hidden]| \phantom{00} & 
    |[hidden]|\phantom{00}  & |[hidden]| \phantom{00} & 
    |[hidden]| \phantom{00} & |[revealed]| 6  & 
    |[hidden]| \phantom{00} & |[hidden]|\phantom{00}  \\
    |[hidden]| \phantom{00} & |[hidden]| \phantom{00} & 
    |[hidden]|\phantom{00}  & |[hidden]| \phantom{00} & 
    |[hidden]| \phantom{00} & |[hidden]|\phantom{00}  & 
    |[hidden]| \phantom{00} & |[hidden]| \phantom{00} & 
    |[hidden]| \phantom{00} & |[hidden]| \phantom{00} & 
    |[hidden]| \phantom{00} & |[hidden]| \phantom{00} \\
};

\node[steplabel] at (-1.0, -3.6) {Step 4};
\matrix (s4) [matrix of nodes, anchor=west,
    row sep=1pt, column sep=1pt,
    nodes={cell}
] at (0.8, -3.75) {
    |[hidden]| \phantom{00} & |[hidden]| \phantom{00} & 
    |[revealed]| 3D & |[hidden]| \phantom{00} & 
    |[hidden]| \phantom{00} & |[hidden]| \phantom{00} & 
    |[hidden]|\phantom{00}  & |[hidden]| \phantom{00} & 
    |[hidden]| \phantom{00} & |[revealed]| 6  & 
    |[hidden]| \phantom{00} & |[hidden]|\phantom{00}  \\
    |[hidden]| \phantom{00} & |[hidden]| \phantom{00} & 
    |[revealed]| 17  & |[hidden]| \phantom{00} & 
    |[hidden]| \phantom{00} & |[incorrect]| 3F  & 
    |[hidden]| \phantom{00} & |[hidden]| \phantom{00} & 
    |[hidden]| \phantom{00} & |[revealed]| 13  & 
    |[hidden]| \phantom{00} & |[hidden]| \phantom{00} \\
};

\node[steplabel] at (-1.0, -4.8) {Step 7};
\matrix (s7) [matrix of nodes, anchor=west,
    row sep=1pt, column sep=1pt,
    nodes={cell}
] at (0.8, -4.95) {
    |[hidden]| \phantom{00} & |[incorrect]| 3F & 
    |[revealed]| 3D & |[hidden]| \phantom{00} & 
    |[revealed]| 14 & |[incorrect]| 36 & 
    |[hidden]|\phantom{00}  & |[hidden]| \phantom{00} & 
    |[hidden]| \phantom{00} & |[revealed]| 6  & 
    |[incorrect]| D & |[revealed]| 37  \\
    |[revealed]| 1D & |[hidden]| \phantom{00} & 
    |[revealed]| 17  & |[hidden]| \phantom{00} & 
    |[revealed]| F & |[incorrect]| 3F  & 
    |[hidden]| \phantom{00} & |[incorrect]| 36 & 
    |[hidden]| \phantom{00} & |[revealed]| 13  & 
    |[incorrect]| 3A & |[revealed]| 12 \\
};

\node[steplabel] at (-1.0, -6.0) {Step 12};
\matrix (s12) [matrix of nodes, anchor=west,
    row sep=1pt, column sep=1pt,
    nodes={cell}
] at (0.8, -6.15) {
    |[revealed]| 3B & |[incorrect]| 3F & 
    |[revealed]| 3D & |[revealed]| 23 & 
    |[revealed]| 14 & |[incorrect]| 36 & 
    |[incorrect]| 23  & |[incorrect]| F & 
    |[revealed]| 18 & |[revealed]| 6  & 
    |[incorrect]| D & |[revealed]| 37  \\
    |[revealed]| 1D & |[incorrect]| 20 & 
    |[revealed]| 17  & |[revealed]| 14 & 
    |[revealed]| F & |[incorrect]| 3F  & 
    |[incorrect]| 1E & |[incorrect]| 36 & 
    |[revealed]| 2B & |[revealed]| 13  & 
    |[incorrect]| 3A & |[revealed]| 12 \\
};


\begin{scope}[shift={(-0.75, -7.2)}]
    \node[revealed, draw=black, line width=0.65pt, minimum width=0.6cm] (leg1) at (0, 0) {};
    \node[font=\scriptsize\sffamily\bfseries, right=2pt of leg1] {Correct};

    \node[incorrect, draw=black, line width=0.65pt, minimum width=0.6cm] (legX) at (2.5, 0) {};
    \node[font=\scriptsize\sffamily\bfseries, right=2pt of legX] {Incorrect};

    \node[hidden, draw=black, line width=0.65pt, minimum width=0.6cm] (leg2) at (4.8, 0) {};
    \node[font=\scriptsize\sffamily\bfseries, right=2pt of leg2] {Masked};

    \node[gtcell, draw=black, line width=0.65pt, minimum width=0.6cm] (leg3) at (7.1, 0) {};
    \node[font=\scriptsize\sffamily\bfseries, right=2pt of leg3] {Ground Truth};
\end{scope}

\coordinate (s0-out) at ($ (s0-1-12.east)!0.5!(s0-2-12.east) $);
\coordinate (s1-in) at ($ (s1-1-12.east)!0.5!(s1-2-12.east) $);

\begin{scope}[shift={(7.2, -1.95)}]
  \node[
      draw=black!100,
    line width=0.95pt,
    inner sep=4pt,
    font=\footnotesize\sffamily\bfseries,
    text=black,
    anchor=west,
    align=left
  ] (iterbox) at (0.5,0)
  {Module A};

\end{scope}

\coordinate (module-in) at ($(iterbox.north west)!0.28!(iterbox.south west)$);
\coordinate (module-out) at ($(iterbox.north west)!0.72!(iterbox.south west)$);
\draw[black, -{Latex[length=2.3mm,width=1.7mm]}, line width=0.95pt]
  (s0-out) -- (module-in);
\draw[black, -{Latex[length=2.3mm,width=1.7mm]}, line width=0.95pt]
  (module-out) -- (s1-in);

\end{tikzpicture}}
\caption{No parity-aware propagation (Module A only)}
\label{fig:tiny_no_mp_visualization}

\end{figure}

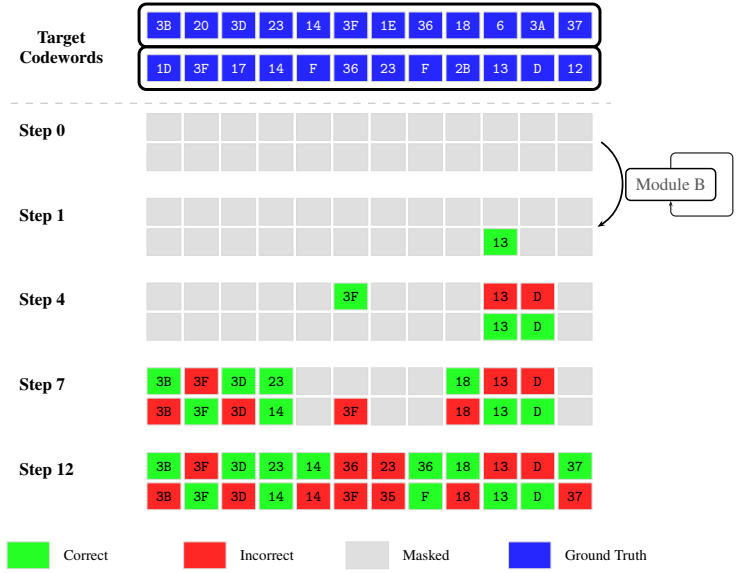
\begin{figure}[!htbp]
\centering
\resizebox{0.9\linewidth}{!}{\usetikzlibrary{backgrounds}

\begin{tikzpicture}[
    cell/.style={
        minimum width=0.48cm, 
        minimum height=0.38cm, 
        font=\scriptsize\sffamily\bfseries,
        draw=black, line width=0.65pt,
        inner sep=1pt,
    },
    revealed/.style={cell, fill=green!85, text=black},
    incorrect/.style={cell, fill=red!85, text=black},
    hidden/.style={cell, fill=gray!25, text=gray!50},
    steplabel/.style={font=\footnotesize\sffamily\bfseries, text=black, anchor=west},
    taulabel/.style={font=\tiny, text=gray, anchor=west},
    acclabel/.style={font=\tiny, text=blue!70, anchor=west},
    gtcell/.style={cell, fill=blue!85, text=white},
]

\node[steplabel, align=center] at (-1.0, 0) {Target\\Codewords};

\matrix (gt) [matrix of nodes, anchor=west,
    row sep=6pt,            
    column sep=1pt,
    nodes={cell}
] at (0.8, 0) {
    |[gtcell]| 3B & |[gtcell]| 20 & |[gtcell]| 3D & |[gtcell]| 23 & |[gtcell]| 14 & |[gtcell]| 3F & |[gtcell]| 1E & |[gtcell]| 36 & |[gtcell]| 18 & |[gtcell]| 6 & |[gtcell]| 3A & |[gtcell]| 37 \\
    |[gtcell]| 1D & |[gtcell]| 3F & |[gtcell]| 17 & |[gtcell]| 14 & |[gtcell]| F & |[gtcell]| 36 & |[gtcell]| 23 & |[gtcell]| F & |[gtcell]| 2B & |[gtcell]| 13 & |[gtcell]| D & |[gtcell]| 12 \\
};

\begin{pgfonlayer}{background}
  \node[
    fit=(gt-1-1)(gt-1-12),
    draw=black,
    line width=1.2pt,
    inner sep=3pt
  ] {};

  \node[
    fit=(gt-2-1)(gt-2-12),
    draw=black,
    line width=1.2pt,
    inner sep=3pt
  ] {};
\end{pgfonlayer}



\draw[black!65, dashed, line width=0.8pt] (-1.0, -0.8) -- (7.5, -0.8);

\node[steplabel] at (-1.0, -1.2) {Step 0};
\matrix (s0) [matrix of nodes, anchor=west,
    row sep=1pt, column sep=1pt,
    nodes={cell}
] at (0.8, -1.35) {
    |[hidden]| \phantom{00} & |[hidden]| \phantom{00} & |[hidden]| \phantom{00} & |[hidden]| \phantom{00} & |[hidden]| \phantom{00} & |[hidden]| \phantom{00} & |[hidden]|\phantom{00}  & |[hidden]| \phantom{00} & |[hidden]| \phantom{00} & |[hidden]|\phantom{00}  & |[hidden]| \phantom{00} & |[hidden]|\phantom{00}  \\
    |[hidden]| \phantom{00} & |[hidden]| \phantom{00} & |[hidden]|\phantom{00}  & |[hidden]| \phantom{00} & |[hidden]| \phantom{00} & |[hidden]|\phantom{00}  & |[hidden]| \phantom{00} & |[hidden]| \phantom{00} & |[hidden]| \phantom{00} & |[hidden]| \phantom{00} & |[hidden]| \phantom{00} & |[hidden]| \phantom{00} \\
};

\node[steplabel] at (-1.0, -2.4) {Step 1};
\matrix (s1) [matrix of nodes, anchor=west,
    row sep=1pt, column sep=1pt,
    nodes={cell}
] at (0.8, -2.55) {
    |[hidden]| \phantom{00} & |[hidden]| \phantom{00} & 
    |[hidden]| \phantom{00} & |[hidden]| \phantom{00} & 
    |[hidden]| \phantom{00} & |[hidden]| \phantom{00} & 
    |[hidden]|\phantom{00}  & |[hidden]| \phantom{00} & 
    |[hidden]| \phantom{00} & |[hidden]| \phantom{00}  & 
    |[hidden]| \phantom{00} & |[hidden]|\phantom{00}  \\
    |[hidden]| \phantom{00} & |[hidden]| \phantom{00} & 
    |[hidden]|\phantom{00}  & |[hidden]| \phantom{00} & 
    |[hidden]| \phantom{00} & |[hidden]|\phantom{00}  & 
    |[hidden]| \phantom{00} & |[hidden]| \phantom{00} & 
    |[hidden]| \phantom{00} & |[revealed]| 13 & 
    |[hidden]| \phantom{00} & |[hidden]| \phantom{00} \\
};

\node[steplabel] at (-1.0, -3.6) {Step 4};
\matrix (s4) [matrix of nodes, anchor=west,
    row sep=1pt, column sep=1pt,
    nodes={cell}
] at (0.8, -3.75) {
    |[hidden]| \phantom{00} & |[hidden]| \phantom{00} & 
    |[hidden]| \phantom{00} & |[hidden]| \phantom{00} & 
    |[hidden]| \phantom{00} & |[revealed]| 3F & 
    |[hidden]| \phantom{00}  & |[hidden]| \phantom{00} & 
    |[hidden]| \phantom{00} & |[incorrect]| 13  & 
    |[incorrect]| D & |[hidden]|\phantom{00}  \\
    |[hidden]| \phantom{00} & |[hidden]| \phantom{00} & 
    |[hidden]|\phantom{00}  & |[hidden]| \phantom{00} & 
    |[hidden]| \phantom{00} & |[hidden]|\phantom{00}  & 
    |[hidden]| \phantom{00} & |[hidden]| \phantom{00} & 
    |[hidden]| \phantom{00} & |[revealed]| 13 & 
    |[revealed]| D & |[hidden]| \phantom{00} \\
};

\node[steplabel] at (-1.0, -4.8) {Step 7};
\matrix (s7) [matrix of nodes, anchor=west,
    row sep=1pt, column sep=1pt,
    nodes={cell}
] at (0.8, -4.95) {
    |[revealed]| 3B & |[incorrect]| 3F & 
    |[revealed]| 3D  & |[revealed]| 23 &  
    |[hidden]| \phantom{00} & |[hidden]| \phantom{00} & 
    |[hidden]| \phantom{00}  & |[hidden]| \phantom{00} & 
    |[revealed]| 18 & |[incorrect]| 13  & 
    |[incorrect]| D & |[hidden]|\phantom{00}  \\
    |[incorrect]| 3B & |[revealed]| 3F & 
    |[incorrect]| 3D  & |[revealed]| 14 & 
    |[hidden]| \phantom{00} & |[incorrect]| 3F  & 
    |[hidden]| \phantom{00} & |[hidden]| \phantom{00} & 
    |[incorrect]| 18 & |[revealed]| 13 & 
    |[revealed]| D & |[hidden]| \phantom{00} \\
};

\node[steplabel] at (-1.0, -6.0) {Step 12};
\matrix (s12) [matrix of nodes, anchor=west,
    row sep=1pt, column sep=1pt,
    nodes={cell}
] at (0.8, -6.15) {
    |[revealed]| 3B & |[incorrect]| 3F & 
    |[revealed]| 3D  & |[revealed]| 23 &  
    |[revealed]| 14 & |[incorrect]| 36 & 
    |[incorrect]| 23  & |[revealed]| 36 & 
    |[revealed]| 18 & |[incorrect]| 13  & 
    |[incorrect]| D & |[revealed]| 37  \\
    |[incorrect]| 3B & |[revealed]| 3F & 
    |[incorrect]| 3D  & |[revealed]| 14 & 
    |[incorrect]| 14 & |[incorrect]| 3F  & 
    |[incorrect]| 35 & |[revealed]| F & 
    |[incorrect]| 18 & |[revealed]| 13 & 
    |[revealed]| D & |[incorrect]| 37 \\
};


\begin{scope}[shift={(-0.75, -7.2)}]
    \node[revealed, draw=black, line width=0.65pt, minimum width=0.6cm] (leg1) at (0, 0) {};
    \node[font=\scriptsize\sffamily\bfseries, right=2pt of leg1] {Correct};

    \node[incorrect, draw=black, line width=0.65pt, minimum width=0.6cm] (legX) at (2.5, 0) {};
    \node[font=\scriptsize\sffamily\bfseries, right=2pt of legX] {Incorrect};

    \node[hidden, draw=black, line width=0.65pt, minimum width=0.6cm] (leg2) at (4.8, 0) {};
    \node[font=\scriptsize\sffamily\bfseries, right=2pt of leg2] {Masked};

    \node[gtcell, draw=black, line width=0.65pt, minimum width=0.6cm] (leg3) at (7.1, 0) {};
    \node[font=\scriptsize\sffamily\bfseries, right=2pt of leg3] {Ground Truth};
\end{scope}

\coordinate (s0-out) at ($ (s0-1-12.east)!0.5!(s0-2-12.east) $);
\coordinate (s1-in) at ($ (s1-1-12.east)!0.5!(s1-2-12.east) $);

\begin{scope}[shift={(7.2, -1.95)}]
  \node[
      draw=black!100,
    line width=0.95pt,
    inner sep=4pt,
    font=\footnotesize\sffamily\bfseries,
    text=black,
    anchor=west,
    align=left
  ] (iterbox) at (0.5,0)
  {Module B};

\end{scope}

\coordinate (module-in) at ($(iterbox.north west)!0.28!(iterbox.south west)$);
\coordinate (module-out) at ($(iterbox.north west)!0.72!(iterbox.south west)$);
\draw[black, -{Latex[length=2.3mm,width=1.7mm]}, line width=0.95pt]
  (s0-out) -- (module-in);
\draw[black, -{Latex[length=2.3mm,width=1.7mm]}, line width=0.95pt]
  (module-out) -- (s1-in);

\end{tikzpicture}}
\caption{No demixing (Module B only)}
\label{fig:tiny_no_slot_visualization}

\end{figure}

\subsection{Auxiliary losses for demixing and parity}
\label{app:aux_losses}
We consider two auxiliary loss terms that impose global consistency through the training objective rather than through architectural constraints. 
Specifically, we evaluate: ($i$) a \emph{demixing loss} that penalizes cosine similarity between slot predictions to discourage slot collapse and promote output diversity; and ($ii$) a \emph{parity loss} that penalizes violations of the parity constraints defined by the parity-check matrix \(H\), i.e., \(Hx=0\) over \(\mathrm{GF}(Q)\).
These losses are weighted by hyperparameters \(\lambda_{\text{d}}\) and \(\lambda_{\text{p}}\), respectively, and are added to the cross-entropy objective during training.
\Cref{tab:app_lambda_sweep} reports results for the MDD baseline under this loss-only formulation.

\begin{table}[h]
\centering
\caption{\cider scales to $K=8$, and PRISM-style remasking further improves accuracy at higher loads.}
\label{tab:larger_K_combined}

\footnotesize
\setlength{\tabcolsep}{2.5pt}
\renewcommand{\arraystretch}{0.95}

\begin{tabular}{c|cc|cc}
\toprule
\multirow{2}{*}{\textbf{$K$}} &
\multicolumn{2}{c|}{\cider} &
\multicolumn{2}{c}{\textbf{CIDER + PRISM}} \\
\cmidrule(lr){2-3}\cmidrule(lr){4-5}
& \textbf{SER ($\downarrow$)} & \textbf{CER ($\downarrow$)}
& \textbf{SER ($\downarrow$)} & \textbf{CER ($\downarrow$)} \\
\midrule
2 & 0.0011 & 0.0073 & --     & --     \\
3 & 0.0006 & 0.0044 & --     & --     \\
4 & 0.0015 & 0.0058 & --     & --     \\
5 & 0.0048 & 0.0141 & --     & --     \\
\midrule
6 & 0.0149 & 0.0349 & \cellcolor{CIDERGreenFill!55}\textbf{0.0064} & \cellcolor{CIDERGreenFill!55}\textbf{0.0163} \\
7 & 0.0441 & 0.1006 & \cellcolor{CIDERGreenFill!55}\textbf{0.0096} & \cellcolor{CIDERGreenFill!55}\textbf{0.0247} \\
8 & 0.1339 & 0.2576 & \cellcolor{CIDERGreenFill!55}\textbf{0.0166} & \cellcolor{CIDERGreenFill!55}\textbf{0.0403} \\
\bottomrule
\end{tabular}
\end{table}

\subsection{Removing the first-reveal stabilization rule}
\label{app:no_first_reveal}

We ablate the first-reveal stabilization rule defined in Appendix~\Cref{app:maskdiff_infer}. Table~\ref{tab:larger_K_no_first} reports performance when we disable this rule and instead run the standard confidence-based unmasking schedule directly from the all-mask initialization. Without first-reveal, SER and CER degrade markedly as $K$ increases. This trend is more clearly seen by comparing Table~\ref{tab:larger_K_combined} with Table~\ref{tab:larger_K_no_first}: for $K=5$, SER/CER worsen from $0.0048/0.0141$ with first-reveal to $0.1697/0.3273$ without first-reveal. These results highlight that how the early reveals are distributed across row hypotheses is critical, and increasingly so in higher-load regimes.

\subsection{\texorpdfstring{Scaling to larger $K$ beyond the main setting}{Scaling to larger K beyond the main setting}}
\label{app:larger_k_results}

Table~\ref{tab:larger_K_combined} extends the evaluation to larger loads at fixed $(Q,L)=(64,12)$. Even in this higher-collision regime, \cider continues to provide meaningful decoding accuracy. For completeness, for $K=7$ and $K=8$ (without PRISM) the error rates are $\mathrm{SER}/\mathrm{CER}=0.0441/0.1006$ and $0.1339/0.2576$, respectively (Table~\ref{tab:larger_K_combined}).

\subsection{Runtime measurement protocol}
\label{app:runtime_protocol}

All wall-clock measurements were conducted using an NVIDIA GeForce RTX 3090 GPU (24GB) and Intel Core i5-14500 CPU.
For the main $K=2$ runtime table, inference time was averaged over 15,000 test samples for learned and BP-based methods, and over 1,000 samples for Top-$J$ exhaustive search. For the higher-load runtime study in Appendix~\Cref{app:runtime_k3}, we use 15,000 samples for \cider and 1,000 samples for FFT-BP and Top-$J$ due to the higher runtime of the classical baselines.

\subsection{\texorpdfstring{Runtime scaling to $K>2$}{Runtime scaling to K>2}}
\label{app:runtime_k3}

To check whether the runtime advantage persists beyond the $K=2$ setting, we repeat the wall-clock comparison at $K=3,4,5$ under the same $(Q,L)=(64,12)$ (Table~\ref{tab:app_full_results_time_k3}). 
We measure the wall-clock time of the FFT-BP variant.
The gap becomes even more pronounced: FFT-BP incurs substantially higher latency due to iterative Tanner-graph message passing, while Top-$J$ exhaustive search becomes quickly impractical even with a small per-slot candidate list. All experiments were conducted using the same environment as above. Inference time was averaged over 15,000 test samples for \cider, 1,000 samples for FFT-BP and Top-$J$ Exhaustive Search (Top 3 and Top 4 for $K=3$).

\subsection{Additional sparse-graph code families: PEG-LDPC and tree codes}
Progressive edge growth (PEG) constructs an LDPC Tanner graph by adding edges sequentially to increase girth and reduce short cycles. This differs in topology from the structured connectivity of the main LDPC benchmark.
\label{app:other_codes}

The main paper focuses on non-binary LDPC codes because Module~B operates on Tanner graphs. To test whether the gains depend on a particular code instance, we additionally evaluate PEG-LDPC and tree codes under the same shared evidence interface.

\begin{table}[h]
\centering
\footnotesize
\setlength{\tabcolsep}{4pt}
\renewcommand{\arraystretch}{1.05}
\caption{PEG-LDPC comparison on Tiny $(Q,L)=(64,12)$, $K=2$. Lower is better.}
\label{tab:app_peg_ldpc}
\begin{tabular}{l c c c}
\toprule
\textbf{Method} & \textbf{SER ($\downarrow$)} & \textbf{CER ($\downarrow$)} & \textbf{Time/sample ($\downarrow$)} \\
\midrule
\cellcolor{CIDERGreenFill!55}\textbf{\cider} & \cellcolor{CIDERGreenFill!55}\textbf{0.0004} & \cellcolor{CIDERGreenFill!55}\textbf{0.0023} & \cellcolor{CIDERGreenFill!55}\textbf{1.39 ms} \\
SIC-BP & 0.0015 & 0.0062 & 6.74 ms \\
FFT-BP & 0.0015 & 0.0062 & 12.04 ms \\
Top-2 & 0.0473 & 0.0497 & 2.27 ms \\
\bottomrule
\end{tabular}
\end{table}

\begin{table}[h]
\centering
\footnotesize
\setlength{\tabcolsep}{4pt}
\renewcommand{\arraystretch}{1.05}
\caption{Tree-code comparison on Tiny $(Q,L)=(64,12)$, $K=2$. Lower is better.}
\label{tab:app_tree_code}
\begin{tabular}{l c c c}
\toprule
\textbf{Method} & \textbf{SER ($\downarrow$)} & \textbf{CER ($\downarrow$)} & \textbf{Time/sample ($\downarrow$)} \\
\midrule
\cellcolor{CIDERGreenFill!55}\textbf{\cider} & \cellcolor{CIDERGreenFill!55}\textbf{0.0001} & \cellcolor{CIDERGreenFill!55}\textbf{0.0007} & \cellcolor{CIDERGreenFill!55}\textbf{1.08 ms} \\
SIC-BP & 0.0017 & 0.0065 & 6.82 ms \\
FFT-BP & 0.0017 & 0.0065 & 11.56 ms \\
Tree-code stitching decoder & 0.0474 & 0.0490 & 1.20 ms \\
Top-2 & 0.0474 & 0.0490 & 2.55 ms \\
\bottomrule
\end{tabular}
\end{table}

\subsection{Failure-mode diagnostics: slot overlap and parity violation}
\label{app:failure_modes}
To directly quantify the two intended failure modes of generic diffusion decoders in this setting, we measure ($i$) slot-overlap rate, i.e., how often different rows claim the same slot symbol evidence, and ($ii$) parity-violation rate, i.e., how often the decoded rows violate the code constraints.
The results in Table~\ref{tab:app_failure_modes} confirm that Module A primarily resolves overlap and Module B primarily resolves parity inconsistency.

\begin{table}[h]
\centering
\footnotesize
\setlength{\tabcolsep}{5pt}
\renewcommand{\arraystretch}{1.05}
\caption{Failure-mode diagnostics on Tiny $(Q,L)=(64,12)$, $K=2$. Rates are percentages over the test set.}
\label{tab:app_failure_modes}
\begin{tabular}{l c c c}
\toprule
\textbf{Model} & \textbf{Slot-overlap rate} & \textbf{Parity-violation rate} & \textbf{SER} \\
\midrule
MDD & 63.5\% & 99.9\% & 0.4201 \\
\cider w/o Module A & 42.6\% & 99.9\% & 0.4127 \\
\cider w/o Module B & 2.9\% & 100.0\% & 0.3991 \\
\cellcolor{CIDERGreenFill!55}\textbf{\cider} & \cellcolor{CIDERGreenFill!55}\textbf{1.5\%} & \cellcolor{CIDERGreenFill!55}\textbf{0.7\%} & \cellcolor{CIDERGreenFill!55}\textbf{0.0011} \\
\bottomrule
\end{tabular}
\end{table}

\subsection{Inference visualizations}
\label{sec:exp_viz}

Figures~\ref{fig:tiny_cider_visualization}--\ref{fig:tiny_no_slot_visualization} visualize intermediate decoded grids during iterative inference on the Tiny setting $(Q,L)=(64,12)$. Green/blue/red/gray indicate correct/ground-truth/incorrect/hidden tokens, respectively.
Figure~\ref{fig:tiny_cider_visualization} shows that \cider progressively resolves ambiguity and converges to two distinct, globally consistent codewords.
Figure~\ref{fig:tiny_mdd_visualization} shows the generic MDD baseline, which often struggles to stabilize refinement under unsourced ambiguity.
Figure~\ref{fig:tiny_no_slot_visualization} (no demixing, i.e., removing Module A) frequently exhibits \emph{duplicate-row collapse}, where both rows lock onto the same high-evidence symbols due to the lack of explicit row competition.
Figure~\ref{fig:tiny_no_mp_visualization} (no parity-aware propagation, i.e., removing Module B) often yields \emph{row swapping/mismatched assembly} across slots, where the two rows repeatedly exchange roles during refinement.

\FloatBarrier
\subsection{Protocol-level scalability: wrapping \cider inside a two-step random access protocol}
\label{app:protocol_scaling}

\paragraph{Motivation (scale up \emph{without} retraining a new monolithic model).}
\cider is a per-bin URA multiuser decoder: it solves \emph{one} URA instance in \emph{one} payload bin, and is trained for a bounded load.
In the single-bin experiments of \Cref{tab:larger_K_combined}, we trained and evaluated $K$-specific models up to
$K_{\max}=8$ active users in that bin.
Our next goal is to scale to \emph{much larger} total frame loads while \emph{reusing the same learned module}
(i.e., the same bank of $K$-specific \cider models, with PRISM for $K\ge 6$), rather than learning a brand-new architecture for large $K$.
To do this, we wrap \cider inside a simple two-step random access protocol that \emph{partitions} users into multiple smaller URA subproblems.

\paragraph{Two-step protocol = random ``binning'' by a preamble.}
At the beginning of each frame, every active device uniformly selects one of $\zeta$ preambles (think of $\zeta$ \emph{bins}) and transmits it in a short preamble phase.
All devices that chose the same preamble are assigned to the same payload bin in the subsequent payload phase.
Thus one frame induces $\zeta$ parallel or time-multiplexed payload bins, indexed by $\chi\in[\zeta]$.

Let $K_{\mathrm{tot}}$ denote the total number of active users in the frame (we use $K_{\mathrm{tot}}$ here to avoid confusion with the
per-slot activity vector $U^{(\ell)}$ in Appendix~\Cref{app:ura_formal_inner}).
The resulting bin loads are
\[
K_\chi \triangleq \big|\{u \in [K_{\mathrm{tot}}] : \text{device }u \text{ chose preamble } \chi\}\big|,
\qquad \sum_{\chi=1}^{\zeta} K_\chi = K_{\mathrm{tot}}.
\]
Under uniform preamble selection, $(K_1,\ldots,K_{\zeta})$ is multinomial with mean $K_{\mathrm{tot}}/\zeta$ per bin.
A convenient one-line approximation for intuition is the Poisson occupancy model:
\[
K_\chi \approx \mathrm{Poisson}(\varkappa),\qquad \varkappa = K_{\mathrm{tot}}/\zeta,
\]
so the \emph{overflow probability} is $p_{\mathrm{ov}}(\varkappa)=\Pr[\mathrm{Poisson}(\varkappa)>K_{\max}]$.
This overflow view organizes the two protocol-level tables. In 
\Cref{tab:protocol_scalability_main2}, the bin count grows with $K_{\mathrm{tot}}$ ($\zeta = \lceil K_{\mathrm{tot}}/4\rceil$), so the expected per-bin load is held near $\mathbb{E}[K_\chi]\approx 4$ and overflow is the dominant residual failure mode at every $K_{\mathrm{tot}}$. The reported CER tracks the user-overflow lower bound up to a small approximately constant additive gap that reflects residual per-bin decoding error.
In contrast, \Cref{tab:protocol_scalability_main} fixes the bin count at $\zeta=25$, so $\mathbb{E}[K_\chi] = K_{\mathrm{tot}}/25$ grows from $0.4$ to $4$ as $K_{\mathrm{tot}}$ sweeps from 10 to 100. At low-to-moderate $K_{\mathrm{tot}}$, overflow is essentially zero (e.g., $\Pr[\mathrm{Poisson}(\varkappa)>K_{\max}] < 10^{-3}$ for $K_{\mathrm{tot}} \le 60$) and the protocol-level CER is instead a mixture of per-load CERs $\mathrm{CER}(K_\chi)$ weighted by the bin-load distribution $K_\chi \sim \mathrm{Binomial}(K_{\mathrm{tot}}, 1/\zeta)$. Because $\mathrm{CER}(K_\chi)$ is itself mildly non-monotone in $K_\chi$ (\Cref{tab:larger_K_combined}), the protocol-level CER inherits a corresponding non-monotonicity in $K_{\mathrm{tot}}$, decreasing as the bin-load distribution shifts away from low loads before the overflow penalty kicks in at high $K_{\mathrm{tot}}$.

\paragraph{Receiver pipeline = decode each bin independently using the same learned module.}
For each bin $\chi$, the receiver runs the standard two-stage pipeline:
($i$) a fixed symbol-wise soft detector (slot-wise AMP--MMSE) produces evidence $S^{(\chi)}\in\mathbb{R}^{L\times Q}$ for that bin, and
($ii$) a multiuser decoder maps $S^{(\chi)}\mapsto \hat X^{(\chi)}\in[Q]^{K_\chi\times L}$.

To focus on protocol partitioning and bounded-load decoding (rather than load-estimation errors), we assume the receiver knows each
bin load $K_\chi$ (e.g., from preamble correlation/energy statistics). Handling errors in estimating $K_\chi$ is an orthogonal systems issue.

Crucially, we do \emph{not} train a single monolithic model that generalizes across variable $K$.
Instead, we deploy a bank of $K$-specific decoders, and in this protocol experiment we enable PRISM for the bin sizes with a trained quality head ($K_\chi\ge 6$):
\[
\text{for } K_\chi \in \{1,\ldots,K_{\max}\},\quad
\hat X^{(\chi)} \leftarrow \big[\mathrm{CIDER}\,(+\mathrm{PRISM})\big]_{K_\chi}\!\left(S^{(\chi)}\right).
\]
If a bin overflows ($K_\chi>K_{\max}$), we declare \emph{bin overflow} and treat that bin as an erasure at the protocol level
(i.e., those users are counted as failures).
The full wrapper pipeline is summarized in \Cref{fig:protocol_scalability_diagram}.

\begin{figure}[t]
\centering
\captionsetup{font=footnotesize}
\begin{tikzpicture}[
  font=\footnotesize\sffamily\bfseries,
  stagebox/.style={draw=black, line width=0.9pt, fill=white, align=center, inner sep=3pt, minimum height=10mm},
  decoder/.style={stagebox, line width=1.2pt, fill=black!8},
  flow/.style={-{Latex[length=2.4mm,width=1.8mm]}, line width=0.95pt, draw=black}
]
\node[stagebox, text width=23mm] (users) {Active users\\$K_{\mathrm{tot}}$};
\node[stagebox, text width=27mm, right=7mm of users] (preamble) {Choose one of $\zeta$ preambles};
\node[stagebox, text width=31mm, right=7mm of preamble] (bins) {$\zeta$ bins\\load $K_\chi$ in bin $\chi$};
\draw[flow] (users) -- (preamble);
\draw[flow] (preamble) -- (bins);

\node[stagebox, text width=25mm, below=14mm of users] (detector) {AMP--MMSE\\detector};
\node[stagebox, text width=22mm, right=5mm of detector] (evidence) {Evidence\\$S^{(\chi)}$};
\node[decoder, text width=31mm, right=5mm of evidence] (cider) {$\big[\mathrm{CIDER}\,(+\mathrm{PRISM})\big]_{K_\chi}$};
\node[stagebox, text width=22mm, right=5mm of cider] (decoded) {Decoded set\\$\hat X^{(\chi)}$};
\draw[flow] (bins.south) -- ++(0,-6mm) -| (detector.north);
\draw[flow] (detector) -- (evidence);
\draw[flow] (evidence) -- (cider);
\draw[flow] (cider) -- (decoded);

\node[stagebox, text width=29mm, below=7mm of decoded] (frame) {Aggregate bins\\frame message set};
\draw[flow] (decoded) -- (frame);
\end{tikzpicture}
\caption{Two-step protocol wrapper. A preamble partitions $K_{\mathrm{tot}}$ active users into $\zeta$ bins. Each supported bin is decoded independently with its load-specific \cider model (with PRISM remasking for $K_\chi\ge 6$); bins above $K_{\max}$ count as erasures.}
\label{fig:protocol_scalability_diagram}
\end{figure}

\FloatBarrier

\begin{table}[h]
\centering
\caption{Stochastic binning scales to large $K_{\mathrm{tot}}$ by decomposing into bounded-load subproblems. Evaluated with a fixed number of bins ($\zeta=25$); SIC-BP style decoding at comparable $K_{\mathrm{tot}}$ would be prohibitively slow.}
\label{tab:protocol_scalability_main}
\footnotesize
\setlength{\tabcolsep}{6pt}
\renewcommand{\arraystretch}{1.0}
\begin{tabular}{c|cc}
\toprule
\textbf{$K_{\mathrm{tot}}$} & \textbf{SER ($\downarrow$)} & \textbf{CER ($\downarrow$)} \\
\midrule
10 & 0.0064 & 0.0417 \\
20 & 0.0043 & 0.0398 \\
30 & 0.0036 & 0.0369 \\
40 & 0.0030 & 0.0258 \\
50 & 0.0030 & 0.0207 \\
60 & 0.0055 & 0.0189 \\
70 & 0.0095 & 0.0205 \\
80 & 0.0174 & 0.0278 \\
90 & 0.0301 & 0.0394 \\
100 & 0.0525 & 0.0619 \\
\bottomrule
\end{tabular}
\end{table}

\begin{table}[h]
\centering
\caption{Stochastic binning scales to many more users by decomposing into bounded-load subproblems, whereas running SIC-style decoding at comparable $K_{\mathrm{tot}}$ would be prohibitively slow. Evaluated for flexible number of bins $\zeta = \lceil K_{\mathrm{tot}} / 4 \rceil$ so that $\mathbb{E}[K_\chi] \approx 4$.}
\label{tab:protocol_scalability_main2}
\footnotesize
\setlength{\tabcolsep}{6pt}
\renewcommand{\arraystretch}{1.0}
\begin{tabular}{c|cc}
\toprule
\textbf{$K_{\mathrm{tot}}$} & \textbf{SER ($\downarrow$)} & \textbf{CER ($\downarrow$)} \\
\midrule
10  & 0.0027 & 0.0108 \\
20  & 0.0225 & 0.0322 \\
30  & 0.0219 & 0.0310 \\
40  & 0.0423 & 0.0517 \\
50  & 0.0389 & 0.0486 \\
60  & 0.0493 & 0.0590 \\
70  & 0.0423 & 0.0513 \\
80  & 0.0502 & 0.0598 \\
90  & 0.0467 & 0.0561 \\
100 & 0.0525 & 0.0619 \\
\bottomrule
\end{tabular}
\end{table}

\paragraph{Experiment setup and metric (protocol-level SER/CER).}
We simulate $K_{\mathrm{tot}}$ active users per frame, assign each user to a bin by uniform preamble selection, and decode each bin using the corresponding $K_\chi$-specific \cider decoder when $K_\chi \le K_{\max}$ (with additional PRISM for $K_\chi\ge 6$). Bins with $K_\chi > K_{\max}$ are treated as overflow erasures. Frame-level SER/CER are computed by aggregating errors over all $K_{\mathrm{tot}}$ users, counting users in overflow bins as errors. We evaluate two binning regimes: \Cref{tab:protocol_scalability_main} reports results with a fixed bin count ($\zeta=25$), and \Cref{tab:protocol_scalability_main2} reports results with $\zeta$ scaled with $K_{\mathrm{tot}}$ to maintain average per-bin load $\mathbb{E}[K_\chi]\approx 4$.

\paragraph{Why this matters.}
This wrapping experiment shows that \cider is not only a stand-alone multiuser decoder, but also a reusable \emph{module} inside a scalable access stack:
we can support $K_{\mathrm{tot}}\gg K_{\max}$ users per frame while keeping the \emph{worst-case per-bin} decoding cost bounded
(by choosing $\zeta$ and enforcing the overflow rule).

\section{Additional controlled evaluations}
\label{app:response_results}
The following experiments use the same within-bin shared-codebook decoding objective and distinguish tests that reuse a checkpoint from those trained for a target setting. They supplement the main fixed-load, calibrated-evidence experiments; the error and latency numbers are tied to their stated implementations.

\subsection{Estimated and mismatched per-bin load}
\label{app:load_robustness}
The Bank mode selects a $K$-specific model using an external energy-based load estimate $\hat K$. Single instead runs the $K=8$ model for every bin, deduplicates and ranks its rows by summed slot-wise log-evidence, and retains the top $\hat K$ rows. Both modes use the same estimator and require no retraining. Missed-detection rate (MDR) and false-alarm rate (FAR) count errors after rectangular row matching. Single uses one checkpoint but always pays the $K=8$ inference cost.
The per-load comparison is reported in the main text (Table~\ref{tab:estimated_load}).

A deliberate one-row error in the load estimate quantifies the asymmetry: $\hat K=K-1$ necessarily misses at least one of $K$ messages, whereas $\hat K=K+1$ necessarily contributes at least one false alarm among $K+1$ reported rows. Rectangular matching counts both errors.
\begin{table}[h]
\centering\small
\setlength{\tabcolsep}{5pt}
\caption{Load-mismatch sweep. Set error is (misses + false alarms)/$(K+\hat K)$.}
\label{tab:load_mismatch}
\begin{tabular}{cccc}
\toprule
True $K$ & $\hat K=K-1$: MDR & $\hat K=K$: MDR & $\hat K=K+1$: MDR/FAR/set error \\
\midrule
3 & .469 & .004 & .001/.250/.143 \\
4 & .291 & .004 & .001/.200/.112 \\
5 & .244 & .017 & .002/.167/.092 \\
6 & .212 & .018 & .002/.143/.078 \\
7 & .205 & .026 & .008/.125/.070 \\
\bottomrule
\end{tabular}
\end{table}
The estimator is therefore part of the receiver rather than an oracle assumption.

\subsection{Evidence, channel, graph, and length shifts}
\label{app:response_robustness}
We test the $L=12,K=2$ checkpoint at $E_b/N_0=10$ dB without weight updates, using 10{,}000 examples per sensing family and 2{,}000 per non-AMP detector. The in-distribution CER is .0073. Non-AMP outputs are converted to the same posterior statistic as the AMP evidence. The measured shifts are reported in \Cref{tab:evidence_shifts}; the uncompensated fading comparison appears in (\Cref{tab:fading_stress}). \par\begin{table}[!htbp]\centering\footnotesize\setlength{\tabcolsep}{4pt}\renewcommand{\arraystretch}{1.02}\caption{Sensitivity of \cider to symbol-wise soft detector mismatch on Tiny $(Q,L)=(64,12)$, $K=2$, without retraining. $I_{\mathrm{AMP}}$ is the number of detector iterations.}\label{tab:detector_mismatch}\begin{tabular}{@{}llcc@{}}\toprule \textbf{Mismatch type} & \textbf{Condition} & \textbf{SER ($\downarrow$)} & \textbf{CER ($\downarrow$)} \\ \midrule None (matched) & SNR $=-0.79$ dB, $I_{\mathrm{AMP}}=10$ & 0.0011 & 0.0073 \\ More AMP iters & SNR $=-0.79$ dB, $I_{\mathrm{AMP}}=20$ & 0.0011 & 0.0075 \\ Reduced AMP iters & SNR $=-0.79$ dB, $I_{\mathrm{AMP}}=5$ & 0.0012 & 0.0078 \\ SNR mismatch & SNR $=-2.79$ dB, $I_{\mathrm{AMP}}=10$ & 0.0063 & 0.0318 \\ SNR mismatch & SNR $=+1.21$ dB, $I_{\mathrm{AMP}}=10$ & 0.0008 & 0.0054 \\ SNR mismatch & SNR $=-4.79$ dB, $I_{\mathrm{AMP}}=10$ & 0.1020 & 0.3087 \\ \bottomrule\end{tabular}\end{table}
\begin{table}[h]
\centering\footnotesize
\setlength{\tabcolsep}{4pt}
\caption{Additive-noise, sensing, detector, and evidence-calibration shifts for one fixed checkpoint. Values in each CER cell correspond to the listed conditions in order.}
\label{tab:evidence_shifts}
\begin{tabular}{p{2.4cm}p{7.2cm}p{3.2cm}}
\toprule
Shift & Conditions & CER \\
\midrule
Sensing & Complex Hadamard / QPSK Hadamard / QR Gaussian & .0113 / .0152 / .0164 \\
Detector & Matched filter / LMMSE / whitened MAP & .0080 / .0070 / .0083 \\
Calibration & Scale $c=.25/.5/2/4$ ($-6/-3/+3/+6$ dB) & .0119 / .0093 / .0088 / .0120 \\
Noise & Gaussian / uniform / Student-$t$ / impulsive & .0073 / .0077 / .0082 / .0090 \\
\bottomrule
\end{tabular}
\end{table}
These moderate shifts preserve a comparable evidence interface; they do not establish robustness to all channel models. \par\begin{table}[!htbp]\centering\footnotesize\setlength{\tabcolsep}{4pt}\renewcommand{\arraystretch}{1.02}\caption{Uncompensated fading with unknown per-user gains at $L=12,K=2$ and $E_b/N_0=10$ dB. No channel estimation or fading-aware retraining is used; lower CER is better.}\label{tab:fading_stress}\begin{tabular}{@{}lccccc@{}}\toprule Decoder & AWGN & Rician $+10$ dB & Rician $0$ dB & Rician $-10$ dB & Rayleigh \\ \midrule SIC-BP & .0078 & .0156 & .0465 & .0792 & .0813 \\ \cider & .0073 & .0343 & .3178 & .4037 & .4070 \\ \bottomrule\end{tabular}\end{table}

The graph and length transfer results are reported in the main text (\Cref{tab:response_transfer}). PEG-LDPC is a progressive-edge-growth Tanner graph, constructed to increase girth and suppress short cycles; the main LDPC benchmark uses structured regular connectivity designed to avoid short cycles and separately sampled nonzero field coefficients.

At $L=72,96$ the transfer latency is 17.57/34.90 ms; the length sweep starts at 3.27 ms for $L=12$. These results support short-block portability with increasing error and compute cost. A separate capacity check on the same $L=12$ task found CER .0011 for a larger $L=48$ model used zero-shot, versus .0073 for the targeted $L=12$ model. This is consistent with capacity being one possible contributor but does not isolate it. The main length sweep compares separately constructed finite graphs and scale-specific models at fixed rate $1/3$; payload grows with $L$ and the exact-row CER criterion becomes stricter, so its non-monotonicity does not isolate a coding-gain effect.

\subsection{Baseline optimization and non-diffusion controls}
\label{app:response_fairness}
We batched and vectorized SIC-BP, FFT-BP, and Top-$J$ on the same RTX 3090 used for \cider. For this comparison, BP and Top-$J$ timing is synchronized and uses FP32 arithmetic, while the deployed \cider is measured in 16-bit mixed precision on the same GPU. BP folds rows into the batch, precomputes extrinsic structure, avoids per-iteration host synchronization, and exits after a valid syndrome. Top-$J$ enumerates in parallel with memory chunking. These measurements are implementation-specific and do not imply an operation-count advantage over FFT-BP.
For \cider, the reported accuracy uses the original sampler, while optimized latency uses a vectorized sampler. The latter can break tied confidence scores differently and is therefore not bit-identical; a 512-frame Tiny check found similar CER despite differing decoded frames.

The per-scale same-GPU latency comparison is reported in the main text (Table~\ref{tab:main_results_classical}).

\begin{table}[h]
\centering\small
\setlength{\tabcolsep}{5pt}
\caption{Amortized batch-128 time (ms/sample) and reserved GPU memory (MB), ordered SIC-BP / FFT-BP / \cider. The memory comparison exposes the larger short-length footprint of the learned decoder.}
\label{tab:response_gpu_batch}
\begin{tabular}{ccc}
\toprule
$L$ & Batch-128 time (ms/sample) & Reserved GPU memory (MB) \\
\midrule
12 & .72 / 1.23 / .67 & 32.7 / 23.2 / 58.4 \\
18 & 1.79 / 3.22 / 1.57 & 57.8 / 48.4 / 83.7 \\
24 & 14.25 / 25.77 / 3.63 & 90.8 / 81.7 / 97.6 \\
48 & 56.07 / 101.97 / 12.26 & 316.6 / 307.2 / 168.7 \\
\bottomrule
\end{tabular}
\end{table}
\begin{table}[h]
\centering\small
\setlength{\tabcolsep}{5pt}
\caption{Reported operating-point latency in ms/sample for CER $\leq .05$. Floor denotes a method that did not reach the target CER; the $>60$ s Top-$J$ entries exceeded the time cap. \cider time includes remasking where deployed.}
\label{tab:response_gpu_operating}
\begin{tabular}{lccc}
\toprule
Setting & SIC-BP & Top-$J$ & \cider \\
\midrule
$K=4,L=12$ & 90.9 & $>60$ s ($J\geq6$) & 10.19 \\
$K=7,L=12$ & 525 & $>60$ s & 31.90 \\
$K=8,L=12$ & Floor: .26 & $>60$ s & 63.00 \\
$K=2,L=18$ & 14.95 & 1197.9 & 3.12 \\
$K=2,L=24$ & 89.81 & $>60$ s ($J\geq3$) & 4.31 \\
$K=2,L=48$ & Floor: .26 & $>60$ s & 10.08 \\
\bottomrule
\end{tabular}
\end{table}

For learned baselines, two training profiles were selected using seed-0 validation SER (CER as tie-break), without test-set selection. NBP chose cosine decay with no warmup; Tanner-Attention chose a 10-epoch warmup followed by cosine decay. Both used the same 10-iteration AMP data at $Q=64,L=12,K=2$, a 70{,}000/15{,}000/15{,}000 train/validation/test split, learning rate $10^{-3}$, batch size 128, and 100 epochs. Weight decay was $10^{-2}$ for no-warmup NBP and Tanner-Attention, and $10^{-4}$ for warmup NBP. NBP, Tanner-Attention, and parameter-matched MDD ($D=192$, four layers and four heads) were evaluated over five seeds on the same test split with Hungarian matching and 16-bit mixed precision. Their mean CERs remain near one; this conclusion is specific to the tested architectures and training profiles.
\begin{table}[h]
\centering\small
\caption{Five-seed neural baseline evaluation (mean $\pm$ standard deviation).}
\label{tab:response_neural}
\begin{tabular}{lcc}
\toprule
Model & SER & CER \\
\midrule
NBP (3.81M parameters) & $.3988\pm.0006$ & $.9995\pm.0002$ \\
Tanner-Attention (3.61M) & $.4022\pm.0008$ & $.9995\pm.0001$ \\
MDD (3.57M) & $.4235\pm.0061$ & $.9995\pm.0001$ \\
\bottomrule
\end{tabular}
\end{table}
The matched-module controls in \Cref{tab:app_full_results_ablation_main} separate masked refinement from structural iteration: one-shot A+B has SER/CER .1013/.3979, one-update iterative A+B .1039/.4093, and twelve-update iterative A+B .3863/.9989, versus .0011/.0073 for \cider. Generic masked diffusion without A+B has .4201/.9996. The existing D3PM-style no-diffusion one-shot control gives .4043/.9996, and loss-only demixing/parity regularization yields CER 1.0000. These results support the joint use of masked refinement and the two structural modules, without claiming that all possible iterative decoders have been ruled out.

\subsection{Propagation depth, remasking cost, and collision failures}
\label{app:response_tradeoffs}
The implementation uses $N_{\mathrm{layer}}=4$ structured layers for Tiny; each layer applies two Tanner passes, giving eight passes per masked-refinement step. The following CER sweep separates this propagation depth from the number $T$ of refinement steps. The default Tiny point is eight Tanner passes and $T=12$.
\begin{table}[h]
\centering\small
\setlength{\tabcolsep}{4pt}
\caption{CER by Tanner passes per refinement step (rows) and masked-refinement steps $T$ (columns).}
\label{tab:response_depth}
\begin{tabular}{c|cccccc}
\toprule
Passes $\backslash$ $T$ & 4 & 8 & 12 & 16 & 20 & 24 \\
\midrule
2 & .5388 & .0497 & .0155 & .0130 & .0126 & .0127 \\
4 & .4390 & .0258 & .0104 & .0099 & .0101 & .0101 \\
8 & .3025 & .0391 & .0073 & .0067 & .0070 & .0074 \\
\bottomrule
\end{tabular}
\end{table}
\begin{table}[h]
\centering\small
\caption{Quality-guided remasking at higher loads. For PRISM-on rows, time includes scoring, row selection, and triggered decoding passes; all times are ms per sample on the RTX 3090.}
\label{tab:response_remasking}
\begin{tabular}{ccccc}
\toprule
$K$ & $T$ & PRISM & SER/CER & Time \\
\midrule
2 & 12 & off & .0011/.0073 & 1.51 \\
3 & 24 & off & .0006/.0044 & 5.47 \\
4 & 42 & off & .0015/.0058 & 10.19 \\
5 & 60 & off & .0048/.0141 & 15.29 \\
6 & 50 & off/on & .0149/.0349 $\to$ .0064/.0163 & 17.40/19.79 \\
7 & 62 & off/on & .0441/.1006 $\to$ .0096/.0247 & 24.31/31.90 \\
8 & 74 & off/on & .1339/.2576 $\to$ .0166/.0403 & 30.77/63.00 \\
\bottomrule
\end{tabular}
\end{table}
The benefit grows with load but adds latency, especially at $K=8$. A separate $K=2$ collision sweep uses 5{,}000 samples per cell and fixes the number of shared symbols between the two transmitted length-12 codewords. \Cref{tab:response_collisions} reports both symbol and codeword errors.
\begin{table}[!htbp]
\centering\small
\setlength{\tabcolsep}{6pt}
\caption{Collision-pattern failure analysis at $K=2,L=12$. Lower error is better.}
\label{tab:response_collisions}
\begin{tabular}{cccc}
\toprule
Shared symbols & Hamming distance & \cider SER/CER & SIC-BP SER/CER \\
\midrule
0 & 12 & .0005/.0030 & .0040/.0155 \\
1 & 11 & .0026/.0248 & .0067/.0252 \\
2 & 10 & .0199/.0823 & .0124/.0543 \\
3 & 9 & .0593/.2870 & .0060/.0347 \\
\bottomrule
\end{tabular}
\end{table}
Thus \cider leads at low overlap, but SIC-BP is stronger when the two codewords share two or three symbols.

\FloatBarrier
\section{Limitations}
\label{app:limitations}
\cider is studied as a shared-codebook URA outer decoder that consumes a slot-wise evidence matrix. The main receiver uses a fixed AMP front end and synthetic sparse-graph codes under a calibrated AWGN model; the detector, sensing, calibration, and additive-noise shifts above are controlled tests rather than an end-to-end learned receiver evaluation. Unknown per-user gains without channel estimation sharply degrade CER in the fading stress test, so a fading-aware front end or retraining would be needed there. High codeword overlap also exposes a failure mode relative to SIC-BP.

The decoder must know or estimate per-bin load. The energy estimator is imperfect, and underestimating load incurs missed messages; the Single $K=8$ checkpoint supports variable loads only with the $K=8$ compute cost. The model uses sparse parity-check structure, and graph-specific or length-specific training still improves accuracy over zero-shot transfer. The $L=72,96$ transfer tests use only 256 samples each and do not establish reliable asymptotic scaling. Protocol-level operation beyond $K_{\max}=8$ relies on stochastic binning and overflow handling.

Our latency and memory comparisons are measurements of the reported implementations on an RTX 3090, not hardware-independent complexity rankings. Quality-guided remasking improves higher-load accuracy but adds scoring, selection, and repeated decoding passes; at $K=8$ its measured time increases from 30.77 to 63.00 ms/sample. The optimized GPU classical baselines narrow some gaps, and \cider uses more reserved GPU memory at the shortest lengths.

\section*{AI Use Statement}
Generative AI tools assisted with drafting and editing this camera-ready revision from the authors' manuscript and rebuttal. The experimental results are author-provided, and the authors take responsibility for checking the final text, data, and citations.

\section{Impact Statement}
This paper addresses the reliable recovery of many short, uncoordinated messages from a single noisy superposition. This decoding problem arises in large-scale wireless access when coordination overhead is undesirable or infeasible. Improving recovery accuracy and latency can reduce retransmissions and access delays, which may translate into better system reliability and lower energy use in battery-constrained deployments. Beyond wireless communication, the approach frames the task as conditional discrete infilling under strong global constraints. This structure also appears in combinatorial inference problems where local evidence must be assembled into globally consistent solutions, so the methodology may generalize to other set-structured or constraint-satisfaction settings. Responsible deployment should pair such methods with appropriate access control, auditing, and governance. Potential negative impacts include misuse in unauthorized access systems or deployment under poorly characterized channels, which could lead to unfair access failures or degraded reliability.

\clearpage
\section*{NeurIPS Paper Checklist}

\begin{enumerate}

\item {\bf Claims}
    \item[] Question: Do the main claims made in the abstract and introduction accurately reflect the paper's contributions and scope?
    \item[] Answer: \answerYes{} 
    \item[] Justification: The abstract and introduction accurately describe \cider as a learned masked-diffusion outer decoder and limit the claims to the evaluated URA/joint multiuser decoding setting under a shared AMP evidence interface. The stated reliability, runtime, module-ablation, and scaling claims are supported by the corresponding experiments.

\item {\bf Limitations}
    \item[] Question: Does the paper discuss the limitations of the work performed by the authors?
    \item[] Answer: \answerYes{} 
    \item[] Justification: The paper includes a separate Limitations section in Appendix~\Cref{app:limitations}. 
    \begin{itemize}
        \item The answer \answerNA{} means that the paper has no limitation while the answer \answerNo{} means that the paper has limitations, but those are not discussed in the paper. 
        \item The authors are encouraged to create a separate ``Limitations'' section in their paper.
        \item The paper should point out any strong assumptions and how robust the results are to violations of these assumptions (e.g., independence assumptions, noiseless settings, model well-specification, asymptotic approximations only holding locally). The authors should reflect on how these assumptions might be violated in practice and what the implications would be.
        \item The authors should reflect on the scope of the claims made, e.g., if the approach was only tested on a few datasets or with a few runs. In general, empirical results often depend on implicit assumptions, which should be articulated.
        \item The authors should reflect on the factors that influence the performance of the approach. For example, a facial recognition algorithm may perform poorly when image resolution is low or images are taken in low lighting. Or a speech-to-text system might not be used reliably to provide closed captions for online lectures because it fails to handle technical jargon.
        \item The authors should discuss the computational efficiency of the proposed algorithms and how they scale with dataset size.
        \item If applicable, the authors should discuss possible limitations of their approach to address problems of privacy and fairness.
        \item While the authors might fear that complete honesty about limitations might be used by reviewers as grounds for rejection, a worse outcome might be that reviewers discover limitations that aren't acknowledged in the paper. The authors should use their best judgment and recognize that individual actions in favor of transparency play an important role in developing norms that preserve the integrity of the community. Reviewers will be specifically instructed to not penalize honesty concerning limitations.
    \end{itemize}

\item {\bf Theory assumptions and proofs}
    \item[] Question: For each theoretical result, does the paper provide the full set of assumptions and a complete (and correct) proof?
    \item[] Answer: \answerNA{} 
    \item[] Justification: The paper does not present new formal theoretical results, such as theorems, lemmas, or propositions requiring proof. The mathematical content consists of problem definitions, model equations, decoding objectives, metric definitions, and computational-complexity derivations. The complexity assumptions and derivations are provided explicitly in Appendix~\Cref{app:complexity_derivation}.

    \item {\bf Experimental result reproducibility}
    \item[] Question: Does the paper fully disclose all the information needed to reproduce the main experimental results of the paper to the extent that it affects the main claims and/or conclusions of the paper (regardless of whether the code and data are provided or not)?
    \item[] Answer: \answerYes{} 
    \item[] Justification: The paper provides the experimental setup, data-generation pipeline, channel model, sensing matrix construction, fixed AMP–MMSE inner detector, LDPC code scales, training and inference schedules, model sizes, evaluation metrics, row-matching procedure, baseline definitions, and runtime measurement protocol in the main text and appendices.

\item {\bf Open access to data and code}
    \item[] Question: Does the paper provide open access to the data and code, with sufficient instructions to faithfully reproduce the main experimental results, as described in supplemental material?
    \item[] Answer: \answerYes{} 
    \item[] Justification: The code is made available through the public repository link in the abstract.

\item {\bf Experimental setting/details}
    \item[] Question: Does the paper specify all the training and test details (e.g., data splits, hyperparameters, how they were chosen, type of optimizer) necessary to understand the results?
    \item[] Answer: \answerYes{} 
    \item[] Justification: The main text summarizes the experimental setting, including the fixed AMP evidence interface, evaluated LDPC scales, compared decoder families, training/inference protocol, and evaluation metrics. Full details are provided in the appendix, including data generation, train/validation/test splits, sensing matrix construction, AMP--MMSE evidence construction, code parameters, optimizer, learning rate schedule, batch size, number of epochs, EMA, precision, mask-ratio sampling, inference schedules, model sizes, and baseline architectures.
    \item[] Guidelines:
    \begin{itemize}
        \item The answer \answerNA{} means that the paper does not include experiments.
        \item The experimental setting should be presented in the core of the paper to a level of detail that is necessary to appreciate the results and make sense of them.
        \item The full details can be provided either with the code, in the appendix, or as supplemental material.
    \end{itemize}

\item {\bf Experiment statistical significance}
    \item[] Question: Does the paper report error bars suitably and correctly defined or other appropriate information about the statistical significance of the experiments?
    \item[] Answer: \answerNo{} 
    \item[] Justification: The current version reports point estimates on held-out test sets but does not include error bars, confidence intervals, or significance tests. Although the claims are supported across multiple experimental settings, statistical uncertainty is not explicitly quantified.

\item {\bf Experiments compute resources}
    \item[] Question: For each experiment, does the paper provide sufficient information on the computer resources (type of compute workers, memory, time of execution) needed to reproduce the experiments?
    \item[] Answer: \answerYes{} 
    \item[] Justification: The paper reports the computing environment and the runtime measurement protocol. Experiments were run on a single NVIDIA GeForce RTX 3090 GPU with 24GB of memory and an Intel Core i5-14500 CPU. Sufficient information about compute is described in Section~\ref{app:compute}.
    \item[] Guidelines:
    \begin{itemize}
        \item The answer \answerNA{} means that the paper does not include experiments.
        \item The paper should indicate the type of compute workers CPU or GPU, internal cluster, or cloud provider, including relevant memory and storage.
        \item The paper should provide the amount of compute required for each of the individual experimental runs as well as estimate the total compute. 
        \item The paper should disclose whether the full research project required more computing than the experiments reported in the paper (e.g., preliminary or failed experiments that didn't make it into the paper). 
    \end{itemize}
    
\item {\bf Code of ethics}
    \item[] Question: Does the research conducted in the paper conform, in every respect, with the NeurIPS Code of Ethics \url{https://neurips.cc/public/EthicsGuidelines}?
    \item[] Answer: \answerYes{} 
    \item[] Justification: The authors have reviewed the NeurIPS Code of Ethics.

\item {\bf Broader impacts}
    \item[] Question: Does the paper discuss both potential positive societal impacts and negative societal impacts of the work performed?
    \item[] Answer: \answerYes{} 
    \item[] Justification: The paper includes an Impact Statement in the Appendix.
    
\item {\bf Safeguards}
    \item[] Question: Does the paper describe safeguards that have been put in place for responsible release of data or models that have a high risk for misuse (e.g., pre-trained language models, image generators, or scraped datasets)?
    \item[] Answer: \answerNA{} 
    \item[] Justification: The paper does not release pretrained generative models, scraped datasets, or other high-risk artifacts for misuse.

\item {\bf Licenses for existing assets}
    \item[] Question: Are the creators or original owners of assets (e.g., code, data, models), used in the paper, properly credited and are the license and terms of use explicitly mentioned and properly respected?
    \item[] Answer: \answerNA{} 
    \item[] Justification:  The paper does not use existing datasets, pretrained models, or third-party research assets as experimental inputs.

\item {\bf New assets}
    \item[] Question: Are new assets introduced in the paper well documented and is the documentation provided alongside the assets?
    \item[] Answer: \answerYes{} 
    \item[] Justification: The paper introduces and releases code for the CIDER decoder and the synthetic URA/LDPC simulation pipeline through the public repository. The repository includes documentation for reproducing the main experiments, including dataset generation, model configurations, training and inference scripts, and evaluation procedures.

\item {\bf Crowdsourcing and research with human subjects}
    \item[] Question: For crowdsourcing experiments and research with human subjects, does the paper include the full text of instructions given to participants and screenshots, if applicable, as well as details about compensation (if any)? 
    \item[] Answer: \answerNA{} 
    \item[] Justification: The paper does not involve crowdsourcing, human-subject experiments, user studies, or data collected from human participants.

\item {\bf Institutional review board (IRB) approvals or equivalent for research with human subjects}
    \item[] Question: Does the paper describe potential risks incurred by study participants, whether such risks were disclosed to the subjects, and whether Institutional Review Board (IRB) approvals (or an equivalent approval/review based on the requirements of your country or institution) were obtained?
    \item[] Answer: \answerNA{} 
    \item[] Justification: The paper does not involve crowdsourcing, user studies, human-subject experiments, or data collected from human participants.

\item {\bf Declaration of LLM usage}
    \item[] Question: Does the paper describe the usage of LLMs if it is an important, original, or non-standard component of the core methods in this research? Note that if the LLM is used only for writing, editing, or formatting purposes and does \emph{not} impact the core methodology, scientific rigor, or originality of the research, declaration is not required.
    \item[] Answer: \answerNA{} 
    \item[] Justification: The core method development and experiments do not use LLMs as an important, original, or non-standard component. Any LLM assistance, if used, was limited to writing, editing, or formatting.

\end{enumerate}

\end{document}